%% file: paper.tex
\documentclass[fleqn,usenatbib]{mnras}

\usepackage[T1]{fontenc}

\DeclareRobustCommand{\VAN}[3]{#2}
\let\VANthebibliography\thebibliography
\def\thebibliography{\DeclareRobustCommand{\VAN}[3]{##3}\VANthebibliography}

\usepackage{graphicx}	% Including figure files
\usepackage{amsmath}	% Advanced maths commands
\usepackage{amssymb}	% Extra maths symbols
\usepackage{ae,aecompl}
\usepackage{lscape}
\usepackage{longtable}
\usepackage{booktabs}
\usepackage{caption}
\input{commands.tex}
\input{properties.tex}

\title[Evolved Star Planet Populations]{Determining the impact of post-main sequence stellar evolution on the transiting giant planet population}

\author[E. M. Bryant et al.]{
\parbox{\textwidth}{
Edward M. Bryant$^{1, 2}$\thanks{E-mail: edward.bryant@ucl.ac.uk}
\& 
Vincent Van Eylen,$^{1}$
}
\\
$^{1}$Mullard Space Science Laboratory, University College London, Holmbury St Mary, Dorking, Surrey, RH5 6NT, UK\\
$^{2}$Department of Physics, University of Warwick, Gibbet Hill Road, Coventry, CV4 7AL, UK
}

\date{Accepted XXX. Received YYY; in original form ZZZ}

\pubyear{2024}
\usepackage{newtxtext,newtxmath}
\usepackage{subfigure}

\begin{document}
\label{firstpage}
\pagerange{\pageref{firstpage}--\pageref{lastpage}}
\maketitle

\begin{abstract}
The post-main sequence evolution of stars is expected to impact the exoplanets residing on close-in orbits around them. Using photometric data from the TESS Full-Frame-Images we have performed a transit search for exoplanets with post-main sequence hosts to search for the imprints of these impacts on the giant planet population. We detect 130 short period planets and candidates, thirty-three of which are newly discovered candidates, from a sample of \Ntotal\ post-main sequence stars spanning the evolutionary stages from the end of the main sequence to the bottom of the red giant branch. We measure an occurrence rate of $0.28\pm0.04\%$ for short period giant planets orbiting post-main sequence stars. We also measure occurrence rates for two stellar sub-populations, measuring values of $0.35\pm0.05\%$ for a sub-population representing the earliest stages of post-main sequence evolution and $0.11^{+0.06}_{-0.05}\%$ for a sub-population of more evolved stars. We show that the giant planet occurrence rate decreases with increasing stellar evolution stage, with a larger occurrence rate decrease observed for shorter period planets. Our results are clear evidence that the population of short period giant planets is being sculpted by the post-main sequence evolution of the host stars, and we conclude that this is most likely through the destruction of these giant planets through the increased strength of planet-star tidal interactions resulting in the rapid tidal decay of the planets' orbits.
\end{abstract}


%%%%%%%%%%%%%%%%%%%%%%%%%%%%%%%%%%%%%%%%%%%%%%%%%%

%%%%%%%%%%%%%%%%% BODY OF PAPER %%%%%%%%%%%%%%%%%%

\section{Introduction}
At the end of their main sequence life span, stars similar in mass to our Sun will undergo a period of evolution. This stellar evolution is predicted to influence the population of planets we observe around post-main sequence host stars \citep[e.g.][]{rasio1996, villaverlivio2009, veras2016postmainsequence}. As the star expands during its post-main sequence evolution, we would expect a large fraction of the of the exoplanets discovered so far, especially those with a semi-major axis smaller than about 1~AU, to be engulfed by the expanding star \citep{Villaver2014}. However, it can be unclear what occurs during the earlier stages of the star's post-main sequence evolution, and exactly at what stage these planets begin to be destroyed by their host stars.

Close-in planets are known to interact with their host stars through tidal interactions \citep{rasio1996}, and these interactions can result in a decay of the planet's orbit, as has been observed for the planet WASP-12\,b \citep{yee2020wasp12}. As host stars evolve off of the main sequence, one of the earliest impacts we expect is an increased rate of orbital decay due to these tidal interactions. The physics governing the dissipation of tides between the host star and an orbiting planet is complex, but the tidal dissipation is characterised as being dominated by two components: the equilibrium tide, in which the dissipation is dominated by turbulent viscosity in the outer convective envelope of the star \citep{zahn1977tides,zahn1989equilibriumtides, esseldeurs2024evolvedstartides}, and the dynamical tide, which arises due to internal stellar oscillations which are excited by the tidal potential \citep{zahn1977tides, zahn2008tides, barkerogilvie2010dynamicaltides}. 

As stars evolve off of the main sequence the stellar structure changes, primarily through the expansion of the outer layers and the deepening of the surface convection zone \citep{nascimento2012subgiants}. The predicted impact of post-main sequence evolution and the resulting change in stellar structure on the strength of planet-star tidal interactions has been widely studied \citep[e.g.][]{barker2020tides, esseldeurs2024evolvedstartides, weinberg2024tides}. Despite the uncertainty around the exact mechanisms which dominate the dissipation of planet-star tidal interactions, these studies all agree that the strength of these interactions is expected to increase during the post-main sequence evolution of the host star.

As such, close-in giant planets are expected to be subject to strong tidal interactions and rapid orbital decay as their host stars evolve through the sub-giant phase and up the red giant branch \citep[e.g.][]{villaverlivio2009, mustillvillaver2012}. We would then expect this rapid tidal decay to leave an imprint on the population of giant planets with evolved host stars. In particular, we expect a reduction in the occurrence rates of giant planets orbiting evolved stars, as compared to the main sequence giant planet population.

There have been a number of studies over recent years specifically targeting the post-main sequence stellar population. Early studies used observations from radial velocity surveys to probe this population. In these studies, the giant stars that were observed covered a wide range of super-solar masses and were used as a proxy to study how giant planet occurrence varies with stellar mass \citep[e.g.][]{johnson2010massmet, reffert2015}. \citet{johnson2010massmet} found the occurrence rate of giant planets to increase monotonically up to a stellar mass of 1.9\,\mstar, with \citet{reffert2015} extending the mass range and finding the giant planet occurrence rate to peak at 1.9\,\mstar\ and then decrease to higher stellar masses. 
However, the stellar masses of the stars surveyed in these studies have been called into question. \citet{lloyd2011, lloyd2013} used stellar evolution models to generate a simulated galactic population, arguing that the expected mass distribution of the evolved star population should be indistinguishable from main sequence F- and G-stars. However, \citet{johnson2013} argued that due to selection effects the evolved star sample studied in the radial velocity surveys is likely subject to a number of biases and so cannot be directly compared to the simulated population from \citet{lloyd2011}. 

Independently, \citet{schlaufman2013} investigated the mass distribution of the evolved star planet hosts by considering the kinematics of the host stars. \citet{schlaufman2013} found the Galactic $UVW$ velocity dispersion for the planet hosting sub-giant and giant stars to be indistinguishable from a sample of F5-G5 main sequence planet hosts. They also found the velocity dispersions for the sub-giant and giant planet hosts to be larger than that for a sample of solar neighbourhood A5-F0 stars. From their results they argued that the evolved star planet hosts have a similar stellar mass distribution to the F5-G5 main sequence stars, and have on average lower stellar masses than the A5-F0 main sequence stars. From these results, \citet{schlaufman2013} concluded that the reduced number of short period giant planets around evolved stars \citep[e.g.][]{bowler2010} is due to tidal orbital decay and not the mass of the host stars. 

Planets detected around giant and sub-giant stars from the early radial velocity surveys were on long orbital periods \citep[$P > 100$\,d; e.g.][]{bowler2010, reffert2015}. While planets at these wide orbital distances are expected to be subject to rapid orbital evolution during the red giant phase \citep{villaverlivio2009} it is planets on much shorter orbital periods which are expected to be the most susceptible to rapid orbital decay due to tides during the early stages of post-main sequence stellar evolution \citep{weinberg2024tides}. 

Space based transit surveys Kepler \citep{borucki2010kepler}, K2 \citep{howell2014K2}, and TESS \citep{ricker2015tess} have contributed to the discovery of a number of giant planets transiting post-main sequence stars with orbital periods shorter than 10\,days \citep[e.g.][]{lillobox2014kepler61, barclay2015kepler61, vaneylen2016k239, chontos2019koi4, grunblatt2022tgt2}. Going beyond the detection of individual planets, \citet{grunblatt2019k2RGB} used K2 photometry to study the occurrence of short period giant planets around 2476 red giant stars. Based on the detection of three planets orbiting these stars, they determined an occurrence rate of $0.49 \pm 0.28 \%$ for planets larger than 1\,\rjup\ with orbital periods in the range $3.5 \leq P \leq 10$ days, for evolved stars with radii $\rstar = 3.5 - 8.0$\,\rsun. They found this to be consistent with the occurrence of planets with similar properties orbiting main sequence stars \citep[see e.g.][]{howard2012kepleroccrates, KunimotoMatthews2020, beleznay2022tessoccrates}. However, the limited stellar sample size available from the K2 photometry reduced the precision they could obtain for the occurrence rates. We also note that the stars targeted by \citet{grunblatt2019k2RGB} are in general more evolved than the majority of stars we study in this work and they did not probe the shortest orbital periods where we might expect to see the largest tidal impacts \citep{weinberg2024tides, esseldeurs2024evolvedstartides}. 

As a result, the question of how the early stages of post-main sequence evolution impact exoplanet systems remains largely unanswered. In particular, it remains unclear if and when planets are likely to undergo rapid orbital decay, and at what stage of their host star's evolutionary lifetime. To address these questions, we use the photometric data from the Full-Frame-Images (FFIs) delivered by the Transiting Exoplanet Survey Satellite \citep[TESS; ][]{ricker2015tess} to measure the occurrence rates of close-in planets around evolved post-main sequence host stars. 
The TESS FFIs provide high precision time series photometry for stars across nearly the entire sky, which have previously been used to study the occurrence of planets as a function of stellar characteristics such as stellar mass to constrain planet formation models \citep{beleznay2022tessoccrates, bryant2023lmstaroccrates, gan2023mdwarfoccrates}.
Here, we study a sample of \Ntotal\ post-main sequence stars with high-precision TESS time series photometry to constrain the occurrence of close-in giant planets, and compare this to their main sequence counterparts and to tidal theory.

We present our post-main sequence selection and sample in Section~\ref{sec:stellar_sample}. The various stages of our planet search, vetting, and analysis pipeline are discussed in Section~\ref{sec:search_and_vetting}, before we present our planet candidate sample in Section~\ref{sec:planets}. We discuss our injection-recovery tests in Section~\ref{sec:injrec} and discuss how we compute the occurrence rates in Section~\ref{sec:occrates}. We present our occurrence rate results and discuss the implications of these results particularly in the context of the impact of post-main sequence evolution on planetary systems in Section~\ref{sec:res_and_disc}, before presenting our conclusions in Section~\ref{sec:conc}.

\section{Defining the TESS Evolved Star Sample}\label{sec:stellar_sample}
\subsection{TESS Photometric Observations}\label{sub:tess}
The Transiting Exoplanet Survey Satellite \citep[TESS; ][]{ricker2015tess} has been monitoring the brightness of stars across the whole sky since its launch in 2018. Situated in an approximately 13\,day orbit around the Earth, TESS observes a strip of the sky, with dimensions of 24\,degrees\,$\times$\,96\,degrees, for two orbits resulting in a total of $\approx$\,27.4\,days of observation for a given area of the sky. This 27.4\,day period of observation is referred to as a sector.

During the first extended mission of TESS observations occurring between 4th July 2020 and 1st September 2022 -- Sectors 27 to 55 -- the TESS Full-Frame-Images were recorded at a cadence of 10\,minutes. Light curves for all stars brighter than a TESS magnitude of 13.5 were produced by the Quick-look Pipeline \citep[QLP; ][]{kunimoto2021qlp1,kunimoto2022qlp2} and made publicly available to the community as a MAST High-Level-Science-Product\footnote{\url{https://archive.stsci.edu/hlsp/qlp}}. Full details of the QLP light curve production are provided in \citet{huang2020qlp}, but we provide some brief details here. The QLP is a difference image photometry pipeline in which a reference image is constructed and then subtracted from each TESS image to produce a difference image. Aperture photometry is then performed on each difference image using circular apertures to compute the differential source brightness. The measured differential fluxes are converted into absolute fluxes using the TESS magnitude provided in the TIC \citep{stassun2019TIC8} for each source along with the TESS instrument zero-point magnitude \citep{vanderspek2018tess}. This step is equivalent to deblending the flux time series and so any transit events in the QLP light curves should not be heavily impacted by dilution from nearby contaminating stars.

As light curves are made for all stars brighter than $T = 13.5\,$mag, the target selection is minimal and likely to not be biased in a way that would impact an occurrence rate determination. The TESS 2\,minute cadence light curve sample for example will be subject to bias due to the inclusion of Guest Observer targets, often planet or candidate host stars. Therefore due to the unbiased nature of the target selection during the production of the QLP light curves and the high quality of the light curves themselves we elect to the QLP data products, for TESS sectors 27 to 55, for the systematic transit search we perform in this work. These sectors constitute the full first extended mission of the TESS satellite, during which FFIs were recorded at a cadence of 10\,minutes.

\subsection{Identifying Post-Main Sequence Stars}\label{sub:evolstars}
\begin{figure*}
    \centering
    \subfigure[]{\includegraphics[width=0.46\linewidth]{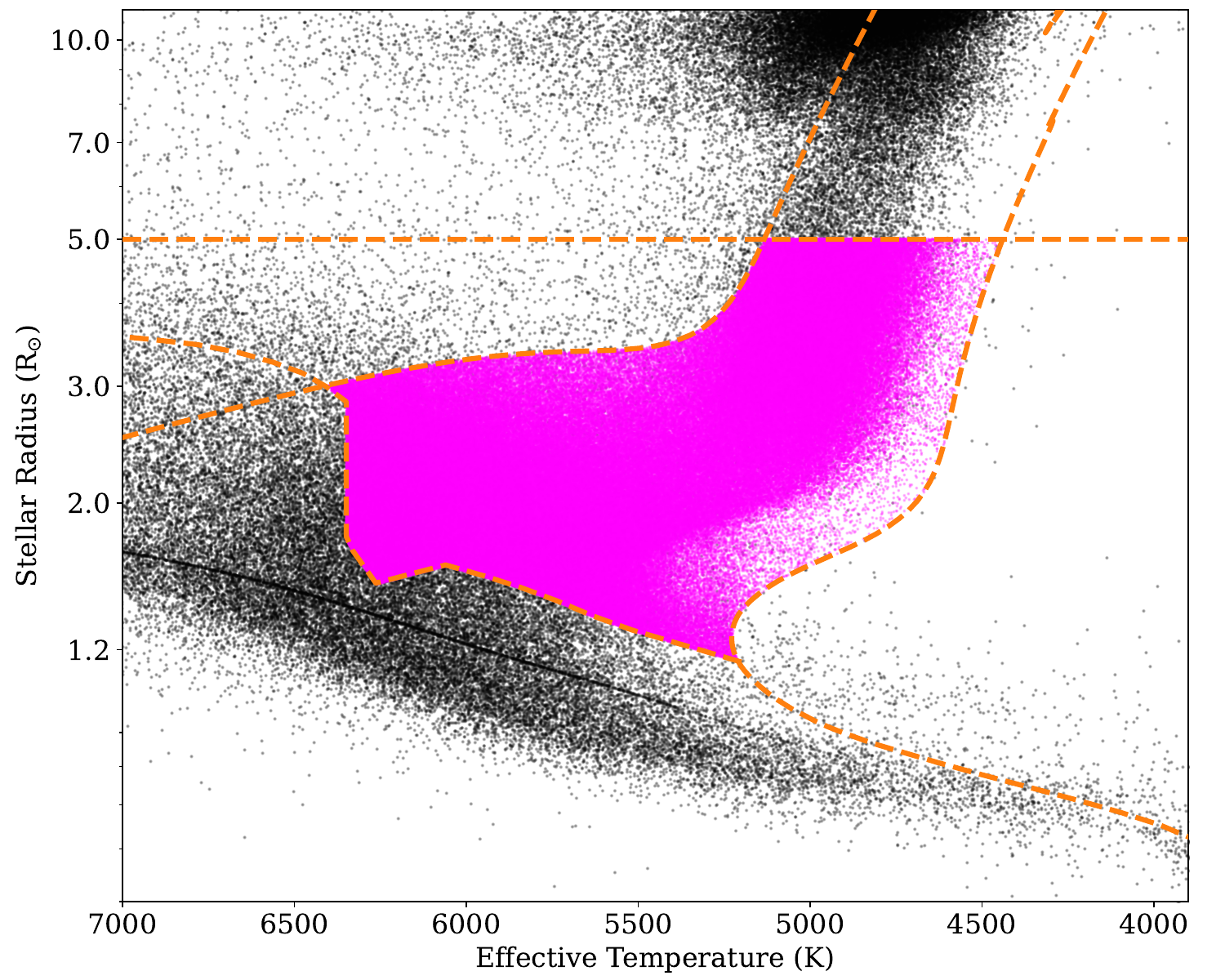}}
    \subfigure[]{\includegraphics[width=0.46\linewidth]{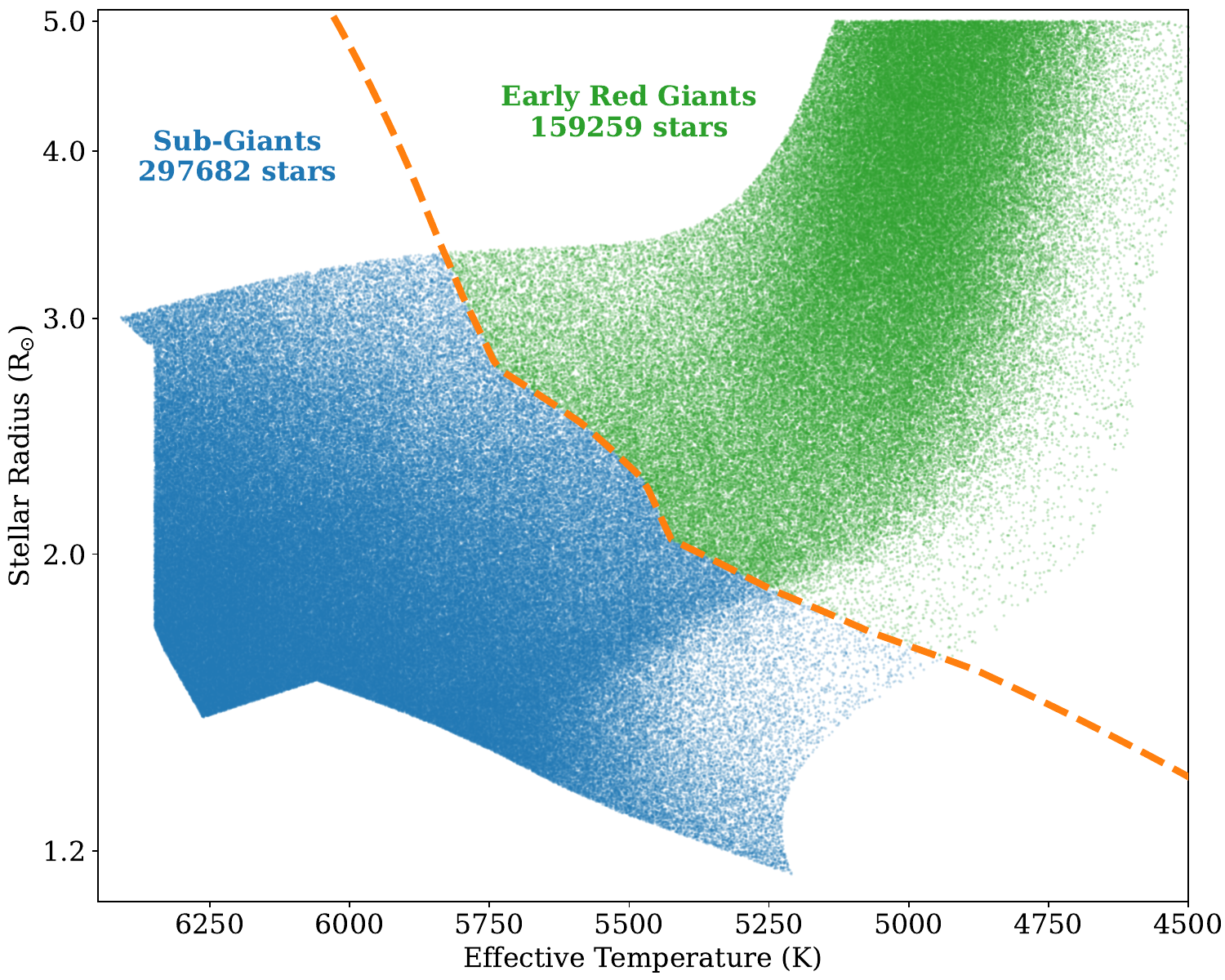}}
    \caption{TESS post-main sequence stellar population used as the input sample for this study. \textbf{Left panel:} The stars included in this survey are plotted as the magenta points. The black points show a selection of stars from the TIC across all evolutionary stages as reference. The orange dashed lines show the various stellar tracks and parameter criteria used to select the sample (see Section~\ref{sub:evolstars}). \textbf{Right panel:} We show the two sub-populations considered in this work. The stars in our sub-giant sample are plotted as the blue points and the stars in our early red giant sample are plotted as the green points. The orange dashed line gives the $\text{EEP} = 465$ boundary used to separate the two samples. See the text in Section~\ref{sub:evolstars} for details on this boundary. The number of stars in each sub-population are given by the annotations.}
    \label{fig:stellar_sample}
\end{figure*}
We use the TESS Input Catalogue \citep[TIC version 8;][]{stassun2019TIC8} to select a sample of post-main sequence stars, using the stellar radius (\rsun) and effective temperature (\teff) values for each TIC source, which are calculated using \textit{Gaia} DR2 photometric information \citep{GaiaDR2}. The combination of \rsun\ and \teff\ are used to identify likely evolved stars using the \textsc{evolstate} package \citep{huber2019evolstate2}. Details of the selection are provided in \citep{berger2018evolstate1} but in short: a PARSEC solar--metallicity stellar evolution track is used to set the terminal age main sequence boundary, and a 15\,Gyr $\feh = +0.5$\,dex MIST isochrone provides a lower bound to exclude main sequence binary stars. In addition to these selections, we introduce an additional MIST evolution track for a $\mstar = 1.4\,\msun$, $\feh = -0.5$\,dex star to set an upper boundary for our sample. This was done to limit our sample to stars with stellar masses comparable to a main sequence sample of  F and G-dwarfs. As a final cut, we impose a limit on the stellar radius of $\rstar \leq 5.0\,\rsun$ and on the apparent magnitude in the TESS band of $T \leq 12$\,mag. These final criteria are motivated by the need of the stars in our sample to be amenable to the detection of transiting exoplanets in the TESS data. Based on these stellar criteria we identify a full sample of 620,244 stars. Crossmatching the stellar sample obtained from this selection with the stars for which there is an available QLP 10\,minute cadence light curve yields a stellar sample of \Ntotal\ post-main sequence stars for our survey, representing 74\% of the full stellar sample. We plot our post-main sequence population in Figure~\ref{fig:stellar_sample}.

In this work we will consider two sub-populations for our full stellar sample. This will enable us to study how the giant planet occurrence rates change through the early stages of post-main sequence evolution. To identify these two sub-populations we use the MIST stellar evolution tracks \citep{Dotter2016MIST, Choi2016MIST} to generate a boundary of equal Equivalent Evolutionary Phase \citep[EEP,][]{Choi2016MIST}. The MIST EEP values denote equivalent stages in the lifetime of a star across different stellar masses. For the purposes of this work, we use a $\text{EEP} = 465$ boundary in $\left[T_\text{eff}, R_\ast\right]$ space to split our sample into the two sub-populations (Figure~\ref{fig:stellar_sample}). Using EEP as the boundary parameter provides us with two sub-populations for which the main distinguishing feature is the evolutionary stage of the stars in each sub-population. The boundary value was chosen empirically so the boundary divides the two regions of high stellar number density in $\left[T_\text{eff}, R_\ast\right]$ space (see Figure~\ref{fig:stellar_sample}).The stars to the lower and left of this boundary in Figure~\ref{fig:stellar_sample} are the less evolved sub-population, which consists of 297,682 stars. We will refer to this population as the sub-giant sub-population. The stars to the upper and right of this boundary in Figure~\ref{fig:stellar_sample} are the more evolved sub-population, which consists of 159,259 stars. We will refer to this population as the early red giant sub-population. 

\section{Planet Search and Vetting}\label{sec:search_and_vetting}
\subsection{Light Curve Pre-Processing}
We access the QLP light curves for our sample from the MAST. For each star, we consider all sectors within the range of Sector 27 to Sector 55 inclusive in which this star was observed together. We also exclude all time stamps with a quality flag $> 0$.

The MIT Quick Look Pipeline includes a detrending step during which an iterative Kepler spline is used to flatten the light curves. We note that the node spacing for these splines is optimized for the detection of short period planets, and so short duration transits \citep{huang2020qlp}. Although we are searching for short period planets, due to the increased stellar radii of post-main sequence stars we expect significantly longer transit durations than would be expected for the same orbital periods around main-sequence stars. Therefore, it is likely that the transit of the planets we are searching for may be significantly distorted by the automated QLP Kepler spline. Instead, we take the undetrended QLP flux time series and employ our own light curve flattening using a biweight-estimation method \citep{Hippke_2019} using a wide window of 1.5\,days. This wide-window custom flattening should enable the detection of the long ($\geq 0.5$\,d in some cases) transits we expect for our sample.

TESS photometry contains multiple data gaps, particularly the regions between subsequent orbits and sectors. These gaps can impact window-based flattening methods such as the one we employ. Therefore, we identify any point within the light curve for a given sector where there exists a spacing of larger than 0.3\,days between subsequent data points. These points are used as the boundaries to split the light curve into multiple portions with each portion being flattened independently. The TESS light curves can often display step changes in the flux across these data gaps, for example due to temperature changes when the satellite passes behind the Earth during its orbit or due to the data gaps being the result of guiding issues. These flux step changes can impact the smooth flattening method we use and so we flatten the individual portions of the light curve independently to avoid the impact of this on the final light curve. We also note that each sector is flattened independently of any other sectors.

\subsection{Transit Event Detection}\label{sub:bls}
\begin{figure}
    \centering
    \includegraphics[width=0.8\linewidth]{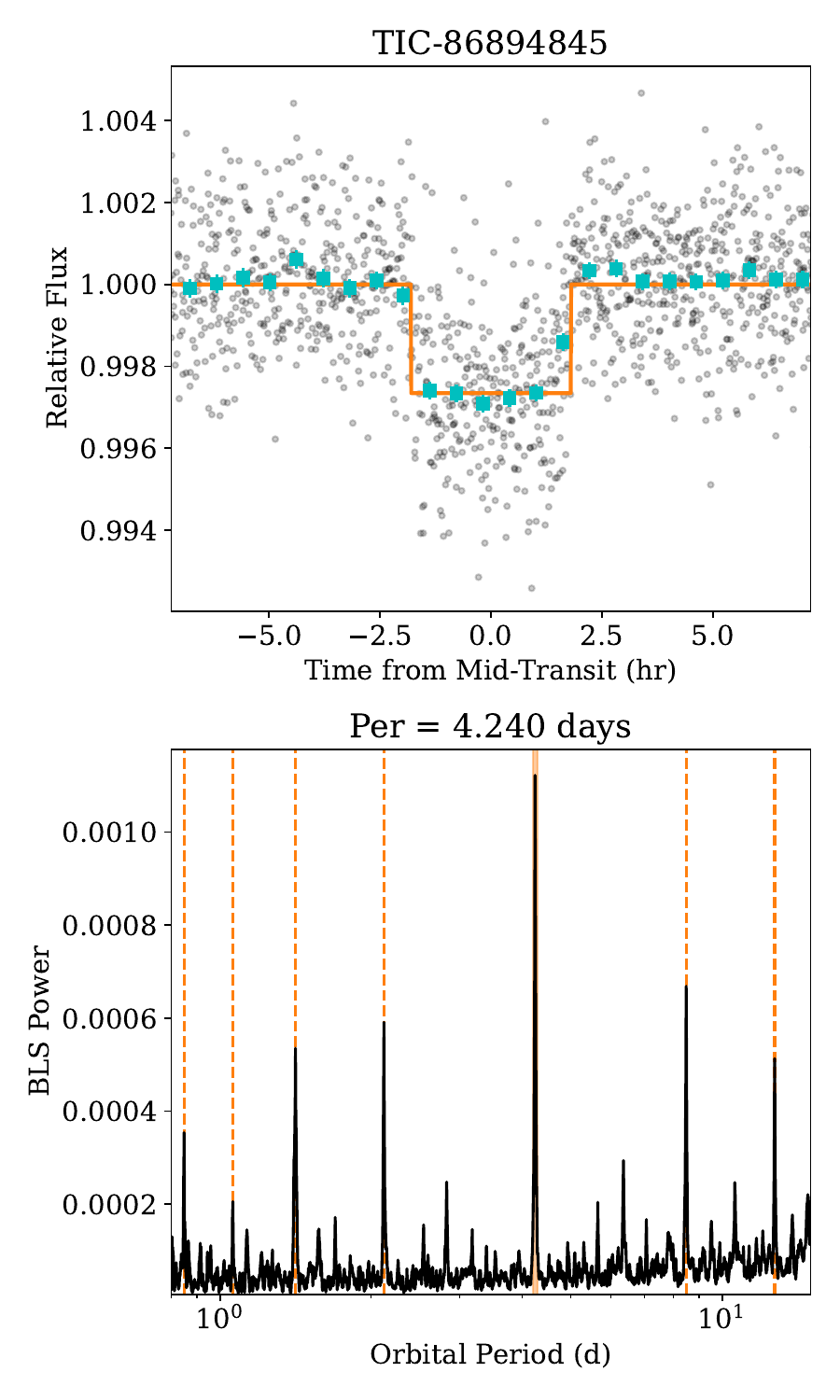}
    \caption{An example BLS candidate detection. \textbf{Top:} TESS QLP flux phase-folded at the orbital period reported by BLS, with the cyan squares showing the data binned in phase. The orange line shows the best-fit BLS box model. Note that we show a zoomed in view around phase 0 for visual clarity of the transit event. The full phase-folded flux data set extends beyond the edges of the plot. \textbf{Bottom:} The BLS periodogram for this candidate. The solid orange vertical line shows the best orbital period reported by BLS, with the dashed lines showing integer multiples and fractions of this period.}
    \label{fig:bls_example1}
\end{figure}
Following the light curve pre-processing we search for periodic transit-like signals using the Box-fitting Least Squares \citep[BLS;][]{kovacs2002BLS} algorithm, as implemented in the \textsc{astropy} Python library. We search initially for transiting events with periods from 0.8 -- 15 days, and subsequently we then exclude any candidate with a period $P < 1$\,day or $P > 12$\,days. From initial analyses these extreme period ranges were found to be dominated by spurious signals. In particular the BLS detections in the longer period range were dominated by spurious detections due to the $13.7$\,day TESS orbit and start--/end--of--orbit systematics.

In order to consider a BLS detection as significant we require a Signal Detection Efficiency of $\mathrm{SDE} \geq 7.3$. We also calculate a transit Signal-to-Noise ratio given by the following equation
\begin{equation} \label{eq:trans_snr}
    {\rm S/N} = \frac{\delta}{\sigma_{\rm LC}}\sqrt{N_{\rm Tobs}}
\end{equation}
where $\delta$ is the transit depth, $\sigma_{\rm LC}$ is the RMS scatter of the out-of-transit region of the light curve, and $N_{\rm Tobs}$ is the number of in-transit flux data points. We require $S/N \geq 8.5$ for a significant detection. To determine this threshold, we consider a subset of 55,185 randomly selected stars from our input sample. For each star, we then calculate the out of transit RMS scatter in the flux. Using this RMS scatter we generate a flat light curve with flux values drawn from a Gaussian distribution with a mean of one and a standard deviation equal to the RMS scatter. We then run our BLS search algorithm on this white noise light curve. This process is repeated twenty times for each star, resulting in 1,103,700 unique simulations. Considering the distribution of $S/N$ values obtained for these flat light curves, the 99.99\% percentile value was 8.42. From these tests, we therefore take 8.5 as the $S/N$ threshold value to limit the false positive rate of BLS yielding detections in flat, white noise dominated light curves to less than 1 in 10,000. We also exclude any BLS detections where the measured duration of the detected event is greater than one quarter the orbital period. Our BLS search yields \NBLShits\ significant detections of transit-like signals. We show an example of a BLS candidate detection in Figure~\ref{fig:bls_example1}.

Transit surveys often impose a requirement of a minimum of three detected transits to consider a BLS detection as a candidate planet signal. We impose a similar requirement for our search, but go one step further and require that we have three detected transits which are further than 1.5 times the transit duration reported by BLS from the start or end of a spacecraft orbit or a spacecraft momentum dump event. Both of these instances are prone to produce transit-like spurious signals due to instrumental trends in the recorded light curve. From our search, \NBLShitsThreeTrans\ transit candidate signals pass this subsequent criterion.

\subsection{Automated False Positive Identification}\label{sec:FPs}
Transit surveys are known to be plagued by astrophysical false positives which can produce signals that mimic the signal of a transiting planet. Primary among these false positives are eclipsing binary systems and variable stars. We therefore employ a set of quantitative vetting checks of the BLS candidates to identify any which are due to such false positive cases based purely on the TESS light curves. Here we provide details of each check performed as well as the number of candidates identified as likely false positives by each check. Note that many candidates are identified as false positives by more than one vetting check. These checks follow those presented in \citet{bryant2023lmstaroccrates} but we have updated the algorithms used in a few cases. 

In order to measure the planet occurrence rates, we employ transit injection and recovery tests in order to quantify the detection efficiency of our transit search pipeline and vetting algorithms. During these simulations, we inject transiting planet signals into real TESS light curves, then passing those simulated transiting planet light curves through our full planet search and candidate vetting pipeline. Full details of these simulations are provided in Section~\ref{sec:injrec}, but here we will provide the percentage of the injected planet signals which are falsely characterised as false positives by each of our light curve vetting checks.

\subsubsection{Transit Depth and Duration Measurement}\label{sub:dep_dur_vet}
The depth and duration of a transit-like event can be a clear identifier that the signal is not being produced by a transiting planet. While the BLS algorithm provides the depth and duration of the best-fitting box model, we note that these box-like models are often imperfect fits to a transit light curve. To get a more accurate assessment of the transit depth and duration we fit a trapezium-shaped model to the phase-folded transit light curve, folding the data using best orbital period and transit epoch from the BLS results. We perform the fit using a least-squares methodology, implemented using the \textsc{lmfit} Python software \citep{lmfit}. We use the transit depth, $\delta$, reported by the \textsc{lmfit} analysis to calculate the planet-to-star radius ratio using $\delta = \left(\rpl/\rstar\right)^2$, noting that for this initial check we do not consider the effect of limb-darkening.  We then use the stellar radius of the host star, as reported in the TIC, to compute the planet radius, excluding any candidate for which we measure $\rpl > 24\,\rearth$. Using the measurement of the transit duration from the trapezium-model fit we also exclude any candidate for which the measured duration is greater than 25\% the orbital period.

A total of \NRplReject\ objects (\RplRejectPerCent\,\%) are identified as likely false positives by the measurement of their transit depth. Only \InjRecRplPerCent\,\% of the injected planet signals were falsely rejected by this check. A total of \NQReject\ objects (\QRejectPerCent\,\%) are identified as likely false positives by the measurement of their transit duration. Only \InjRecQPerCent\,\% of the injected planet signals were falsely rejected by this step.

\subsubsection{Stellar Density Matching}\label{sub:rho_vet}
The duration of an exoplanet transit is closely related to the stellar density of the host star. Therefore, if the eclipse duration for one of our candidates is too short or long to be physically plausible for a planetary orbit around the host star, we can exclude this candidate from our sample. To do this, we use the transit parameters obtained from the BLS and vetting steps in Sections~\ref{sub:bls}~and~\ref{sub:dep_dur_vet} to calculate a transit stellar density, $\rho_{\rm trans}$, using Kepler's Third Law and assuming an equatorial and circular orbit. We note that for transiting planets this value will likely differ from the host star density as a result of the planet's orbit having a non-zero impact parameter or eccentricity. For the stars in our sample, many do not have reported stellar masses in the TIC and for those that do the reported stellar masses are unreliable \citep{stassun2019TIC8}. Therefore, for each star we use the reported TIC stellar radius, \rstar, to set an upper and lower boundary for the allowable range for $\rho_{\rm trans}$. 

Considering the lower boundary, we calculate the corresponding stellar density value for the host star radius assuming a lower stellar mass limit of 0.9\,\msun. An eccentric orbit can alter the transit duration and impact the stellar density measurement from the transit \citep{dawson2012photoeccentric, vaneylen2015photoeccentric}. We therefore exclude any candidate for which $\rho_{\rm trans}$ is less than 10\% of the lower boundary stellar density value, to make sure we do not falsely exclude true planets with eccentric orbits. A total of \NRhoTransRejectLow\ objects (\RhoTransRejectLowPerCent\,\%) are excluded by this assessment. We use an upper stellar mass limit of 1.4\,\msun\ to compute the upper boundary stellar density value. For this calculation we are assuming an equatorial orbit, and so an impact parameter $b = 0$. Orbits with $b > 0$ will result in shorter transit durations, and so this calculation is likely to yield an overestimated stellar density value. We therefore exclude any candidate for which $\rho_{\rm trans} > 100 \times \rho_{\ast; \text{ upper}}$, excluding \NRhoTransRejectUpp\ such candidates (\RhoTransRejectUppPerCent\,\%). Overall, through this stellar density analysis we exclude \NRhoTransRejectTot\ candidates as likely false positives (\RhoTransRejectTotPerCent\,\%). Only \InjRecRhoTransRejectPerCent\,\% of the injected planet signals were falsely rejected by this step.

\subsubsection{Secondary Eclipse Check}
A clear signifier that the transit-like event has been produced by an eclipsing binary is the presence of a secondary eclipse -- a second, shallower eclipse event seen in the light curve at the same period as the primary eclipse. To search for the presence of a secondary eclipse a series of phase values are determined which span the out-of-transit region of the light curve with a spacing of 0.005 between each phase value. We search a range of phase values in order to identify eclipsing binaries on eccentric orbits, for the which the secondary eclipse will not necessarily fall at phase 0.5. At each phase value, a box-like model is fit to the phase-folded light curve, with the duration fixed to the best BLS duration. The depth of the deepest box model is recorded and used to calculate a secondary eclipse S/N using Eq.\ref{eq:trans_snr}. Any candidate for which the secondary eclipse S/N is greater than 8 is excluded as a likely false positive. Short period giant planets may also exhibit secondary eclipses that are visible in the TESS light curves \citep[e.g.][]{kabath2022toi1181}, although the secondary eclipses produced by planets will in general be shallower than those from eclipsing binaries. Giant planets on sufficiently close orbits to produce detectable secondary eclipses are also very likely to have had their orbits circularised through tidal interactions \citep[e.g.][]{jackson2008} and so their secondary eclipses will be close to phase 0.5. Therefore, in order to avoid incorrectly rejecting real planets, any object with detected secondary eclipse shallower than 1\,ppt and within a phase range of 0.45-0.55 is not excluded from our candidate list, as these cases may be planets.

A total of \NSecEclipseReject\ objects (\SecEclipseRejectPerCent\,\%) are identified as likely false positives by the identification of a likely secondary eclipse. Only \InjRecSecEclipsePerCent\,\% of the injected planet signals were falsely rejected by this step.

\subsubsection{Odd-Even Eclipse Depth Difference}
If the primary and secondary eclipses of a binary system are similar in depth then the BLS algorithm can often misidentify both sets of eclipses as primary transit events, reporting half the true orbital period as the best period. In this case, the odd and even eclipses will be different depths, and we can use this depth difference to identify these eclipsing binary scenarios. For this we again fit a box-like model to the odd events and even events independently, fixing the box duration and centre time to the values reported by BLS. We then determine a depth difference S/N value using a modified version of Eq.\ref{eq:trans_snr} given by
\begin{equation}
    {\rm S/N} = \frac{|\delta_{\rm odd} - \delta_{\rm even}|}{\sigma_{\rm LC}}\frac{1}{\sqrt{\frac{1}{N_{\rm Tobs;\,odd}} + \frac{1}{N_{\rm Tobs;\,even}}}}
\end{equation}
where $\delta_{\rm odd}$ and $\delta_{\rm even}$ are the odd and even transit depths, $N_{\rm Tobs;\,odd}$ and $N_{\rm Tobs;\,even}$ are the number of in-transit data points for the odd and even transit events, and $\sigma_{\rm LC}$ is the standard deviation of the out-of-transit portion of the light curve. We identify any candidate with a S/N value greater than 8 as a likely false positive.

A total of \NOddEvenReject\ objects (\OddEvenRejectPerCent\,\%) are identified as likely false positives by the measurement of a likely difference in the depths of the odd and even events. Only \InjRecOddEvenPerCent\,\% of the injected planet signals were falsely rejected by this step.

\subsubsection{Inter-Sector Depth Difference}\label{sub:sec_depth_diff}
Due to the large size of the TESS pixels, TESS light curves suffer from contamination from nearby stars. During the QLP light curve generation process the transit features are undiluted, assuming the transit is occurring on the target star. However, if one of the nearby stars is itself an eclipsing binary, then it will impose a transit-like signal onto the target star light curve. If the level of contamination of this nearby eclipsing binary changes between sectors then the observed transit depth will also change. This change in contamination can be due to the rotation of the TESS spacecraft meaning the stars are observed by a different camera and are located on a different part of a CCD. This can result in the PSF changing between sectors and so the contamination level also changes.

To assess the depth differences between multiple sectors, we fit the phase folded light curve for each individual sector of a given candidate using a box-like model with the timing and duration fixed to the BLS value. We then fit this box-like model to each sector to assess the transit depth. To assess whether there is a significant inter-sector depth difference we calculate a signal-to-noise value for the depth difference between the deepest and shallowest sector depths given by
\begin{equation}
    {\rm S/N} = \frac{\delta_{\rm max} - \delta_{\rm min}}{\sigma_{\rm LC}}\frac{1}{\sqrt{\frac{1}{N_{\rm Tobs;\,max}} + \frac{1}{N_{\rm Tobs;\,min}}}}
\end{equation}
where $\delta_{\rm max}$ and $\delta_{\rm min}$ are the maximum and minimum individual sector transit depths, $N_{\rm Tobs;\,max}$ and $N_{\rm Tobs;\,min}$ are the number of in-transit data points for these two transit events. We identify any object with a S/N value greater than 8 as a likely false positive.

A total of 2,913 objects (13.96\%) are identified as likely false positives by the measurement of a significant difference between the eclipse depths measured in different sectors. Only 1.25\% of the injected planet signals were falsely rejected by this step.

\subsubsection{Eclipse Phased Variability}
This check searches for variability in the light curve in phase with the orbit of the companion. Such phased variability is a signifier the transit-like event is caused by a stellar companion. We determine the $R^2_{\rm Harmonic}$ value, following \cite{montalto2020diamante}, but see also Eqs.~1 and 2 of \cite{bryant2023lmstaroccrates} for details. We exclude any object with $R^2_{\rm Harmonic} > 0.6$ as a false positive. A total of \NHarmonicReject\ objects (\HarmonicRejectPerCent\,\%) are identified as likely false positives by the identification of variability phased with the eclipse events. Only \InjRecHarmonicPerCent\,\% of the injected planet signals were falsely rejected by this step.

\subsubsection{Lomb-Scargle Analysis}
Continuously varying stars can often be erroneously identified by the BLS algorithm as periodic transit-like signals. We employ a number of techniques to identify these objects. These algorithms are the same as those employed in \citet{bryant2023lmstaroccrates}, and so we refer the reader there for details on the implementation. Here we simply report on the thresholds used in this instance and the numbers of objects identified by each check.

As a first check to identify variable stars we compute a Generalised Lomb-Scargle \citep[GLS;][]{lomb1976,scargle1982} periodogram of the out-of-transit portion of the light curve, excluding any object which yields a GLS power greater than 0.3 as a false positive. A total of \NGLSReject\ objects (\GLSRejectPerCent\,\%) are identified as likely false positives by the presence of a significant peak in the Lomb-Scargle periodogram of the out-of-transit portion of the light curve. Only \InjRecGLSPerCent\,\% of the injected planet signals were falsely rejected by this step.

\subsubsection{Excess Scatter Analysis}
For the out-of-transit portion of the phase folded light curve we compare the RMS flux scatter to the RMS of the point-to-point flux scatter to determine the $\sigma_{\rm excess}$ value as defined in Eq.3 of \citet{bryant2023lmstaroccrates}. A value of $\sigma_{\rm excess}$ different to unity signifies the presence of continuous variability in the light curve. We exclude any object with $\sigma_{\rm excess} < 0.4$ or $\sigma_{\rm excess} > 1.3$ as a false positive. A total of \NExcessRMSReject\ objects (\ExcessRMSRejectPerCent\,\%) are identified as likely false positives by the measurement of an increased root mean square scatter of their light curves in comparison to the point-to-point light curve scatter. Only \InjRecExcessRMSPerCent\,\% of the injected planet signals were falsely rejected by this step.

\subsubsection{Light Curve Asymmetry Test}
We determine the asymmetry of the phase folded light curve following the prescription of \citet{bryant2023lmstaroccrates}. For this we compare the RMS point-to-point scatter of the phase-folded light curve running from phase -0.5 to 0.5, with the light curve folded using the absolute values of the phase, to run from 0.0 to 0.5. We calculate the ratio of scatters of the absolute phase folded light curve to the standard folded light curve, excluding any object for which this ratio is less than 0.8 or greater than 1.3 as a false positive.  A total of \NPtPSymmReject\ objects (\PtPSymmRejectPerCent\,\%) are identified as likely false positives by the measurement of a significant asymmetry in their out-of-transit light curves. Only \InjRecPtPSymmPerCent\,\% of the injected planet signals were falsely rejected by this step.

\subsubsection{Increased Scatter During Transit}
In addition to astrophysical false positive scenarios, the BLS algorithm can also detect spurious signals from non-astrophysical sources as periodic signals. Occasionally the BLS algorithm can incorrectly identify regions of high scatter, in particular systematic decreases in flux, as transit events. To identify these cases, first we compute the flux residuals to the best fit trapezium model. We calculate the RMS point-to-point scatter of the flux residuals for the in-transit, pre-transit, and post-transit portions of the light curve. We then calculate the ratios of the in-transit scatter value to pre-transit value and post-transit values. We exclude any object for which either of these two ratios is greater than 5.

A total of \NPtPInOutReject\ objects (\PtPInOutRejectPerCent\,\%) are identified as likely false positives by the measurement of an increased scatter during the in-transit portion of the light curve. Only \InjRecPtPInOutPerCent\,\% of the injected planet signals were falsely rejected by this step.

\subsubsection{Near data gap event check}
Transit searches in TESS observations in particular are impacted by systematics which arise at the start and end of data gaps due to the spacecraft orbit. These systematics often arise from increased scattered light on the detector, imperfect flux detrending at these points, or a combination of both these effects. We therefore determine whether the BLS phase-fold signal is dominated by events near these data gaps, and so are likely dominated by spurious signals. To do this we extract the log-likelihood values for each transit event as computed by the BLS algorithm. We then calculate the total log-likelihood contributions from the events near the data gaps, $\ln L_{\rm edge}$, and those which are firmly within the TESS data chunks, $\ln L_{\rm body}$. If $\ln L_{\rm body} \leq 0$ we automatically exclude the object, and if $\ln L_{\rm edge} \leq 0$ the object automatically passes this check. If both values are positive and non-zero we then calculate the ratio $\ln L_{\rm edge} / \ln L_{\rm body}$ excluding any object for which this ratio is greater than 3.

A total of \NLnLRatioReject\ objects (\LnLRatioRejectPerCent\,\%) are identified as likely false positives due to the fact that their BLS detections are dominated by events which occur close to a data gap or momentum dump event. Only \InjRecLnLRatioPerCent\,\% of the injected planet signals were falsely rejected by this step.

\subsubsection{Overall Light Curve Vetting Results}
Overall, out of the \NBLShitsThreeTrans\ objects containing periodic signals identified as transit-like candidates using BLS, our light curve vetting algorithms identify \NTotalVetFPs\ (\TotalVetFPsPerCent\,\%) of them as likely false positive scenarios. This leaves \NTotalVetPass\ objects which remain as possible transiting planet candidates. For the transiting planet signals we simulate for the injection and recovery tests (see Section~\ref{sec:injrec} for details) just \InjRecTotalPerCent\,\% of them are rejected in total by our light curve vetting algorithms. Therefore, while there will be a small number of genuine planets misclassified as false positives by these checks, the majority of the \NTotalVetFPs\ objects rejected will be true false positives.

\subsection{Transit Fitting Analysis}\label{sec:fitting}
\subsubsection{Transit Model}\label{sub:transit_model}

\begin{table}
    \caption{Parameter Priors for Transit Fitting}
    \input{fitting_priors}
    \label{tab:priors}
\end{table}

The next stage of the candidate selection pipeline is to perform a transit fitting analysis for the candidates which pass the light curve vetting stage. We perform this analysis to obtain a more accurate estimate of the radius and impact parameter of the companion. For this we use the \textsc{emcee} package \citep{emcee} to perform a Markov Chain Monte Carlo (MCMC) analysis.

We use the \textsc{batman} package \citep{batman} to generate the transit models at each step of the MCMC sampling using the following as free parameters: the reference mid-transit time, \tc, the orbital period, $P$, the planet-to-star radius ratio, \rprs, the semi-major axis of the orbit scaled to the stellar radius, \ars, and the orbital inclination, $i$. For the stellar limb darkening we use the quadratic law fitting for the $q_{\rm 1}$ and $q_{\rm 2}$ parameters following the \citet{kipping2013LD} parameterisation for the limb-darkening coefficients. We also fit for an out-of-transit flux offset, $f_{\rm 0}$, such that the out-of-transit flux baseline is equal to $1 + f_{\rm 0}$. We have a total of eight free parameters for this analysis. For all parameters we use uniform priors such that we ensure physically realistic transit models but do not bias the results of the analysis in any way. We provide the exact prior ranges used in Table~\ref{tab:priors}. We run 20 independent chains for a burn-in phase of 5,000 steps per chain followed by a sampling phase of 50,000 steps per chain, resulting in a total of 1 million posterior samples for each candidate. 

From this transit analysis we apply a further set of candidate selection criteria, using the posterior distributions. First, we identify the radius ratio which gives the highest log likelihood value, and use this along with the TIC stellar radius to calculate the best-fit planet radius $R_\text{P; Best}$. We then exclude any candidate with $R_\text{P; Best} > 22\,\rearth$ as a likely eclipsing binaries. We also include a check to exclude any candidate with $R_\text{P; Best} = 0$, although we note that the lowest $R_\text{P; Best}$ value for any candidate is $1.36\,\rearth$, and so no candidates are excluded by this lower limit. We also compute the duration of best-fitting transit model, and exclude any candidate for which this transit duration is greater than 25\% of the best-fit orbital period. Next, we use the posterior distributions of \ars\ and $i$ to compute a posterior distribution for the impact parameter $b$. We then exclude any candidate for which either the median or best-fit $b$ values are greater than 0.9. Finally, we use our parameter posterior distributions to compute a posterior distribution for the transit stellar density, $\rho_{\rm trans}$. For each object we also compute a lower and upper stellar density value for each object using the same method as in Section~\ref{sub:rho_vet}. If the 95$^\text{th}$ percentile value of the $\rho_{\rm trans}$ posterior is less than $0.1 \times \rho_{\ast; \text{ lower}}$ or if the 5$^\text{th}$ percentile value is greater than $10 \times \rho_{\ast; \text{ upper}}$ we exclude the candidate. Note the tighter upper boundary as compared to when considering $\rho_{\rm trans}$ in the light curve vetting stage. This is because while were are still assuming a circular orbit we are no longer assuming an equatorial orbit (i.e. $b = 0$) in the calculation of $\rho_{\rm trans}$. From this transit fitting analysis we identify \NTransitFitReject\ of our candidates as likely false positives, leaving a sample of \NTransitFitPass\ planet candidates. For the 10,000 injected transiting planets we apply our transit fitting analysis to 6.4\,\% of them were falsely labelled as false positives.

\subsubsection{Flat Line, Sinusoidal, and Odd-Even Eclipse Models}
We perform additional modelling analyses to assess whether a planet transit model best describes the light curves of each object. To assess this we fit each light curve with additional models. The first of these is a flat line model, governed by a single parameter, $C_\text{flat}$, to assess whether the data warrants the inclusion of a transit model. We fit for just a single free parameter: the constant flux level, $C_\text{flat}$.  The second is a sinusoidal model, to further assess whether the light curve is better explained by a transit model or a continuous variability like model. For the sinusoid model, the phase of the sinusoid is fixed such that a minimum of the sinusoid occurs at phase 0. We consider five possible sinusoid periods for each object: the detected BLS period for the candidate, along with the second, third, fourth, and fifth harmonics of this period. For each sinusoid period, we fit for the semi-amplitude of the sinusoid, $A$, as well as the flux baseline level, $C$, as free parameters, selecting the period which yields the highest log likelihood value as the best sinusoidal model.

For all analyses performed here we sample the parameters using an MCMC process using the \textsc{emcee} package, running eight independent walkers for 3,000 burn-in steps, followed by 15,000 samples per walker. We then use the Bayesian Information Criterion (BIC) to assess whether a transit-like model is warranted by the light curve over either the flat line or sinusoidal models. The BIC for a given model is calculated using the equation
\begin{equation}
    \text{BIC} = k \ln{n} - 2 \ln{\mathcal{L}} ,
\end{equation}
where $k$ is the number of parameters, $n$ is the number of data points, and $\mathcal{L}$ is the maximum log likelihood value obtained for a given model. For the transit model we use $k_{\rm trans} = 8$, for the sinusoidal model we use $k_{\rm sinusoid} = 4$, and for the flat line model we use $k_{\rm flat} = 1$. We note that the model which is preferred yields the lowest BIC value, with a BIC difference of greater than 10 denoting strong evidence in favour of a particular model \citep{neath2012bic}. As such, for an object to remain in our candidate sample it must satisfy these two criteria

\begin{align}
    \text{BIC}_\text{FlatLine} - \text{BIC}_\text{Transit} \geq 10 \text{ ,}\\
    \text{BIC}_\text{Sinusoid} - \text{BIC}_\text{Transit} \geq 10 \text{ .}
\end{align}

We also consider the scenario in which the light curve is better explained by an eclipsing binary scenario in which the odd and even eclipses are different depths. To assess this we perform a further round of transit fitting analysis using the same models and modeling prescription as in Section~\ref{sub:transit_model}. The one difference is an alteration to the prior used for \tc, where for this analysis we use a uniform prior between \tc$_\text{Best} \pm 0.5 \times \text{Dur}_\text{Best}$, where \tc$_\text{Best}$ and $\text{Dur}_\text{Best}$ are the best fit mid-transit time and transit duration from the analysis performed in Section~\ref{sub:transit_model}. 

During this analysis we perform two MCMC sampling processes, one for the odd eclipse events and one for the even eclipse events. We then take the sum of the maximum log likelihood values from these two sampling processes along with a number of parameters $k_{\rm odd-even} = 16$ to compute the BIC for the Odd-Even eclipse model. In this instance, we consider $\Delta \text{BIC} = \text{BIC}_\text{Transit} - \text{BIC}_\text{OddEven}$ and exclude any object with $\Delta \text{BIC} > 10$ as a likely false positive. We also extract the best fit radius ratio values for each eclipse, \rprs$_\text{Odd}$ and \rprs$_\text{Even}$, and compute the ratio of these two values. We then further exclude any object for which this ratio is less than 0.5 or greater than 2. \\

From these further modelling analyses we identify a further 657 candidates as likely false positives, leaving a planet candidate sample of 707 candidates. Of the 9,360 injected planets labelled as planet candidates by the transit fitting analysis 4.84\,\% are incorrectly labelled as false positives by the additional modelling analyses.

\subsubsection{Individual Sector Transit Fitting}\label{sub:indsecfit}
In order for any of the 707 planet candidates to remain as high quality transiting planet candidates, the transit signal must persist across all TESS sectors. To assess this for our candidates we fit all available QLP light curves for each candidate, fitting each sector individually. We downloaded the QLP light curves for each star using the \textsc{lightkurve} software \citep{lightkurve}. The fitting process used is the same as in Sector~\ref{sub:transit_model}, except we now use a burn-in phase of 3,000 steps per chain and a sampling phase of 20,000 steps per chain. 

From this analysis, we employ two checks to identify likely false positives. First, at each step in the sampling process we calculate the transit depth, $\delta$, of the model. Any objects for which more than 20\% of the MCMC samples from any individual sector yield $\delta == 0$ are rejected as false positives. These scenarios are most likely cases where systematic events during one or a small number of sectors, such as a high amount of scattered light, produces a spurious signal which in the phase fold can be mistaken for a transit signal. However, such a signal would not be seen in all sectors. 

For the second check, we identify objects for which the measured transit depth from the MCMC varies significantly between sectors. To do this we calculate the following quantity
\begin{equation}
    \text{Median}\left( \frac{|\delta_\text{i} - \text{Median}\left(\delta_\text{i}\right)|}{\sigma_{\delta\text{i}}} \right),
\end{equation}
where $\delta_\text{i}$ is the transit depth measured for an individual sector and $\sigma_{\delta\text{i}}$ is the transit depth uncertainty. In words, we calculate the absolute difference of each individual sector transit depth from the median of all the individual sector transit depth values. This difference is then weighted by the uncertainty in the individual transit depth. We then take the median of these values as our metric. Any candidate for which this metric is greater than three we take as showing a high level of transit variability characteristic of a false positive scenarios. This second check can identify nearby blended eclipsing binary scenarios (see Section~\ref{sub:sec_depth_diff}), spurious signals such as those caused by high scattered light events, or signals produced by stellar variability, for which the variability signal and amplitude can vary on a timescale of months to years.

Using this analysis we identify a further 260 false positives from our candidate list, leaving a sample of 447 planet candidates. Considering the injection and recovery simulations, 3.2\% of the injected planets for which the individual sector fitting analysis was performed were falsely labelled as false positives.

\subsection{Final Candidate Checks}\label{sec:final_checks}
\subsubsection{Blend Scenario Final Analysis}
The TESS pixels are 21\,\arcsec\ on each edge. Therefore, light curves generated from the TESS full frame images can suffer from contamination from nearby stars. If these nearby stars are themselves eclipsing binaries, this can result in apparent transit events being observed for the target. A crucial step in TESS planet searches is to identify and remove these nearby blend cases.

To do this, we use the \textsc{transit-diffImage} tool\footnote{\url{https://github.com/stevepur/transit-diffImage}} to produce difference images for the transit events observed for our candidates. We use \textsc{transit-diffImage} to determine the centroid location of the difference image for each sector using a Pixel Response Function (PRF) modeling approach. We then compare the PRF centroid position of the difference image to the catalog pixel position of the target at the time of the TESS observations. Any object for which the PRF centroid position is offset from the target position by greater than 1 pixel is excluded as a likely nearby blend. Out of the 447 planet candidates which pass the transit fitting analysis, 236 are identified and excluded as likely nearby blends based on the PRF centroid analysis, leaving a sample of 211 planet candidates.

Nearby blended scenarios can also be identified using the \textsc{TESSPositionalProbability} code \citep{Hadjigeorghiou2024tessposprob}, which quantifies the probability that each star in the TESS image surrounding the target could be the real source of the observed signal. However, currently \textsc{TESSPositionalProbability} only works using light curves from the SPOC pipeline. For this study, 171 of the remaining 211 candidates do have a TESS-SPOC FFI light curve available \citep{caldwell2020tessspoc} and so we run \textsc{TESSPositionalProbability} for this subset of our candidates. We reject any object for which the target star does not have the highest probability as a likely false positive, thereby identifying and excluding a further 34 candidates as likely nearby blends.

Combined, these nearby blend checks identify a total of 270 likely blended false positive scenarios, leaving us with 177 high quality, likely on target, transit planet candidates.

\subsubsection{Gaia Non-Single Stars}
The \textit{Gaia} mission \citep{GaiaDR2,GaiaDR3} released a catalogue of non-single stars (NSS) as a part of its third data release\footnote{\url{https://gea.esac.esa.int/archive/documentation/GDR3/Gaia_archive/chap_datamodel/sec_dm_non--single_stars_tables/}}. We cross-matched our target list with this catalogue to identify whether any of our candidates have been found to be stellar multiple systems. We find thirteen objects identified as NSS, although two of these (TIC-29119552 and TIC-148340346) are identified with periods longer than 100 days and a further three (TIC-155858369, TIC-386699314, and TIC-397510904) simply show long term astrometric or spectroscopic trends indicative of a wide stellar companion. Such stellar multiples will not mimic the observed transit signals and will not preclude these target stars from also hosting short period planets, and so we do not remove these from our candidate list. However, seven of our candidates (TIC-46627823, TIC-72556406, TIC-165493409, TIC-231279168, TIC-231630147, TIC-289539327, and TIC-449050248) are listed as NSS with the \textit{Gaia} NSS period matching the period of our planet candidate. We therefore identify these systems as eclipsing binaries and remove them from our candidate list. One further candidate (TIC-231630147) is listed as an NSS with a period of 44.85 days. While this does not match the period of the planet candidate, a stellar companion at such a period would very likely prohibit the formation and stable orbit of any planet at a close orbital distance, and so we also remove this candidate from our sample. Overall, we exclude eight candidates from our sample based on the \textit{Gaia} NSS information, leaving 169 planet candidates.

Further to this, thirteen of our candidates are reported as eclipsing binary or stellar multiple false positives by the TESS Follow-Up Team via ExoFoP\footnote{\url{https://exofop.ipac.caltech.edu/tess/}}. We remove these thirteen from our candidate list, leaving 156 planet candidates.

\subsubsection{Visual Light Curve Inspection}\label{sec:visual_inspect}
As a last check, we now manually inspect the light curves of the remaining objects, to determine if any are clear false positives which evaded the automatic checks deployed in this work. We first visually inspect the 10\,min cadence QLP light curves for our candidates. We find ten of our candidates to be likely false positives, and so exclude them from our sample. The light curves of these ten candidates display either continuous variability, odd-even depth differences, or secondary eclipses which show the objects to be variable stars or eclipsing binaries. While these signals are visible to the eye they do not meet the signal-to-noise thresholds to be identified by our automated light curve vetting. For one further candidate - TIC-45047401 - our visual inspection reveals the presence of just a single clear transit event. While this may still be a true planet, the true orbital period will be significantly longer than our 12\,day upper limit. Therefore we also excluded TIC-45047401 from our candidate sample.

\begin{figure}
    \centering
    \includegraphics[width=0.985\linewidth]{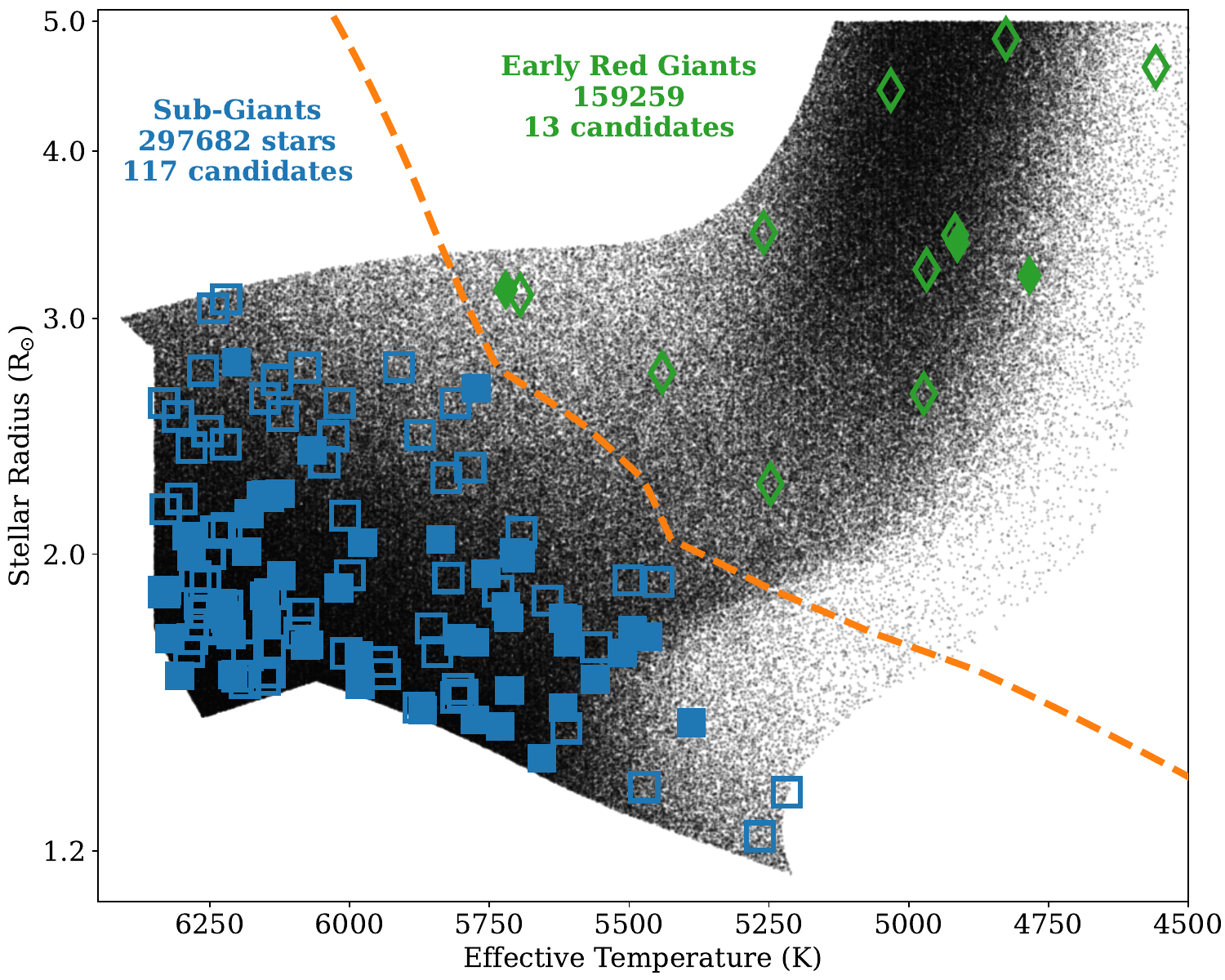}
    \caption{Host stars of the 130 planet candidates detected using our planet search and vetting pipeline. The 117 host stars within our sub-giant sub-population are plotted as the blue squares, and the 13 host stars within our early red giant sub-population are plotted as the green diamonds. Filled in symbols highlight host stars of confirmed planets (according to the NASA exoplanet archive), and open symbols highlight those stars which host planet candidates which are yet to be confirmed. Our full stellar sample is plotted as the black points and the orange line shows the $\text{EEP} = 465$ boundary we use to separate the two sub-populations (see Section~\ref{sub:evolstars}).}
    \label{fig:planet_sample}
\end{figure}

After this we now visually inspect all the QLP, TESS-SPOC FFI, and SPOC 2\,minute cadence light curves which exist for our remaining 145 candidates. Similar to the analysis performed in Section~\ref{sub:indsecfit} we are primarily checking to ensure a consistent transit signal is observed across all sectors and light curve extraction pipelines. Through this inspection, we identify fifteen of our candidates as likely false positives, and remove them from our candidate sample. This leaves us with a final sample of 130 planet candidates. We plot the host stars of our candidates in relation to our overall input stellar sample in Figure~\ref{fig:planet_sample}. 

\section{Planet Candidate Sample}\label{sec:planets}
\begin{figure}
    \centering
    \includegraphics[width=\linewidth]{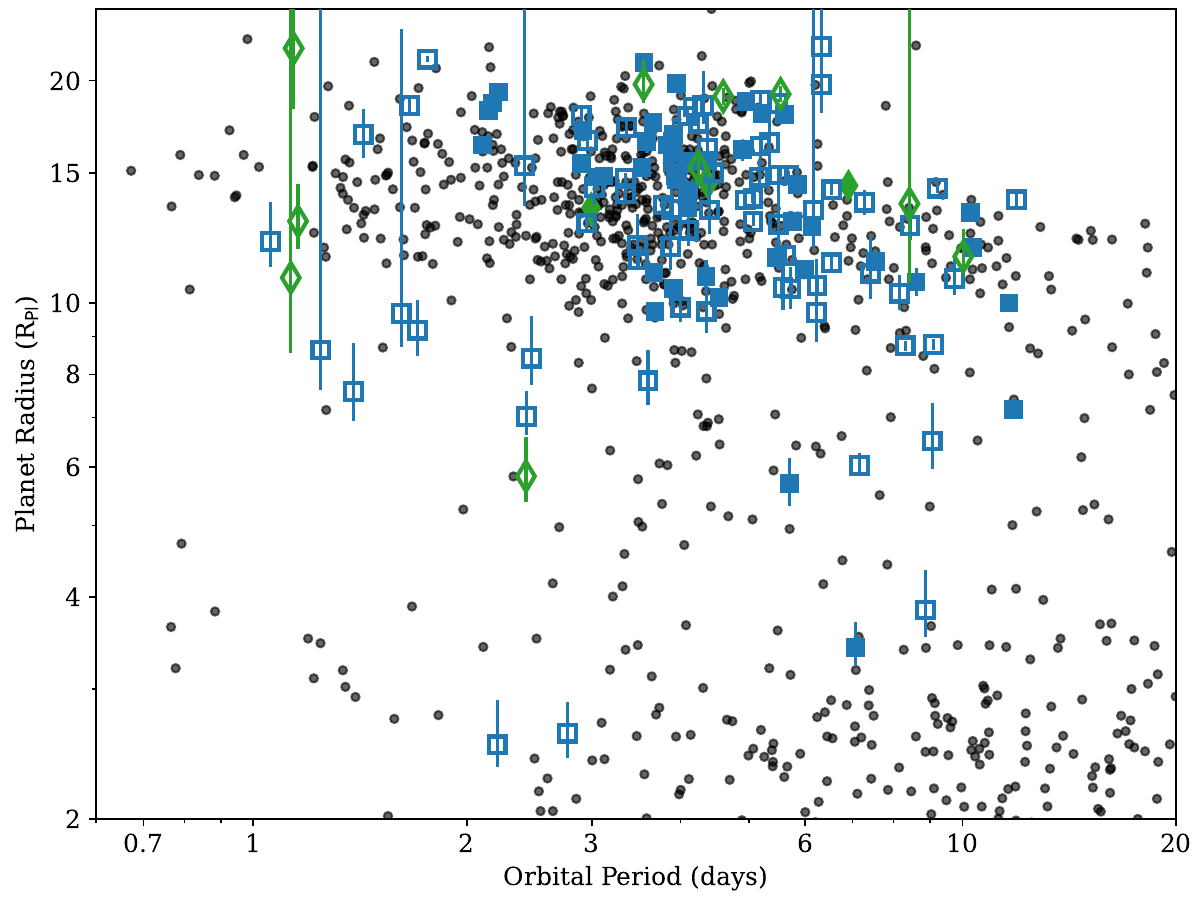}
    \caption{Our post-main sequence sample of planets and planet candidates. As with Figure~\ref{fig:planet_sample}, planets and candidates around stars in our sub-giant sub-population are plotted as the blue squares and those around stars in our early red giant sub-population are plotted as the green triangles. Confirmed planets (according to the NASA exoplanet archive) are plotted as the filled symbols and open symbols show as yet unconfirmed planet candidates. The black points show the population of known planets from the NASA exoplanet archive with a planet mass and radius measured to better than 40\,\% precision.}
    \label{fig:planets}
\end{figure}
Out of our 130 planet candidates, a selection of 48 of our candidates are already known planets present in the NASA exoplanet archive\footnote{\url{https://exoplanetarchive.ipac.caltech.edu/index.html}; accessed on 4th March 2025}. A further 49 are reported as candidates in the TESS Objects of Interest (TOI) catalogue \citep{guerrero2021tois}. This leaves 33 new candidates not yet reported to the community. We plot the phase-folded QLP light curve for each of the 33 new planet candidates we detect in Figure~\ref{fig:newCandLCs1}.
We plot the orbital periods and radii of all 130 of our planet candidates in Figure~\ref{fig:planets} and provide details on them and their host stars in Tables~\ref{tab:planet_new},~\ref{tab:planet_toi},~\ref{tab:planet_kp},~and~\ref{tab:planet_small}. From Figure~\ref{fig:planets} we can see that the bulk of our detected planets and candidates sit at orbital periods longer than approximately 3\,days, with only a few planet candidates seen at short orbital periods $P \leq 2\,\mathrm{days}$. This is in contrast to the current known population of primarily main sequence stars for which there is a decent sized population of giant planets at these short periods. This already is suggesting that strong tidal interactions may be at play for our post-main sequence population \citep[e.g.][]{barker2020tides, esseldeurs2024evolvedstartides, weinberg2024tides}, although we will need to compute the occurrence rates before drawing strong conclusions about the population demographics. 

\subsection{False Positive Probabilities}\label{sub:fpps}
Despite our extensive vetting of our planet candidates, there will remain some false positives within our planet candidate sample. For the calculation of the planetary occurrence rates, we therefore now estimate a False Positive Probability (FPP) for each of our candidates. For the candidates in our sample which have been independently published as confirmed objects, we assign $\text{FPP} = 0$. For the remaining candidates, we estimate the FPP values using the \textsc{triceratops} software \citep{triceratops1_2020,triceratops2_2021}. \textsc{triceratops} is a Python code which simulates a range of possible transiting planet or eclipsing binary scenarios which could produce the observed transit signal, and then calculates the relative probabilities of each scenario for a given target star. Using these scenarios and probabilities, each candidate is then assigned a FPP value. For the 82 candidates for which we calculate a \textsc{triceratops} FPP value we obtain a mean FPP value of 0.41. Of these 82 candidates, 30 have a value of FPP > 0.5.

\section{Injection-Recovery Simulations}\label{sec:injrec}
\begin{figure*}
    \centering
    \includegraphics[width=\linewidth]{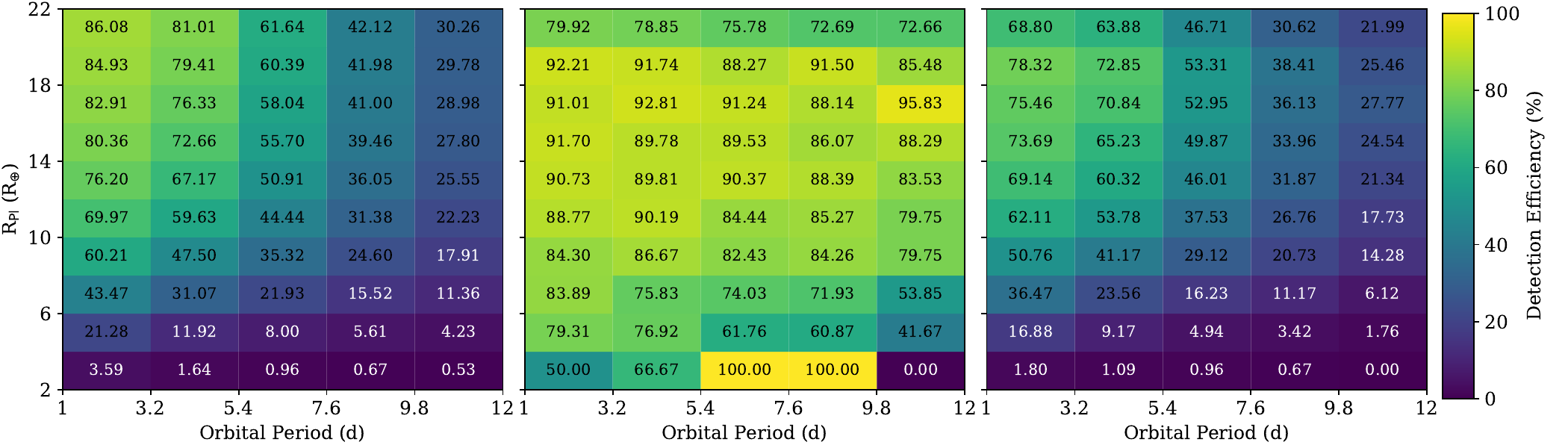}
    \caption{Detection efficiency of stages of our planet search and vetting pipeline. The three panels show the detection efficiencies for: \textbf{left:} our full planet search and automated light curve vetting (see Sections~\ref{sub:bls}~and~\ref{sec:FPs}); \textbf{middle:} our complete transit fitting analysis (see Section~\ref{sec:fitting}); \textbf{right:} the combined performance of the complete planet search, vetting, and fitting analysis pipeline. The colour of the grid cells gives the detection efficiency (\%) for planets within the cell. The numbers in each box also give the percentage detection efficiency within each grid cell.}
    \label{fig:detect_eff}
\end{figure*}
\begin{figure}
    \centering
    \includegraphics[width=\linewidth]{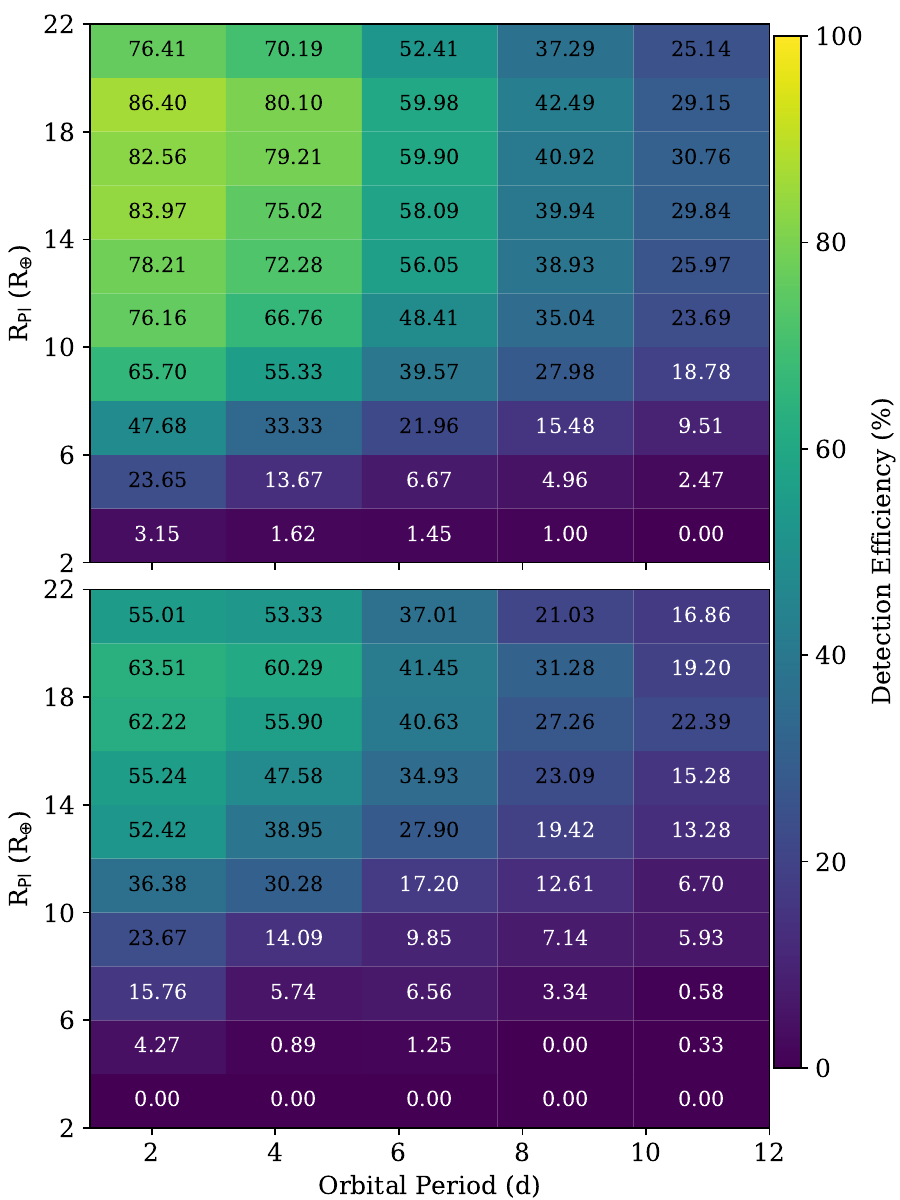}
    \caption{Detection efficiency of our full planet search and vetting pipeline for our two stellar sub-populations (see Section~\ref{sub:evolstars} and Figure~\ref{fig:stellar_sample}). The detection efficiency for our sub-giant sub-population in the top panel and for our early red giant sub-population in the bottom panel. As for Figure~\ref{fig:detect_eff} the colour of the grid cells gives the detection efficiency (\%) for planets within the cell and the numbers in each box also give the percentage detection efficiency within each grid cell. The information shown on each panel is analogous to the right panel of Figure~\ref{fig:detect_eff}.}
    \label{fig:detect_eff_2SubPops}
\end{figure}

We utilise planet injection-recovery simulations to quantify the detection efficiency of our pipeline. This is a vital step in converting our population of planet candidates into an overall occurrence rate estimation. As has been performed during previous planet population studies \citep[e.g.][]{vansluijs2018whitedwarfs,bryant2023lmstaroccrates} we simulate and inject planet transits into real TESS light curves. We use as our base light curves the 408,758 stars from our initial input sample which did not yield a significant BLS detection. For each of these stars we simulated four transiting planet signals, giving us a total of 1,635,032 injected planet light curves.

We applied our full transit search and light curve vetting algorithms (Sections~\ref{sub:bls}~and~\ref{sec:FPs}) to all simulated light curves. Due to computational constraints we are unable to perform a full transit fitting analysis on all 626,330 simulated objects which pass the BLS and light curve vetting stages. However, to assess the performance of our light curve fitting stages we instead randomly select a set of 10,000 simulated light curves, performing our full transit fitting analysis (Section~\ref{sec:fitting}) on this subset. 

We provide the results of these tests for our full stellar sample in Figure~\ref{fig:detect_eff} and for our two sub-populations independently in Figure~\ref{fig:detect_eff_2SubPops}, plotting our pipeline detection efficiency as a function of planet period and radius.  From these results we can see that the pipeline detection sensitivity decreases to lower radii, and is significantly reduced for small radii, $R_P \leq 8$\rearth, especially for the early red giant sub-population. Therefore, while before now we have not introduced a lower limit on the planet radius in our planet search, when determining the occurrence rates we focus on the giant planet population, considering only planets with $\rpl \geq 8\,\rearth$. 

Of our 130 planets and planet candidates, for twelve we measure $\rpl \leq 8\,\rearth$, including three previously known planets: HD~89345\,b \citep[TIC-350020859; ][]{vaneylen2018hd89345}, TOI-329\,b \citep[TIC-169765334; ][]{polanski2024toi320}, and TOI-1736\,b \citep[TIC-408618999; ][]{akanamurphy2023toi1736}, and four TOI candidates: TOI-1291.01 (TIC-198186769), TOI-5069.01 (TIC-381360757), TOI-5177.01 (TIC-350033870), and TOI-5645.01 (TIC-366804698). While we do not include these twelve planets and candidates in our occurrence rate analysis we do include details on them and their host stars in Table~\ref{tab:planet_small}. 

\section{Occurrence Rate Determination}\label{sec:occrates}
\begin{figure*}
    \centering
    \includegraphics[width=0.9\linewidth]{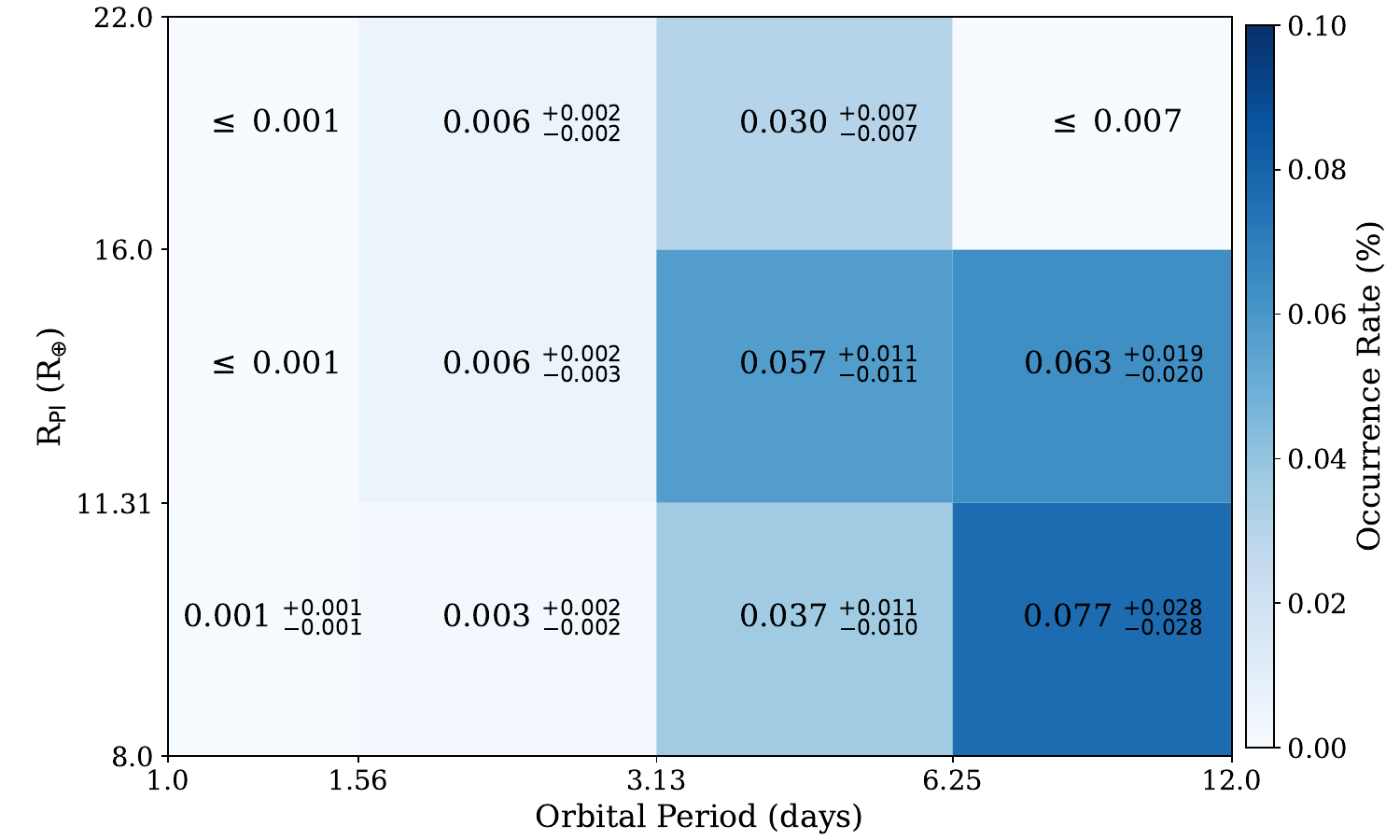}
    \caption{Occurrence rates measured in this study for post-main sequence host stars as a function of the planetary radius, \rpl, and orbital period. The colour of each grid cell gives the occurrence rate in that grid cell, and the numbers quoted also provide a measure of the occurrence rate (\%) within the corresponding grid cell. In the grid cells for which the measured occurrence rate is zero we report the 68$^{\rm th}$ percentile upper limit.}
    \label{fig:occ_rate_grid}
\end{figure*}
We now determine the planet occurrence rates for our post-main sequence stellar population using our sample of detected planet candidates and the detection efficiencies we determine using our injection-recovery tests. We calculate the occurrence rate, $\focc = n_\text{pl} / N^{\prime}$, where $n_\text{pl}$ is the number of detected planets and $N^{\prime}$ is the number of stars amenable to the detection of an exoplanet \citep[see][]{bayliss2011hjFreq,gan2023mdwarfoccrates,bryant2023lmstaroccrates}. We use our estimated false positive probability (FPP) for each candidate (see Section~\ref{sub:fpps}) to compute $n_\text{pl}$ as
\begin{equation}
    n_\text{pl} = \Sigma_{i} \left(1 - \text{FPP}_{i}\right).
\end{equation}
$N^{\prime}$ is calculated using the results of our injection-recovery simulations as
\begin{equation}
    N^{\prime} = N_{\ast} \frac{\Sigma_{i=1}^{N_{\ast \rm ; \ sim}} \delta_{{\rm det}, i} \mathcal{P}_{{\rm tr}, i}}{N_{\ast \rm ; \ sim}} \frac{\Sigma_{j=1}^{N_{\ast \rm ; \ fit}} \delta_{{\rm fit}, j}}{N_{\ast \rm ; \ fit}} ,
\end{equation}

where $N_\ast$ is the number of stars in the input sample, $N_{\ast \rm ; \ sim}$ is the total number of simulated planet light curves, $N_{\ast \rm ; \ fit}$ is the number of simulated light curves included in the transit fitting analysis, $\delta_{{\rm det}, i}$ is a detection delta function equal to one if a given simulated planet was detected as a candidate by the BLS (Section~\ref{sub:bls}) and labelled as a planet candidate by the light curve vetting (Section~\ref{sec:FPs}) or zero otherwise, $\delta_{{\rm fit}, i}$ is a similar detection delta function but for the transit fitting analysis (Section~\ref{sec:fitting}), and $\mathcal{P}_{{\rm tr}, i}$ is the geometric probability that a given injected planet would transit. This transit probability is given as

\begin{equation}
    \mathcal{P}_{{\rm tr}, i} = 0.9 \times \frac{R_{\ast}}{a} ,
\end{equation}
where $R_\text{P}$ is the radius of the injected planet, $R_{\ast}$ is the stellar radius of the host, and $a$ is the semi-major axis of the orbit. The factor of $0.9$ is applied as in our search we only consider as a planet any candidate for which our transit fitting analysis yields an impact parameter of $b \leq 0.9$.

We know that giant planet occurrence rates are dependent on the orbital period, $P$, and radius of the candidate, \rpl, \citep[e.g.,][]{KunimotoMatthews2020} and from Figure~\ref{fig:detect_eff} we see that our detection efficiency also depends on $P$ and \rpl.
Therefore, when computing the occurrence rates we split up the planet candidate and injected planet samples into a grid in orbital period and planet radius. For each $\left[P, R_\text{P}\right]$ grid cell we then compute number of detected planets, their false positive probabilities, and the $N^{\prime}$ value. 

To assess the uncertainty in our occurrence rate calculations, we perform a bootstrapping analysis. For this, we randomly draw 1000 sets of stars from our input sample. These randomly drawn stellar samples consist of \Ntotal\ stars, the same size as the initial stellar sample, and are drawn with replacement. We then compute the occurrence rates using our approach above for each random sample. From the set of occurrence rate samples we then compute the 16$^\text{th}$ and 84$^\text{th}$ percentile values, taking the difference between these values and our measured occurrence rate value as the 1$\sigma$ uncertainty value on the occurrence rate. 

We note that these calculations are performed independently for each $\left[ P, \rpl \right]$ grid cell. For any grid cell in which we have no detected planet candidates, or only planet candidates with $1 - \mathrm{FPP} < 10^{-4}$, we instead take the occurrence rate value to be equal to zero with an upper uncertainty obtained by calculating the 68$^\text{th}$ percentile upper limit. We calculate these upper limits following \citet{vansluijs2018whitedwarfs} who note that the occurrence rate upper limit can be calculated using
\begin{equation}
    f_\text{max}\left(P, \rpl\right) = 1 - (1 - C)^{\frac{1}{N^{\prime}\left(P, R_\text{P} \right) + 1}},
\end{equation}
where $C$ is the confidence level for the upper limit. 

\section{Results and Discussion}\label{sec:res_and_disc}
\subsection{Post-Main Sequence Occurrence Rates}\label{sub:pms_occrates}

{\renewcommand{\arraystretch}{1.3}
\begin{table*}
    \centering
    \begin{tabular}{|l|c|c|c|c|}
    \toprule
    \textbf{Stellar Host Population} & \textbf{Period Range (days)} & \textbf{Planet Radius Range (\rearth)} & \textbf{Occurrence Rate (\%)} & \textbf{Comments}\\
    \toprule
     \textbf{Full Post-Main Sequence} & $1 \leq P \leq 12$ & $8 \leq \rpl \leq 22$ & $0.28\pm0.04$ & Figure~\ref{fig:occ_rate_grid} \\
      & $1 \leq P < 6$ & $8 \leq \rpl \leq 22$ & $0.12 \pm 0.02$ & \\
      & $6 \leq P \leq 12$ & $8 \leq \rpl \leq 22$ & $0.17 \pm 0.04$ & \\
      & $1 \leq P \leq 10$ & $8 \leq \rpl \leq 22$ & $0.23 \pm 0.03$ & $\ast$ \\
     & & & & \\
     \textbf{Sub-Giant Sub-Population} & $1 \leq P \leq 12$ & $8 \leq \rpl \leq 22$ & $0.35\pm0.05$ & Figure~\ref{fig:occ_grid_subPops} \\
      & $1 \leq P < 6$ & $8 \leq \rpl \leq 22$ & $0.17\pm0.02$ & Figure~\ref{fig:occrate_vs_Per_2SS} \\
      & $6 \leq P \leq 12$ & $8 \leq \rpl \leq 22$ & $0.19\pm0.05$ & Figure~\ref{fig:occrate_vs_Per_2SS} \\
      & $1 \leq P \leq 10$ & $8 \leq \rpl \leq 22$ & $0.30^{+0.05}_{-0.04}$ & Figure~\ref{fig:occrate_vs_MainSeq} \\
      & $1 \leq P \leq 10$ & $9 \leq \rpl \leq 22$ & $0.27 {\pm 0.04}$ & $\dagger$ ; Figure~\ref{fig:occrate_vs_Mass} \\
      & & & & \\
     \textbf{Early Red Giant Sub-Population} & $1 \leq P \leq 12$ & $8 \leq \rpl \leq 22$ & $0.11^{+0.06}_{-0.05}$ & Figure~\ref{fig:occ_grid_subPops} \\
      & $1 \leq P < 6$ & $8 \leq \rpl \leq 22$ & $0.02\pm^{+0.02}_{-0.01}$ & Figure~\ref{fig:occrate_vs_Per_2SS} \\
      & $6 \leq P \leq 12$ & $8 \leq \rpl \leq 22$ & $0.11^{+0.14}_{-0.07}$ & Figure~\ref{fig:occrate_vs_Per_2SS} \\
      & $1 \leq P \leq 10$ & $8 \leq \rpl \leq 22$ & $0.06^{+0.08}_{-0.03}$ & Figure~\ref{fig:occrate_vs_MainSeq} \\
      & $1 \leq P \leq 10$ & $9 \leq \rpl \leq 22$ & $0.06^{+0.09}_{-0.03}$ & $\dagger$ ; Figure~\ref{fig:occrate_vs_Mass} \\
      \toprule
      \multicolumn{5}{l}{$\ast$ - Same parameter range as used for the comparison to previous main sequence studies in Figure~\ref{fig:occrate_vs_MainSeq}} \\
      \multicolumn{5}{l}{$\dagger$ - Direct comparison to \citet{beleznay2022tessoccrates}}
    \end{tabular}
    \caption{Summary of the occurrence rates derived in this work for different planet parameter ranges and for different stellar populations. We provide occurrence rates for our full stellar sample as well as for our two sub-populations, which we distinguish using a boundary of equal Evolutionary Equivalent Phase (EEP = 465) in $\left[T_\text{eff}, R_\ast\right]$ space (see Section~\ref{sub:evolstars} and Figure~\ref{fig:stellar_sample} for more details).}
    \label{tab:occ_rates_all}
\end{table*}
}

Using the results from our planet search and vetting we measure a short period, giant planet ($1 \leq P \leq 12$\,d and $8 \leq \rpl \leq 22$\,\rearth) occurrence rate value of $0.28\pm0.04\%$ for our full post-main sequence stellar population. We also compute occurrence rate values of $0.35\pm0.05\%$ and $0.11^{+0.06}_{-0.05}\%$ for our sub-giant and early red giant sub populations respectively. Full details of how we define these sub-populations are provided in Section~\ref{sub:evolstars}, but in brief we distinguish the two populations using a boundary of equal Evolutionary Equivalent Phase (EEP = 465) in $\left[T_\text{eff}, R_\ast\right]$ space (see Figure~\ref{fig:stellar_sample}), such that our early red giant sub-population consists of stars which are more evolved than those in our sub-giant sub-population. The occurrence rate we measure for our early red giant sub-population is lower than the occurrence rate we measure for our sub-giant sub-population at a level of $3.1\sigma$.

\begin{figure*}
    \centering
    \includegraphics[width=\linewidth]{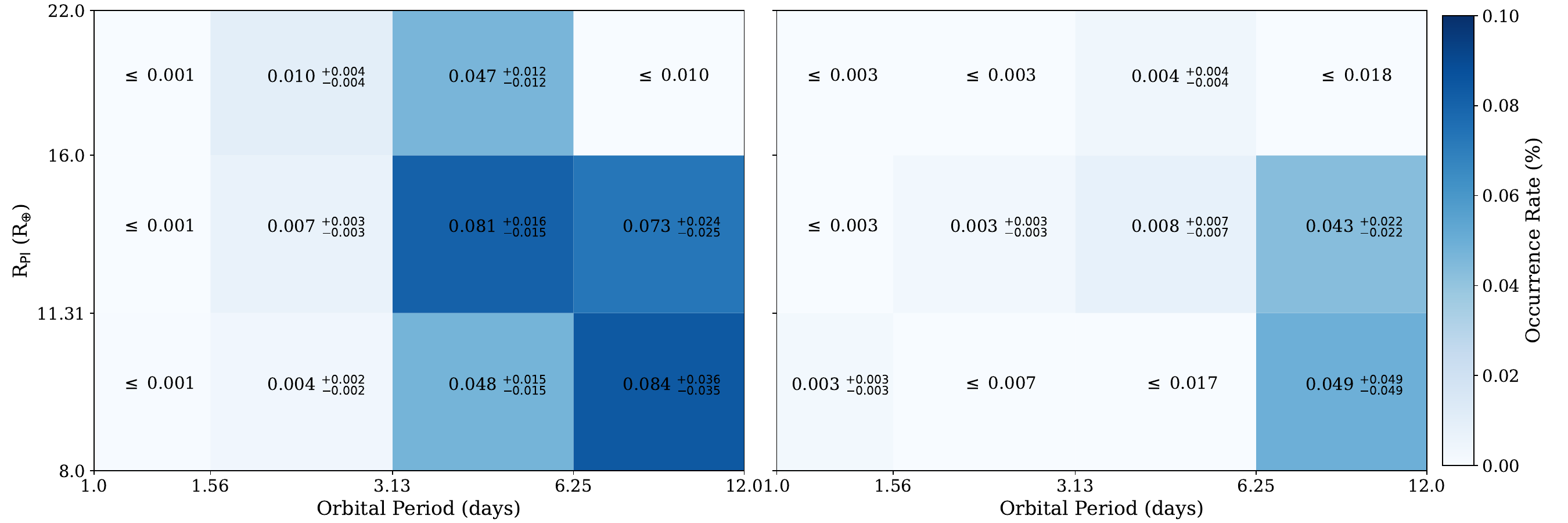}
    \caption{Occurrence rates measured for the two sub-populations considered in this study, with the sub-giant sub-population (EEP $< 465$) on the left and the more evolved, early red giant sub-population (EEP $\geq 465$) on the right. The colour scale and quoted numbers are as in Figure~\ref{fig:occ_rate_grid}.}
    \label{fig:occ_grid_subPops}
\end{figure*}

We plot our occurrence rate measurements as a function of planet period and radius for our full post-main sequence sample in Figure~\ref{fig:occ_rate_grid}. The full occurrence rate results for the two sub-populations considered are plotted in Figure~\ref{fig:occ_grid_subPops}. We provide a summary of the giant planet occurrence rates we measure in this work in Table~\ref{tab:occ_rates_all}.

\begin{figure}
    \centering
    \includegraphics[width=\linewidth]{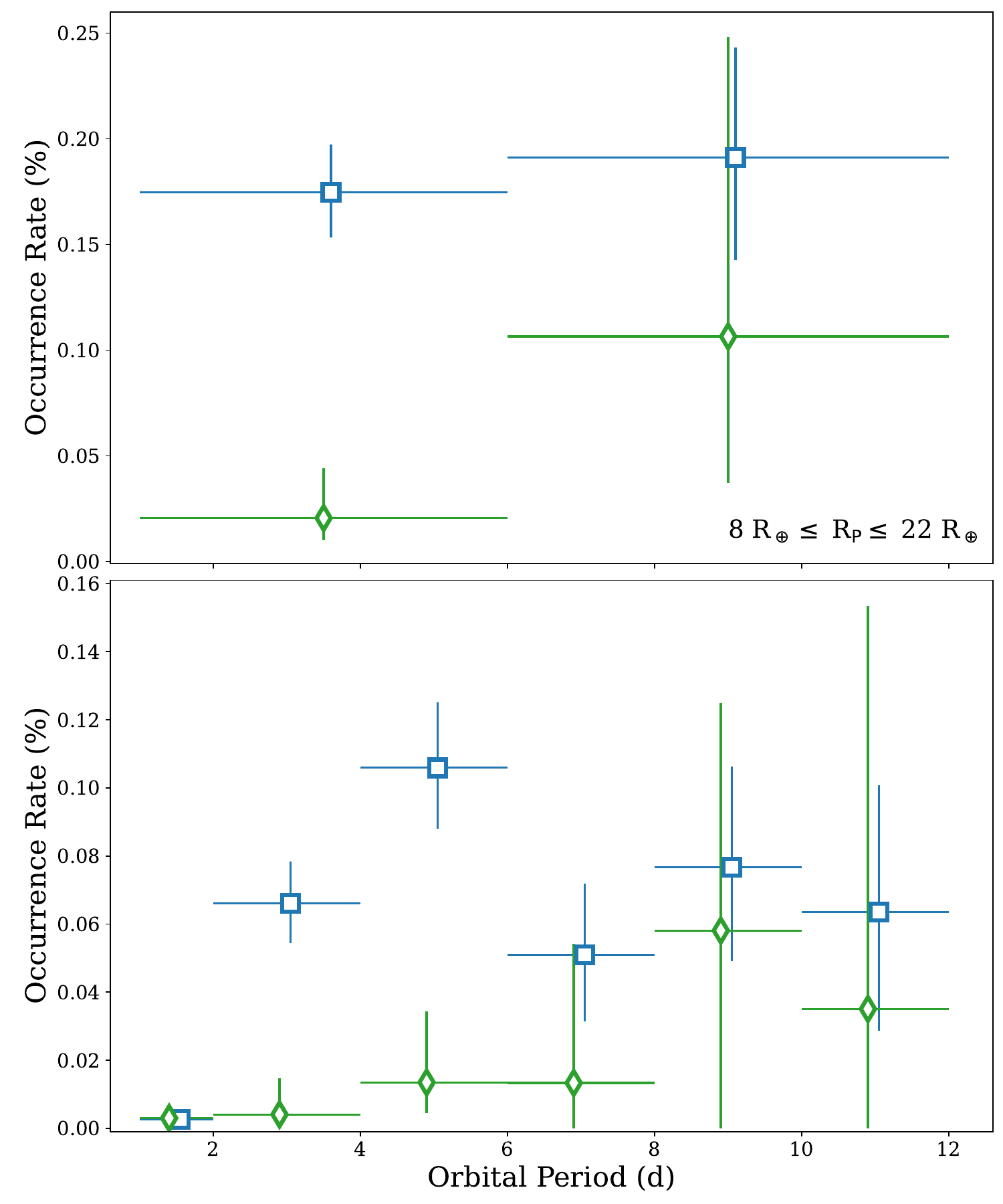}
    \caption{Period dependence of the giant planet occurrence rates we measure for the two post-main sequence sub-populations we consider (see Figure~\ref{fig:stellar_sample}), plotting the occurrence rates for the sub-giant sub-population as the blue squares, and for the early red giant sub-population as the green diamonds.}
    \label{fig:occrate_vs_Per_2SS}
\end{figure}

Our results show a clear decrease in the occurrence rate of short period giant planets for the more evolved host star sub-population. This suggests that the population of these giant planets decreases in number during these early stages of post-main sequence stellar evolution. From our occurrence rate results shown in Figure~\ref{fig:occ_grid_subPops} we see that the occurrence rates for both sub-populations show a strong dependence with the orbital period of the planet. We compare the period dependence of the occurrence rates in Figure~\ref{fig:occrate_vs_Per_2SS}, where we can clearly see that the difference in occurrence rates between these two sub-populations is also dependent on the period of the planet. For moderate periods ($6 \leq P \leq 12\,\mathrm{days}$) we measure occurrence rates of $0.19\pm0.05\%$ for the sub-giant host star and $0.11^{+0.14}_{-0.07}\%$ for the more evolved early red giant host stars. The occurrence rates for these moderate periods are consistent between the two sub-populations within the uncertainties. However, for shorter orbital periods ($1 \leq P < 6\,\mathrm{days}$) the occurrence rate is significantly lower for the more evolved host stars. We measure occurrence rates of $0.17\pm0.02\%$ and $0.02^{+0.02}_{-0.01}\%$ for the sub-giant and early red giant sub-populations respectively finding an occurrence rate reduction for the more evolved early red giant sub-population at a level of $5.3\sigma$.

Using a finer period spacing of 2 days shown in the bottom panel of Figure~\ref{fig:occrate_vs_Per_2SS} we see that the largest measured occurrence rate differences are within the period range $2\,\mathrm{days} \leq P < 6\,\mathrm{days}$, with consistent occurrence rate measurements for the shortest orbital periods, $P < 2\,\mathrm{days}$, for which we find an occurrence rate for our sub-giant sub-population of $0.003 \pm 0.002$\,\% and an occurrence rate for our early red giant sub-population of $0.003 \pm 0.003$\,\%. However, we note that the reduced number of planet detections for the early red giant host stars when considering more period bins in this way results in larger occurrence rate uncertainties. Therefore, at this stage we cannot say whether the agreement between our occurrence rate measurements for $P < 2\,\mathrm{days}$ is a real astrophysical result or is simply due to the limited statistical significance of this measurements.

Our results clearly and significantly yield two key results about the population of short period giant planets with post-main sequence host stars. The first is that these occurrence rates are significantly lower for the more evolved early red giant host star sub-population, compared to the less evolved sub-giant host star sub-population. The second is that this occurrence rate reduction shows a strong period dependence, with shorter period planets showing a larger occurrence rate reduction.

\subsection{Comparison to previous works}\label{sub:prev_comp}
Only a few studies have previously measured planetary occurrence rates for post-main sequence host stars. Radial velocity surveys of giant stars have been performed \citep[e.g.][]{bowler2010, reffert2015} although due to the low numbers of stars studied these surveys only yielded detections of planets on much wider orbits than those we study in this work. For example, \citet{bowler2010} report the results of a survey of just 31 evolved stars and out of all their detected planets the lowest orbital semi-major axis is 0.68\,AU. Whereas all our detected planets and candidates are at much closer separations of $a \leq 0.2\,\mathrm{AU}$. As these works probe a much different parameter space they do not provide a good comparison to our results, however we do note that this observed lack of planets at close orbital distances motivated suggestions that these planets had been destroyed by tidal interactions \citep[e.g.][]{schlaufman2013}. 

One study to which we can directly compare our results is that of \citet{grunblatt2019k2RGB} who used photometry from the \textit{K2} mission to measure an occurrence rate of short period giant planets ($3.5 \leq P \leq 10\,\mathrm{days}$; $\rpl \geq 1\,\rjup$) around low-luminosity red giant branch stars ($\rstar = 3 - 8\,\rsun$) of $0.49\pm0.28\%$. We note that this stellar sample likely consists of some stars of a similar evolution stage to the more evolved stars in our early red giant sample, and some stars which are more evolved than those we study in this work. Due to the large uncertainty of the \citet{grunblatt2019k2RGB} measurement, we find a difference between this value and the measured occurrence rate of $0.11^{+0.06}_{-0.05}\%$ for our early red giant sub-population of just $1.3\,\sigma$. The \citet{grunblatt2019k2RGB} measurement is based on the detection of just three planets, with orbital periods of 8.4, 9.2, and 9.3 days. Therefore, their results also show evidence of a large reduction in occurrence rates for shorter orbital periods for evolved host stars, in qualitative agreement with our results. 

Considering the null detection by \citet{grunblatt2019k2RGB} for giant planets with $3.5 \leq P \leq 8\,\mathrm{days}$ we compute the occurrence rate for our early red giant sub-population for the period range $3.5 \leq P \leq 8$, measuring a value of ${0.026^{+0.045}_{-0.016}\%}$. The stellar sample surveyed by \citet{grunblatt2019k2RGB} consisted of 2476 stars. Therefore, considering our occurrence rate measurements we would expect just ${0.64^{+1.11}_{-0.40}}$ giant planets with $3.5 \leq P < 8\,\mathrm{days}$ around these stars. Geometric transit probabilities for these systems will be on the order of 10-20\%, and so the non-detection of any giant planets with $P \leq 8\,\mathrm{days}$ by \citet{grunblatt2019k2RGB} is to be expected from our occurrence rate measurements. Therefore, our results for the population of giant planets with post-main sequence stars are not inconsistent with the previous findings from \citet{grunblatt2019k2RGB}, but they are significantly more precise due to the much larger stellar sample made available to us by the TESS Full-Frame-Images.

At the end of their post-main sequence evolution the stars we studied in this work will become white dwarfs, and so it is interesting to contrast our results with the current investigations into the planetary populations around white dwarf stars. There have been a few studies probing the occurrence of planets around white dwarfs \citep{vansluijs2018whitedwarfs, robert2024whitedwarfs}. Due to null detections of any planets \citet{vansluijs2018whitedwarfs} could only place an upper limit on the short period giant planet occurrence rate of $< 1.5\%$. The search of \citet{robert2024whitedwarfs} also did not yield the detection of any transiting giant-planet sized bodies. These null detections of any giant planets is in agreement with the rapid decrease in short period giant planet occurrence rates we observe in the post-main sequence stars we study. However there have been a few individual detections of planets orbiting white dwarfs \citep[e.g.][]{gansicke2019wdplanet, manser2019WDplanet, vanderburg2020wdplanet}. This suggests that a small number of planets may survive to the white dwarf stage of stellar evolution. Although these planets are expected to have resided on much wider orbits ($a > 1\,\mathrm{AU}$) during the stellar evolution stages we probe here \citep{vanderburg2020wdplanet} and so again these works do not provide a direct contradiction of our results.

\subsection{Comparison to the Main Sequence population}
\begin{figure}
    \centering
    \includegraphics[width=0.95\linewidth]{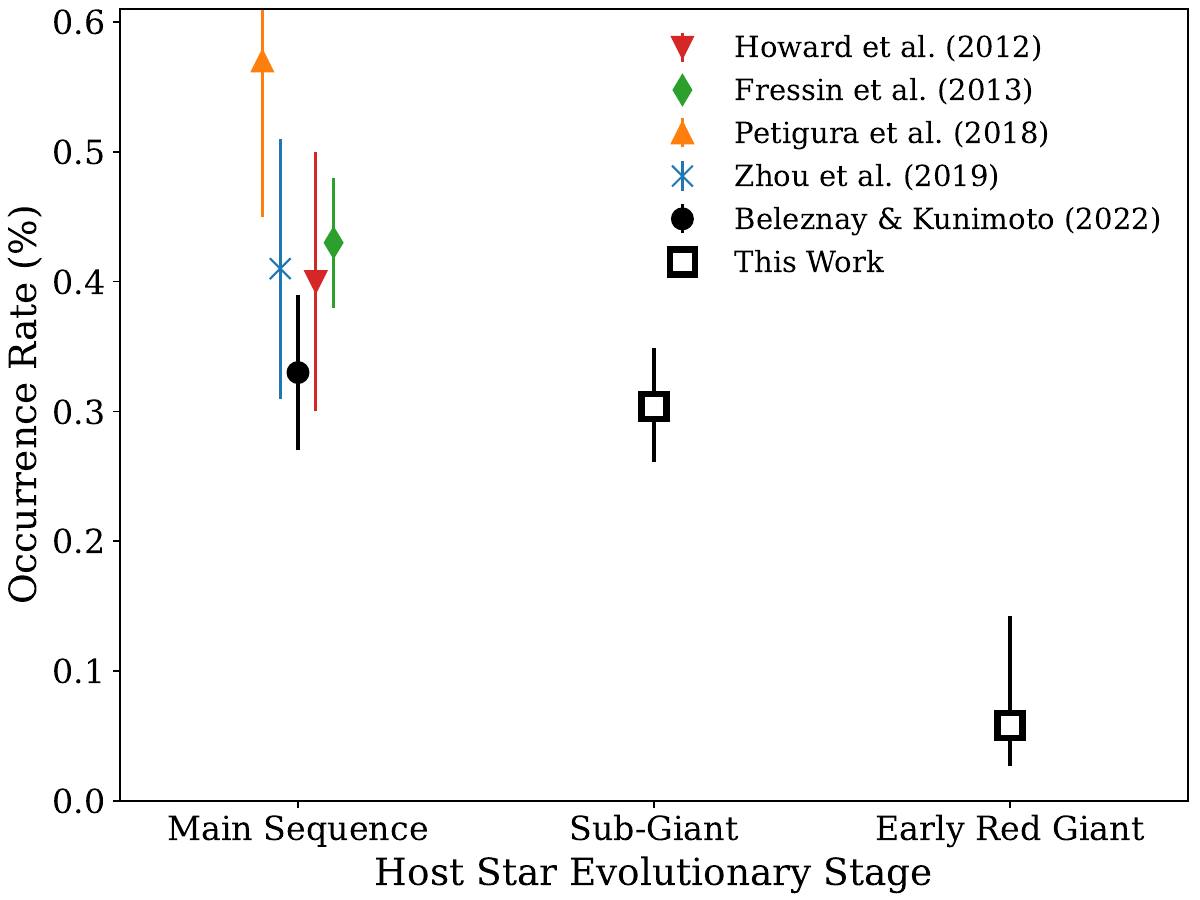}
    \caption{We compare our giant planet occurrence rate measurements (open squares) for the two sub-populations (see Figure~\ref{fig:stellar_sample}) with the main sequence occurrence rate results from previous studies using results from \textit{Kepler} and TESS. Our occurrence rate measurements shown in this plot were calculated using the following parameter ranges: $1 \leq P \leq 10\,\mathrm{days}$ and $8 \leq \rpl \leq 22\,\rearth$. The period range was chosen to most closely match the period ranges used by the previous main sequence population studies (see Section~7.3 for more details on the parameter ranges considered by the other studies included here).}
    \label{fig:occrate_vs_MainSeq}
\end{figure}
We now seek to compare our planet occurrence rates for giant planets orbiting evolved stars, to occurrence rates for similar planets that orbit stars which formed with similar properties but still reside on the main sequence. A number of previous works have studied the occurrence rates of short period giant planets around solar-like main sequence stars. 
Using results from the \textit{Kepler} mission \citep{borucki2010kepler} for a sample of main sequence FGK type stars \citet{howard2012kepleroccrates}  measured an occurrence rate for planets in the ranges $P < 10\,\mathrm{days}$ and $8 \leq \rpl \leq 32\,\rearth$ of $0.4 \pm 0.1\%$ and \citet{Fressin_2013} measured an occurrence rate of $0.43\pm0.05\%$ ($0.8 \leq P \leq 10\,\mathrm{days}$ and $6 \leq \rpl \leq 22\,\rearth$). Combining the \textit{Kepler} results for FGK main sequence stars with spectroscopic monitoring, \citet{Petigura_2018} measured an occurrence rate of $0.57^{+0.14}_{-0.12}\%$, for planets with periods $1 \leq P \leq 10\,\mathrm{days}$ and radii $8 \leq \rpl \leq 24\,\rearth$. \citet{KunimotoMatthews2020} used the results from the complete \textit{Kepler} DR25 release for main sequence FGK stars to measure an occurrence rate of $0.77^{+0.16}_{-0.14}\%$ ($P = 0.78 - 12.5\,\mathrm{days}$ and $\rpl = 8 - 16\,\rjup$). Using TESS photometry and considering giant planets with $0.9 \leq P \leq 10\,\mathrm{days}$ and $0.8 \leq \rpl \leq 2.5\,\rjup$ \citet{Zhou_2019} measured an occurrence rate of $0.41\pm0.10\%$ for a sample of main sequence AFG stars ($0.8 < \mstar \leq 2.3\,\msun$). Expanding upon this work and also using TESS data, \citet{beleznay2022tessoccrates} measured an occurrence rate of $0.33\pm0.04\%$ considering the same stellar and planetary ranges as \citet{Zhou_2019}. They also measure occurrence rates of $0.29\pm0.05\%$, $0.36\pm0.06\%$, and $0.55\pm0.14\%$ for A ($1.4 < \mstar \leq 2.3\,\msun$), F ($1.05 < \mstar \leq 1.4\,\msun$), and G ($0.8 < \mstar \leq 1.05\,\msun$) stars respectively. We summarise these results and compare them to our post-main sequence occurrence rate results in Figure~\ref{fig:occrate_vs_MainSeq}.

All these previous works yield consistent results for the occurrence rate of short period giant planets around main sequence stars. In general these occurrence rates show agreement with our occurrence rate measurements for our sub-giant sub-population and are greater than our measured occurrence rates for our more evolved early red giant sub-population. However, all the main sequence occurrence rates consider slightly different stellar and planetary parameter ranges from each other. These parameters are all known to influence the measured occurrence rates and so a direct comparison to each previous study simultaneously cannot be robustly performed. Therefore, we select the results of \citet{beleznay2022tessoccrates} for a detailed comparison to our results, primarily due to the fact that their results are derived from the largest stellar sample and they provide occurrence rate measurements for distinct stellar mass ranges. We do note that due to the overall agreement of the different main sequence occurrence rate results, comparing to the results of just \citet{beleznay2022tessoccrates} is unlikely to impact our conclusions.

\subsection{The occurrence rate reduction is not due to formation}
\begin{figure}
    \centering
    \includegraphics[width=0.95\linewidth]{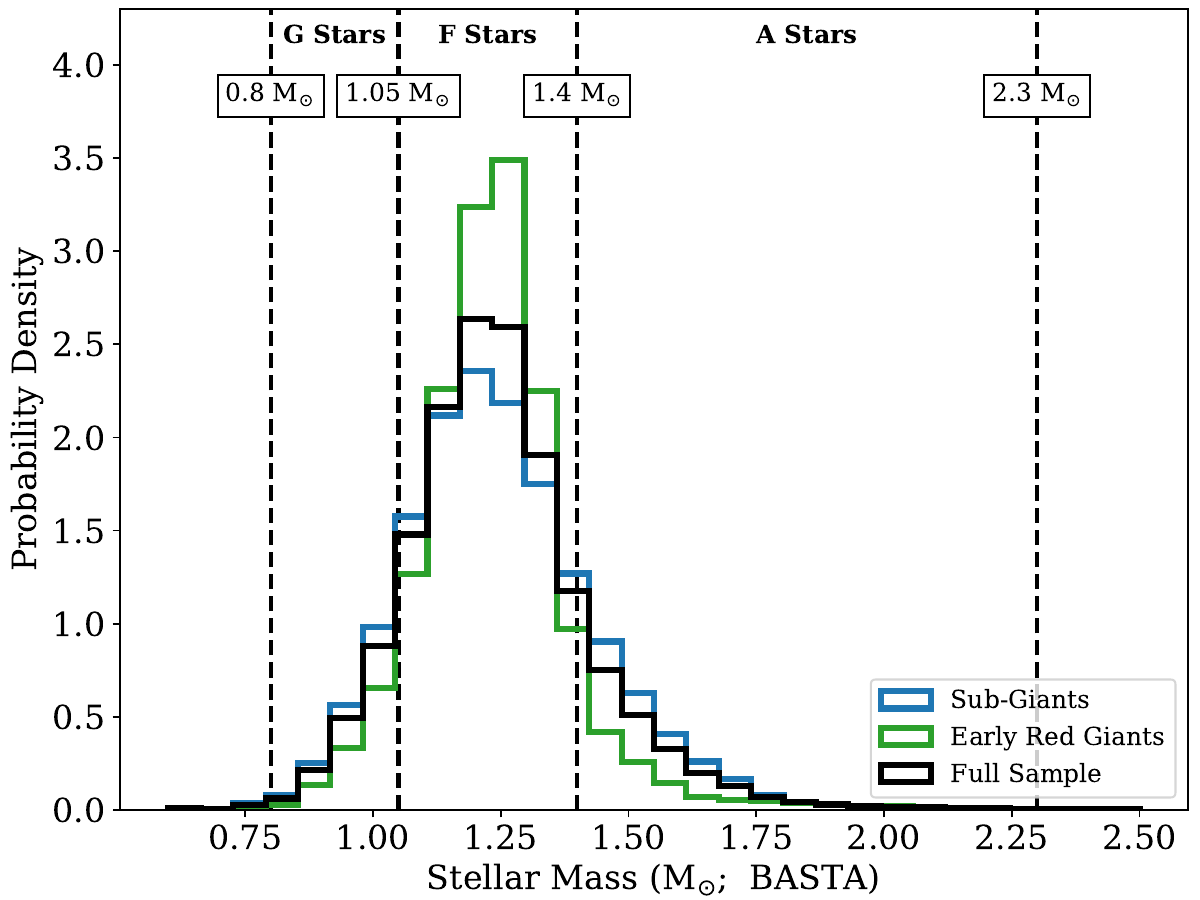}
    \caption{Mass probability density distribution for our post-main sequence sample. The vertical dashed lines denote the location of the stellar mass boundaries used by \citet{beleznay2022tessoccrates} to distinguish main-sequence G-stars ($0.8 \leq \mstar \leq 1.05\,\msun$), F-stars ($1.05 \leq \mstar \leq 1.4\,\msun$), and A-stars ($1.4 \leq \mstar \leq 2.3\,\msun$)}
    \label{fig:basta_mass}
\end{figure}

\begin{figure}
    \centering
    \includegraphics[width=0.95\linewidth]{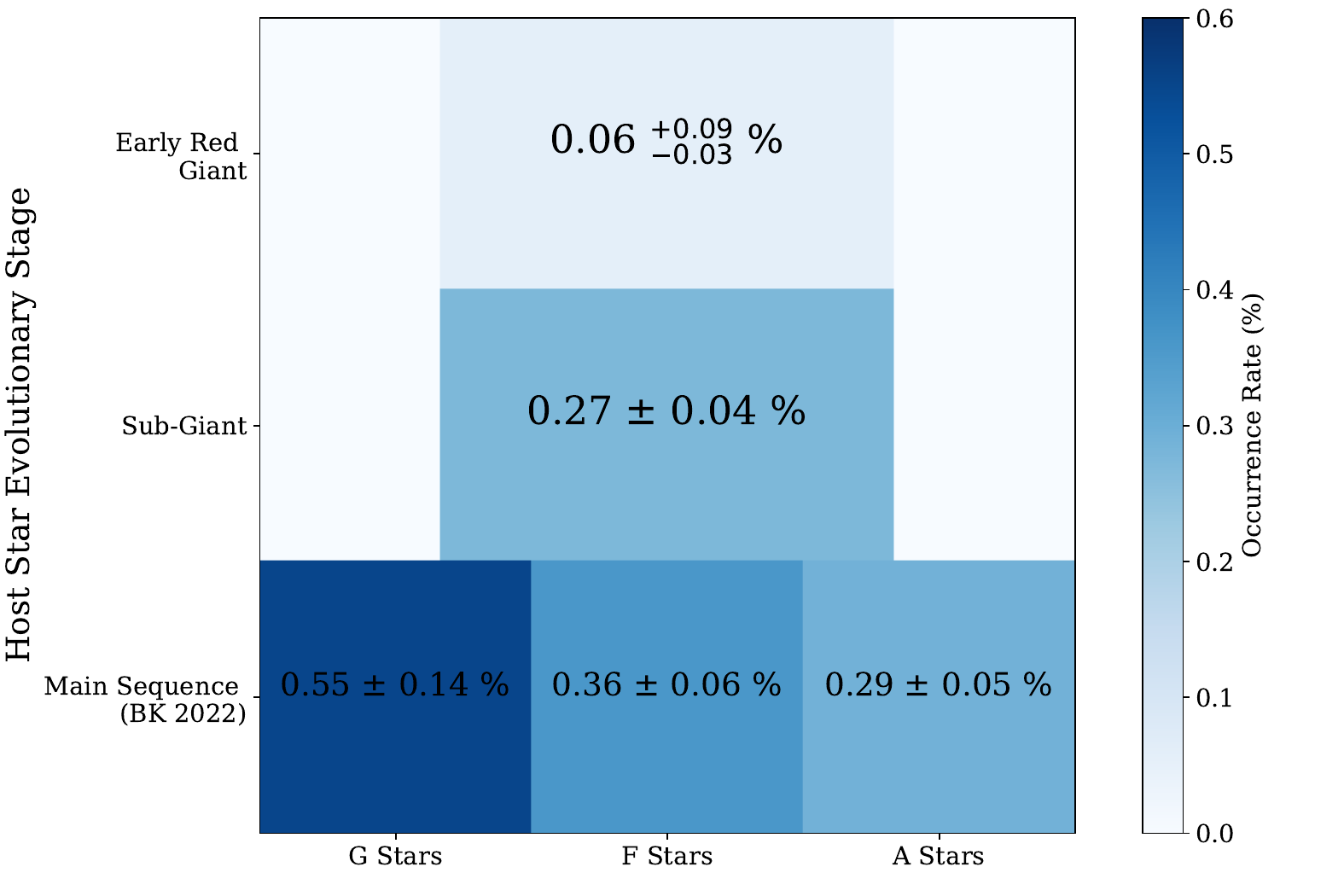}
    \caption{Comparison of the occurrence rate of giant planets as a function of the mass and evolutionary state of the host star sample. See the caption of Figure~\ref{fig:basta_mass} for the corresponding mass ranges used for G, F, and A stars.}
    \label{fig:occrate_vs_Mass}
\end{figure}
The occurrence rate of short period giant planets is known to scale with the mass of the host star \citep{reffert2015, KunimotoMatthews2020, beleznay2022tessoccrates, bryant2023lmstaroccrates}. This has been linked to stars of different masses hosting protoplanetary discs of differing masses and lifetimes, which then influences the ability of a giant planet to form \citep[e.g.][]{kennedy2008, alibert2011, johnston2024}. From our target selection criteria, we expect our sample to be dominated by stars in the mass range of $0.9\msun \leq \mstar \leq 1.4\msun$, corresponding to stars which were F and G-dwarfs when on the main sequence. This is due to the $1.4\,\msun$ stellar evolution track used to define the upper boundary of our stellar sample (see Figure~\ref{fig:stellar_sample}), and the fact that lower mass stars are in general expected to have main sequence lifetimes longer than the age of the universe. In previous studies, there has been some debate over the stellar parameters for post-main sequence host stars \citep[e.g.][]{lloyd2011, johnson2013, schlaufman2013}. We note that for the stars in our sample the TIC either does not provide a mass or the quoted mass is unreliable \citep{stassun2019TIC8}. Therefore, while we do expect our sample to be dominated by stars in the mass range equivalent to main sequence F and G stars, we look to obtain an independent assessment of the mass range covered by our sample.

To do this we use the Bayesian Stellar Algorithm \citep[\textsc{BASTA}; ][]{basta2022}. BASTA uses stellar isochrones and evolution tracks, along with observable astrophysical stellar properties, to determine stellar parameters, and has been used widely to date for the characterisation of exoplanet host stars \citep[e.g.][]{knudstrup2023BASTAeg1, persson2022BASTAeg2, osborne2024BASTAeg3, weeks2024BASTAeg4}. For such an analysis, it is important to have an estimate of the stellar metallicity, $[M/H]$. However, obtaining a robust metallicity for each individual star in our sample requires careful observations and analysis which are beyond the scope of this work. Instead, we use the \textit{Gaia} DR3 \citep{GaiaDR3} results to obtain metallicity measurements for stars in our sample. Specifically we use the GSP-Spec metallicity values, which we calibrate following \citet{recioblanco2023GSPSpecCal}. We use these metallicities as input to BASTA, along with the stellar effective temperature as reported in the TIC, and the parallax, and \textit{Gaia} G, G$_{Bp}$, and G$_{Rp}$ magnitudes from \textit{Gaia} DR3. The RA and dec coordinates for each star are also provided to obtain an estimate of the dust reddening. 

Using BASTA we are then able to obtain a stellar mass estimate for 403,504 stars in our sample with an available GSP-Spec metallicity. We provide the BASTA mass distribution for our sample in Figure~\ref{fig:basta_mass}. The BASTA mass distribution results suggest that our post-main sequence stellar sample is indeed dominated by stars which were F-stars on the main sequence. The 95$^\text{th}$ percentile range for the BASTA results spans the mass range from 0.93\,\msun\ to 1.67\,\msun, with 71.3\% of the stars in our post-main sequence sample lying in the main sequence F-star mass range of $1.05\,\msun \leq \mstar \leq 1.4\,\msun$ used by \citet{beleznay2022tessoccrates}. Following the mass ranges used by \citet{beleznay2022tessoccrates}, our post-main sequence stellar sample is most likely dominated by stars with a mass similar to main sequence F-stars, and also includes a few stars with a mass similar to main sequence early-G and late-A stars. We also note that we have no evidence for differing stellar mass distributions between our two sub-populations.

We compare our post-main sequence occurrence rates to the individual results for main sequence stars from \citet{beleznay2022tessoccrates} in Figure~\ref{fig:occrate_vs_Mass}. From this figure we can see that the occurrence rate we measure for our sub-giant population agrees with the main sequence F-star and A-star occurrence rates at a level of $1.25\,\sigma$ and $0.31\,\sigma$ respectively. For the early red giant sub-population, we see that our measured occurrence rates are significantly lower than the main sequence occurrence rates, regardless of the stellar mass, at a level of $2.9\,\sigma$, $2.8\,\sigma$, and $2.2\,\sigma$ as compared to the main sequence G, F, and A-stars respectively. 

\begin{figure}
    \centering
    \includegraphics[width=0.95\linewidth]{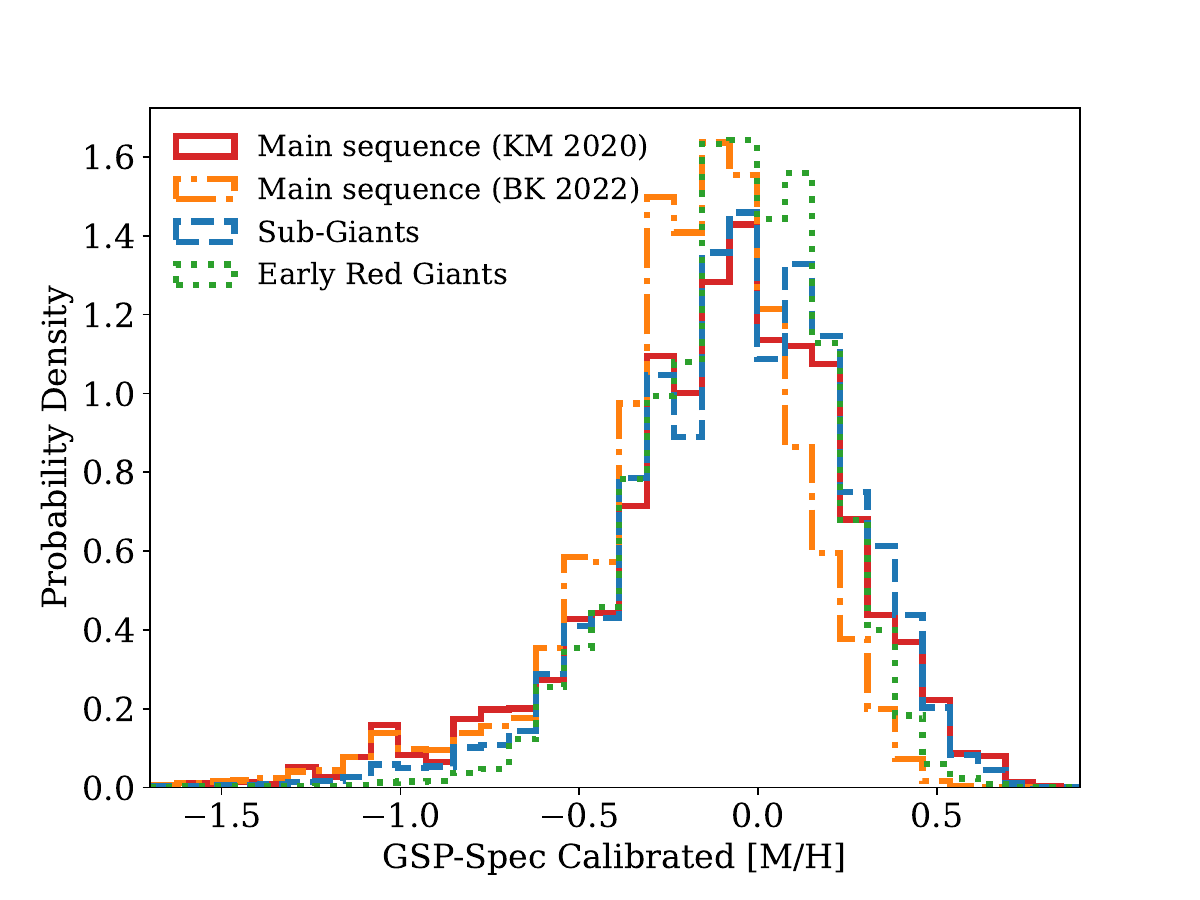}
    \caption{Probability density distributions of the GSP-Spec metallicities for our post-main sequence sample and our main sequence comparison samples. The metallicity distribution for our sub-giant sub-population is shown as the blue dashed line, and the distribution for our early red giant sub-population as the green dotted line. The GSP-Spec metallicity distributions we produce as representative comparisons for the \citet{KunimotoMatthews2020} and \citet{beleznay2022tessoccrates} main sequence samples are shown as the solid red line and dash-dotted orange line respectively.}
    \label{fig:met_dist}
\end{figure}

The bulk metallicity of the host star can also impact the formation of giant planets \citep{IdaLin2004,fischervalenti2005pmc,ercolanoclarke2010metallicity}. \citet{Petigura_2018} studied the variation of planet occurrence rates with stellar metallicity. As a rough estimate from their results we might expect the occurrence rate to increase by a factor of two for a metallicity increase of 0.1\,dex. We obtain GSP-Spec metallicity values, again calibrated following \citet{recioblanco2023GSPSpecCal}, for the stars included in the \citet{KunimotoMatthews2020} main sequence sample. We note that there are available GSP-Spec metallicity values for only 5,297 of the 79,183 F- and G-stars they survey, primarily due to the intrinsic brightness of the \textit{Kepler} stellar sample resulting in the majority of the stars being too faint for a GSP-Spec metallicity measurement. To obtain a representative comparison to the \citet{beleznay2022tessoccrates} sample, we follow the selection criteria they provide in their Section~2.1, selecting stars from the TESS Candidate Target List \citep[CTL;][]{stassun2019TIC8} with $T < 10.5\,\mathrm{mag}$, $0.8\,\msun \leq \mstar \leq 2.3\,\msun$, and $\log g > 4.1$. This gives us a sample of 192,247 stars, 147,637 of which have an available GSP-Spec metallicity measurement. We compare the metallicity distributions of these stellar samples in Figure~\ref{fig:met_dist}, from which we do not see large discrepancies between the metallicity distributions. Both our post-main sequence sub-populations display metallicity distributions that show good agreement with each other, as well as the \citet{KunimotoMatthews2020} main-sequence sample. Fitting Gaussian models to the metallicity distributions we find mean values of $-0.03\,\mathrm{dex}$ and $-0.04\,\mathrm{dex}$ for our sub-giant and early red giant sub-populations respectively, as well as a value of $-0.05\,\mathrm{dex}$ for the stars from the \citet{KunimotoMatthews2020} sample. 

The metallicity distributions for our sub-populations also show reasonable agreement with the \citet{beleznay2022tessoccrates} comparison metallicity distribution, for which we find a mean value of $-0.14\,\mathrm{dex}$ from fitting a Gaussian model, with our post-main sequence samples showing a larger amount of super-solar metallicity stars. We therefore do not expect the stellar metallicities to be significantly biasing our occurrence rate measurements, and if anything we would expect our measured post-main sequence occurrence rates to be biased to slightly higher values as compared to the \citet{beleznay2022tessoccrates} results. The true occurrence rate differences as a function of the stellar evolutionary state could therefore be slightly larger than we observe, although we note that as we are not comparing to the exact sample surveyed by \citet{beleznay2022tessoccrates} we cannot make any strong conclusions. 

From this analysis, we expect our post-main sequence stellar sample to be dominated by stars with masses similar to main sequence F-stars, and we see that our post-main sequence stars show similar bulk metallicity distributions to the main sequence stellar populations. Therefore, it is unlikely that the significant reduction we measure for the giant planet occurrence rates for our early red giant sub-population compared to the giant planet occurrence rates for main sequence stars is a result of reduced planet formation influenced by stellar mass or metallicity. Therefore, we have compelling evidence that the reduction in occurrence rates we observe for our post-main sequence stellar sample is the result of a process which occurs during the post-main sequence evolution of the host star, as a result of this evolution. We also have evidence that this process does not occur immediately at the onset of post-main sequence evolution, as our sub-giant population shows consistent giant planet occurrence rates with the main sequence F and A-stars, but instead occurs continuously throughout the early stages of the post-main sequence lifetime of the star. It is possible that this process is rapid orbital decay due to increased tidal interactions.

\subsection{The reduction in giant planet occurrence is dependent on orbital period}

\begin{figure}
    \centering
    \includegraphics[width=\linewidth]{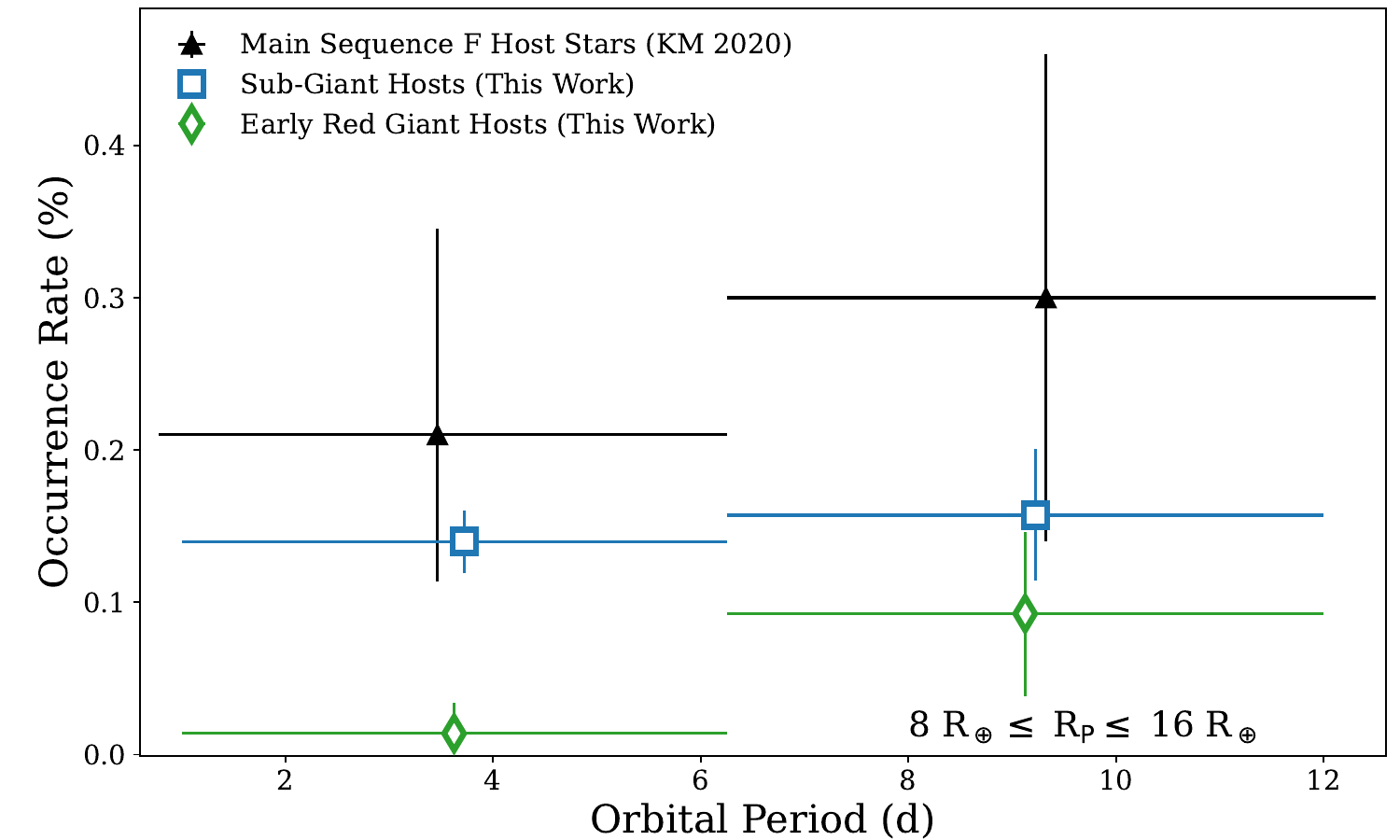}
    \caption{Variation of giant planet occurrence rates with orbital period for planets around evolved host stars (this work) and main sequence host stars \citep{KunimotoMatthews2020}. For the evolved host stars we plot our results for our sub-giant sub-population as the blue squares and for our early red giant sub-population as the green diamonds. We compare to occurrence rates for main sequence F-stars (black triangles). The post-main sequence occurrence rates shown in these plots were computed using a planet radius range of $8 \leq \rpl \leq 16\,\rpl$. This radius range is used for this plot to provide a like-for-like comparison to the \citet{KunimotoMatthews2020} occurrence rates, which were computed using the same planet radius range (see Table~\ref{tab:occ_comp_KM1} for details on these occurrence rate values).}
    \label{fig:per_occrate_vs_MS}
\end{figure}
{\renewcommand{\arraystretch}{1.3}
\begin{table*}
    \centering
    \begin{tabular}{|l|c|c|r|}
    \toprule
       \textbf{Stellar Host Population} & \textbf{Period Range (days)} & \textbf{Occurrence Rate (\%)} & \textbf{Reference} \\
    \toprule
       Main Sequence F-stars & $0.78 \leq P \leq 6.25$ & $0.21^{+0.14}_{-0.10}$ & \citet{KunimotoMatthews2020} \\
       Main Sequence F-stars & $6.25 < P \leq 12.5$ & $0.3\pm0.16$ & \citet{KunimotoMatthews2020} \\
       Sub-Giant Stars & $1.0 \leq P \leq 6.25$ & $0.14\pm0.02$ & \textit{This Work} \\
       Sub-Giant Stars & $6.25 < P \leq 12$ & $0.16{\pm0.04}$ & \textit{This Work} \\
       Early Red Giant Stars & $1.0 \leq P \leq 6.25$ & $0.014^{+0.020}_{-0.008}$ & \textit{This Work} \\
       Early Red Giant Stars & $6.25 < P \leq 12$ & $0.09\pm0.05$ & \textit{This Work}  \\
    \toprule
    \end{tabular}
    \caption{Occurrence rates computed in this work for post-main sequence host stars the planet radius range $8 \leq \rpl \leq 16\,\rearth$ and for main sequence F-stars taken from \citet{KunimotoMatthews2020}. \citet{KunimotoMatthews2020} define their F-star sample using the effective temperature range $6000 \leq T_\text{eff} < 7300\,\mathrm{K}$.}
    \label{tab:occ_comp_KM1}
\end{table*}
}
We already saw that the occurrence rate difference between our two sub-populations is dependent on the planet orbital period (see Section~\ref{sub:pms_occrates}). We would like to compare this effect to planets orbiting main sequence stars, however, \citet{beleznay2022tessoccrates} do not provide their occurrence rate measurements as a function of orbital period. Instead, in order to compare the period dependence of our occurrence rates to the main sequence population, we use the results of \citet{KunimotoMatthews2020}. We make this comparison in Figure~\ref{fig:per_occrate_vs_MS}, comparing to the \citet{KunimotoMatthews2020} results for main sequence F-stars, as these are likely to most closely match the stellar mass distribution of our stellar sample. We note that \citet{KunimotoMatthews2020} consider planets in the range of $8\,\rearth \leq \rpl \leq 16\,\rearth$. As such, we compute a new set of occurrence rate measurements using these radii limits for this comparison, to provide a more robust comparison. We provide these occurrence rates, along with those from \citet{KunimotoMatthews2020} in Table~\ref{tab:occ_comp_KM1}.

We can see that the giant planet occurrence rate measured for our sub-giant sub-population is consistent with the main sequence F-stars for both short periods ($P \leq 6.25\,\mathrm{days}$) and moderate periods ($P > 6.25\,\mathrm{days}$), at levels of $0.69\,\sigma$ and $0.85\,\sigma$ respectively for the two period ranges. For our early red giant population, for the moderate planet period range we find the occurrence rate is consistent with the main sequence F-star occurrence rate at a level of $1.25\,\sigma$. For the short period range we find a lower giant planet occurrence rate for the early red giant sub-population with a significance of $1.9\,\sigma$. We note that the significance with which we can distinguish the main sequence and post-main sequence occurrence rates is primarily limited by the uncertainty on the main sequence occurrence rates from \citet{KunimotoMatthews2020}. Revisiting the period dependence of main sequence giant planet occurrence rates with the larger stellar sample available from the TESS mission, as compared to the \textit{Kepler} mission, would enable these main sequence occurrence rates to be measured to a greater precision. This would enable the orbital period and stellar evolutionary stage at which the giant planet population begins to diminish, and the extent by which this occurs, to be determined to greater significance.

From the occurrence rates we measure in this work, and from the comparison of them to previous population studies for main sequence host stars, we have shown that the population of short period ($P \leq 12\,\mathrm{days}$) giant planets diminishes during the post-main sequence evolution of their host stars. We have also shown that the decrease in giant planet occurrence rates is larger for shorter orbital periods. This suggests that the giant planet population is being diminished by an evolutionary effect during the post-main sequence lifetime of the host stars, and it is very likely that tidal interactions between the planet and star are the source of this evolutionary influence.

\subsection{Tidal Decay and Destruction}\label{sub:tidaldecay}
Tidal theories predict that interactions between stars and the planets orbiting them can lead to the planet's orbit shrinking and the planet spiralling in towards its host star \citep[e.g.][]{rasio1996}. The tidal interactions between stars and their close-in planets are believed to be dominated by two components: the equilibrium tide and the dynamical tide \citep{zahn1977tides, zahn1989equilibriumtides, zahn2008tides}. While the exact mechanism dominating these interactions is unclear, the strength of these tidal interactions is also expected to depend strongly on the planet's semi-major axis with a smaller semi-major axis, and so a shorter orbital period, resulting in much stronger interactions and much faster orbital decay \citep[e.g.][]{GoldreichSoter1966, rasio1996}. Therefore, the fact that we observe a larger occurrence rate reduction for shorter orbital periods around more evolved stars is highly consistent with the picture in which these planet-star tidal interactions are sculpting the post-main sequence giant planet population.

The strength of these interactions is expected to increase during the post-main sequence evolution of the host stars. \citet{esseldeurs2024evolvedstartides} studied how tidal dissipation within stars changes throughout the evolution of the star, predicting an increase in the tidal dissipation for both the equilibrium and dynamical components from the beginning of the post-main sequence evolution phase. For stars with a convective envelope while on the main sequence, they predict a gradual increase in the in the tidal dissipation, while for stars with a radiative envelope on the main sequence the predicted tidal dissipation increase is instantaneous. \citet{barker2020tides} also investigated the expected variation in tidal dissipation throughout the evolution of stars. They find that dynamical tides due to the dissipation of internal waves may become highly, if briefly, efficient as the star evolves off the main sequence. Their results also predict the dissipation of equilibrium tides to be impactful for evolved stars and negligible during the main sequence. 

\citet{weinberg2024tides} also study the predicted tidal dissipation in sub-giant stars and the orbital decay time for planets in orbit around them. For stars with $\mstar \leq 1.0\,\msun$ they find a slight increase in the orbital decay time as the star evolves. For more massive stars ($\mstar \geq 1.2 \msun$) they predict minimal tidal decay while the stars are on the main sequence, but they find that the orbital decay time decreases significantly as the star evolves off the main sequence. On the sub-giant branch, \citet{weinberg2024tides} find that the orbital decay time can be faster than 10\,Myr around these higher mass stars for the shortest orbital periods, $P \leq 2\,$days. Even for slightly longer orbital periods, $P \leq 4\,$days, they find orbital decay timescales on the order of 100\,Myr for the older sub-giant models they consider. \citet{weinberg2024tides} note that this timescale is comparable to the evolutionary timescale of these sub-giant stars. 

From our results we find that the short period giant planet occurrence rates in general are consistent between the main sequence stellar population and our sub-giant sub-population, which consists of stars that have only just begun their post-main sequence evolution. We also find significantly reduced giant planet occurrence rates for our sub-population of more evolved early red giant host stars, noting that this more evolved stellar population still represents an early stage relative to the full post-main sequence evolution of these stars. Therefore, our results are in agreement with these theoretical predictions that these tidal interactions increase in strength as the stars evolve off of the main sequence, with rapid tidal decay occurring during the early stages of the post-main sequence lifetime of these stars.

Our results therefore provide strong evidence that the population of short period giant planets is heavily impacted by the post-main sequence evolution of their host stars. The increased strength of the planet-star tidal interactions, as a result of the changing stellar structure during the early stages of post-main sequence evolution, is the most likely driving force impacting the giant planet population. 

Recent works have claimed that the occurrence rate of short period giant planets with main sequence host stars decreases with increasing stellar age \citep{chen2023hjageoccrates, miyazaki2023hjageoccrates}. These works have suggested that such a decrease may be due to these tidal effects having an impact on the giant planet population even during the main sequence lifetime of their host stars \citep{chen2023hjageoccrates, miyazaki2023hjageoccrates}. 

Moreover, while we are limited by the small number of candidate planets at these short orbital periods, we do not see a difference between the measured giant planet occurrence rates for $P \leq 2\,\mathrm{days}$ for our sub-giant and early red giant sub-populations (see Figure~\ref{fig:occrate_vs_Per_2SS} and Section~\ref{sub:pms_occrates}). We note that we cannot rule out this simply being a result of the limited statistical significance of our results due to a lack of detected planets at these very short orbital periods. However if this is a real astrophysical feature of the population then it would be in keeping with the prediction from \citet{weinberg2024tides} that planets at these short orbital periods exhibit very short tidal decay timescales around post-main sequence stars, and so full orbital decay is to be expected before their host stars evolve to the bottom of the red giant branch. 

\subsection{Studying post-main sequence population demographics with our detected planet sample}
Beyond the occurrence rates we measure in this work, the sample of 118 giant planets and planet candidates with post-main sequence host stars we detected using our homogeneous planet search and vetting pipeline represents a statistically unbiased planet population with which to study the demographics of the post-main sequence planet population. For example, it has previously been suggested that the population of planets with evolved host stars may reside on more eccentric orbits than planets with main sequence host stars, potentially as a result of the tidal interactions between the planets and their host stars \citep[e.g.][]{Villaver2014, Grunblatt_2018}. In addition, it has been predicted that the increased irradiation of these planets due to the increase in radius and luminosity of their evolving host stars can cause these planets to undergo a level of re-inflation \citep[e.g.][]{grunblatt2017reinflation, Thorngren_2021}. Moreover, recent obliquity measurements for hot Jupiters orbiting sub-giants suggest that the tidal interactions in these systems may also lead to highly efficient damping of the orbital obliquity of these systems \citep{saunders2024}. 

However, the specific study of these demographics is beyond the scope of this paper, primarily as further spectroscopic observations of our planet candidates will be required. This is because having an accurate measure of the planet mass is crucial for understanding the tidal interactions as the strength of them is predicted by tidal theory to depend on the planet to star mass ratio, $\mpl / \mstar$, with higher mass ratios resulting in stronger tidal interactions \citep{ogilvie2014, Patra_2020}. The planet mass is also a key component for understanding the level of planet radius inflation observed \citep{sestovic2018}. Fortunately, as our candidate planets have relatively bright host stars ($T \leq 12\,\mathrm{mag}$) and are likely to exhibit large radial velocity semi-amplitudes on the order of tens or even hundreds $\mathrm{m\,s}^{-1}$ the measurement of their masses should be easily within the capabilities of existing spectrographs, such as HARPS \citep{mayor2003harps} and HARPS-N \citep{harpsN}. Precisely monitoring the radial velocity variation for these stars will also yield a measurement of the orbital eccentricity of these systems. 

Through these radial velocity observations we will also be able to refine our occurrence rate measurements, which are calculated using false positive probability (FPP) estimates computed using the \textsc{triceratops} software package \citep{triceratops1_2020}. While, these FPP estimates are a key aspect of measuring planet occurrence rates and \textsc{triceratops} has been widely used for that purpose \citep[e.g.][]{bryant2023lmstaroccrates, vach2024, giacalone2025}, radial velocity follow up observations would enable us to confirm or reject the planet candidates in our sample as bona-fide planets, thus refining our knowledge of the overall population.

Finally, of particular interest from the 82 new candidates and TOIs we detect using our pipeline are eleven with short orbital periods $P \leq 2\,\mathrm{days}$, shorter than any currently confirmed giant planets around sub-giant and red giant stars. As shorter period planets are expected to exhibit more rapid tidal decay \citep[e.g.][]{rasio1996} then if these candidates are confirmed as real planets they would therefore become key puzzle pieces for understanding how post-main sequence stars impact their close-in giant planets. 

\section{Conclusions}\label{sec:conc}
We conduct a systematic transit search for giant planets ($8\,\rearth \leq \rpl \leq 22\,\rearth$) on short orbital periods ($1 \leq P \leq 12~\mathrm{days}$) around a sample of \Ntotal\ post-main sequence stars, using TESS Full-Frame-Image data produced by the QLP pipeline. From this search we detected a sample of 118 giant planets and planet candidates, and a further twelve planets and candidates with smaller radii, $\rpl < 8\,\rearth$. Using injection-recovery simulations to quantify the detection efficiency of our planet search and candidate vetting pipeline we measure the short period giant planet occurrence rate of this population, deriving a value of $0.28\pm0.04$\,\% across our full post-main sequence stellar sample. 

We also divide our stellar sample using a boundary of equal Equivalent Evolutionary Phase, $\mathrm{EEP} = 465$ \citep{Choi2016MIST} into two stellar sub-populations: our sub-giant sub-population, which consists of stars in the earliest stages of post-main sequence evolution, and our early red giant sub-population of more evolved stars (see Section~\ref{sub:evolstars} and Figure~\ref{fig:stellar_sample}). We measure occurrence rates for our two sub-populations, finding values of $0.35\pm0.05\%$ for our sub-giant sub-population and $0.11^{+0.06}_{-0.05}\%$ for our early red giant sub-population, finding a significant decrease in giant planet occurrence rates for more evolved stars. We also show that the decrease in occurrence rate is dependent on the planet orbital period with a larger occurrence rate difference observed for shorter period planets. 

Comparing our results with previous occurrence rate measurements for main sequence host stars \citep{howard2012kepleroccrates, Fressin_2013, Petigura_2018, Zhou_2019, KunimotoMatthews2020, beleznay2022tessoccrates}, we again find a significant decrease for our early red giant sub-population as compared to the main sequence, again with a greater occurrence rate decrease for shorter period planets. For our sub-giant sample we find our measured giant planet occurrence rate shows agreement with the main sequence planet population. By considering the mass and metallicity distributions of the main sequence and post-main sequence stellar samples we show that a differing level of giant planet formation due to variations in these parameters is unlikely to be the cause of the occurrence rate trends we observe.

We conclude that the reduction in giant planet occurrence rates for more evolved stars we observe is due to these planets being impacted by the post-main sequence evolution of their host stars. Due to the period dependence observed for the occurrence rate decrease we conclude that the dominant impact is most likely rapid tidal decay of the planets' orbits due to the increasing strength of the planet-star tidal interactions as the host stars evolve \citep[e.g.][]{rasio1996, barker2020tides, esseldeurs2024evolvedstartides, weinberg2024tides}.

Our homogeneously detected sample of planets and planet candidates with post-main sequence host stars also represents a quality resource with which to study wider demographics of the post-main sequence, such as giant planet re-inflation \citep[e.g.][]{grunblatt2017reinflation, Thorngren_2021} or the orbital eccentricity distribution of these systems \citep[e.g.][]{Villaver2014, Grunblatt_2018}. Accurate and precise individual mass measurements for all our planet candidates and host stars are required before studying these aspects of the post-main sequence giant planet population. Observing the host stars spectroscopically, for example with spectrographs such as HARPS, will provide the spectra necessary to better understand the nature of the stars and determine the mass of the orbiting companions. Such spectroscopic observations would also refine our occurrence rate measurements through the confirmation of our planet candidates and would also enable to us to better understand the tidal mechanisms at play in these system. 

One further limit to our ability to constrain the impact of post-main sequence evolution on the giant planet population is our knowledge of the main sequence giant planet population and how this varies with orbital period. Revisiting these occurrence rates, and how they vary with stellar mass and orbital period, with the full extent of the TESS Full-Frame-Image observations will enable us to see any differences between the main-sequence and post-main sequence more clearly. 

Obtaining a better characterisation of the stars surveyed would also improve our understanding of the impact of post-main sequence evolution on the giant planet population. An improved understanding of the evolutionary stage of each of the stars in our sample would allow us to better constrain how the population varies during this evolution. Obtaining better constraints of the mass and metallicities of the stars in our sample as well as the stars in any main sequence comparison sample would also enable us predict the impact of these on the giant planet occurrence rates, thereby allowing us to better disentangle these formation effects from the evolutionary effects.

The PLATO mission \citep{rauer2014plato} can play a large role in the future in understanding the post-main sequence planet population. PLATO will deliver much higher precision photometry than TESS which will enable us to detect planets and measure occurrence rates for more evolved larger radius planets on the red giant branch. PLATO will also enable us to accurately characterise these stars through asteroseismology, providing much better constraints on the masses and evolution states for all the stars studied.

\section*{Acknowledgements}

We are grateful to Angharad Weeks for useful discussions and support around the use of BASTA to obtain stellar mass estimates. The contributions at the Mullard Space Science Laboratory by E.M.B. have been supported by STFC through the consolidated grant ST/W001136/1. V.V.E. has been supported by UK's Science \& Technology Facilities Council through the STFC grants ST/W001136/1 and ST/S000216/1. This research has made use of the NASA Exoplanet Archive, which is operated by the California Institute of Technology, under contract with the National Aeronautics and Space Administration under the Exoplanet Exploration Program. This paper includes data collected by the TESS mission, which are publicly available from the Mikulski Archive for Space Telescopes (MAST). Funding for the TESS mission is provided by NASA’s Science Mission directorate. We acknowledge the use of public TOI Release data from pipelines at the TESS Science Office and at the TESS Science Processing Operations Center. 
%%%%%%%%%%%%%%%%%%%%%%%%%%%%%%%%%%%%%%%%%%%%%%%%%%
\section*{Data Availability}
The TESS Quick Look Pipeline light curves used in this work are all publicly available as a MAST High Level Science Product. They can be accessed from \url{https://archive.stsci.edu/hlsp/qlp}.



\bibliographystyle{mnras}
\bibliography{paper} % if your bibtex file is called example.bib

%\begin{thebibliography}{99}
%\end{thebibliography}

%%%%%%%%%%%%%%%%%%%%%%%%%%%%%%%%%%%%%%%%%%%%%%%%%%

%%%%%%%%%%%%%%%%% APPENDICES %%%%%%%%%%%%%%%%%%%%%

\appendix
\clearpage
\onecolumn
\section{Planet Candidate Tables}
\begin{landscape}
\begin{center}
{\renewcommand{\arraystretch}{2.0}
\tiny
\begin{longtable}{| c | c | c | c | c | c | c | c | c | c | c | c |}
    \caption{Stellar and planetary properties for the new planet candidates we detect with our planet search pipeline and their host stars} \label{tab:planet_new} \\
    
    \hline 
    TIC & \textit{T} (mag) & \rstar\ (\rsun; TIC) & \teff\ (K; TIC) & Sub-Population$^\ast$ & $P$ (days) & $T_\text{C} - 2457000$ & \rprs & \rpl (\rearth) & \ars & $i$ (degrees) & $b$ \\
    \hline 
    \endfirsthead

    \multicolumn{12}{c}
    {{\tablename\ \thetable{} -- continued from previous page}} \\
    \hline 
    TIC & \textit{T} (mag) & \rstar\ (\rsun; TIC) & \teff\ (K; TIC) & Sub-Population & $P$ (days) & $T_\text{C} - 2457000$ & \rprs & \rpl (\rearth) & \ars & $i$ (degrees) & $b$ \\
    \hline 
    \endhead

    \hline \multicolumn{12}{|l|}{$\ast$ - SG: sub-giant sub population ;  ERG: early red giant sub-population} \\ \hline
    \endfoot
    
    \input{planet_param_table_new2}
    
\end{longtable}
\newpage
\begin{longtable}{| c | c | c | c | c | c | c | c | c | c | c | c | c |}
    \caption{Stellar and planetary properties for the TOI planet candidates we independently detect with our planet search pipeline and their host stars}\label{tab:planet_toi} \\
    
    \hline 
    TIC & TOI & \textit{T} (mag) & \rstar\ (\rsun; TIC) & \teff\ (K; TIC) & Sub-Population$^\ast$ & $P$ (days) & $T_\text{C} - 2457000$ & \rprs & \rpl (\rearth) & \ars & $i$ (degrees) & $b$ \\
    \hline 
    \endfirsthead

    \multicolumn{13}{c}
    {{\tablename\ \thetable{} -- continued from previous page}} \\
    \hline 
    TIC & TOI & \textit{T} (mag) & \rstar\ (\rsun; TIC) & \teff\ (K; TIC) & Sub-Population$^\ast$ & $P$ (days) & $T_\text{C} - 2457000$ & \rprs & \rpl (\rearth) & \ars & $i$ (degrees) & $b$ \\
    \hline 
    \endhead

    \hline \multicolumn{6}{|l}{$\ast$ - SG: sub-giant sub population ;  ERG: early red giant sub-population} & \multicolumn{7}{r|}{{Continued on next page}} \\ \hline
    \endfoot

    \hline \multicolumn{13}{|l|}{$\ast$ - SG: sub-giant sub population ;  ERG: early red giant sub-population} \\ \hline
    \endlastfoot
    
    \input{planet_param_table_TOI2}
    
\end{longtable}

\begin{longtable}{| c | c | c | c | c | c | c | c | c | c | c | c | c | c |}
    \caption{Stellar and planetary properties for the known planets we independently detect with our planet search pipeline and their host stars} \label{tab:planet_kp} \\
    
    \hline 
    TIC & Planet & \textit{T} (mag) & \rstar\ (\rsun; TIC) & \teff\ (K; TIC) & Sub-Population$^\ast$ & $P$ (days) & $T_\text{C} - 2457000$ & \rprs & \rpl (\rearth) & \ars & $i$ (degrees) & $b$ & Reference \\
    \hline 
    \endfirsthead

    \multicolumn{14}{c}
    {{\tablename\ \thetable{} -- continued from previous page}} \\
    \hline 
    TIC & Planet & \textit{T} (mag) & \rstar\ (\rsun; TIC) & \teff\ (K; TIC) & Sub-Population$^\ast$ & $P$ (days) & $T_\text{C} - 2457000$ & \rprs & \rpl (\rearth) & \ars & $i$ (degrees) & $b$ & Reference \\
    \hline 
    \endhead

    \hline \multicolumn{7}{|l}{$\ast$ - SG: sub-giant sub population ;  ERG: early red giant sub-population} & \multicolumn{7}{r|}{{Continued on next page}} \\ \hline
    \endfoot

    \hline \multicolumn{14}{|l|}{$\ast$ - SG: sub-giant sub population ;  ERG: early red giant sub-population} \\ \hline
    \endlastfoot
    
    \input{planet_param_table_KP2}
    
\end{longtable}

\begin{longtable}{| c | c | c | c | c | c | c | c | c | c | c | c | c | c |}
    \caption{Stellar and planetary properties for the small planets ($\rpl \leq 8\,\rearth$) we detect with our planet search pipeline, but do not include in our occurrence rate measurements, and their host stars} \label{tab:planet_small} \\
    
    \hline 
    TIC & TOI & Planet & \textit{T} (mag) & \rstar\ (\rsun; TIC) & \teff\ (K; TIC) & Sub-Population$^\ast$ & $P$ (days) & $T_\text{C} - 2457000$ & \rprs & \rpl (\rearth) & \ars & $i$ (degrees) & $b$ \\
    \hline 
    \endfirsthead

    \multicolumn{14}{c}
    {{\tablename\ \thetable{} -- continued from previous page}} \\
    \hline 
    TIC & Planet & \textit{T} (mag) & \rstar\ (\rsun; TIC) & \teff\ (K; TIC) & Sub-Population$^\ast$ & $P$ (days) & $T_\text{C} - 2457000$ & \rprs & \rpl (\rearth) & \ars & $i$ (degrees) & $b$ \\
    \hline 
    \endhead

    \hline \multicolumn{7}{|l}{SG: sub-giant sub population ;  ERG: early red giant sub-population} & \multicolumn{7}{r|}{{Continued on next page}} \\ \hline
    \endfoot

    \hline \multicolumn{14}{|l|}{SG: sub-giant sub population ;  ERG: early red giant sub-population} \\ 
    \multicolumn{14}{|l|}{Planet references: TOI-329~b \citet{polanski2024toi320};   HD~89345~b \citet{vaneylen2018hd89345};   TOI-1736~b \citet{akanamurphy2023toi1736}} \\ \hline
    \endlastfoot
    
    \input{planet_param_table_small_V3}
    
\end{longtable}

}
\end{center}
\end{landscape}
\clearpage

\section{Newly Detected Planet Candidate Light Curves}
\begin{figure}
    \centering
    \includegraphics[width=\linewidth]{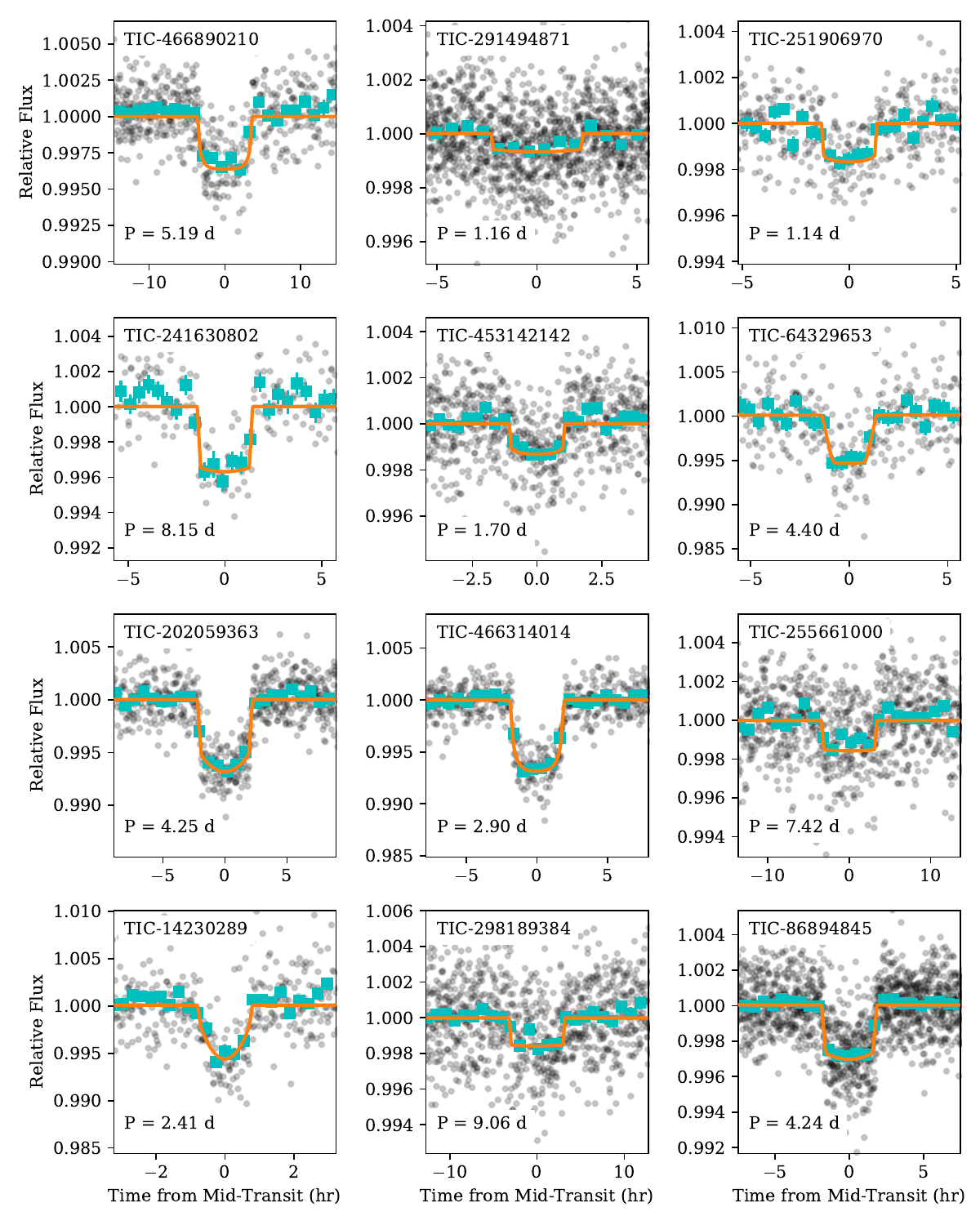}
    \caption{Phase-folded QLP LCs of the new planet candidates we detect in this work. The data is phase-folded using the best fit ephemerides from the transit fitting presented in Section~\ref{sec:fitting} and the orange line shows the best fit transit model from the same analysis. The cyan squares show the data binned in phase. The bin width is determined for each object such that there are 6 binned points per transit duration. The annotations on each axis give the TIC ID of the host star and orbital period of each candidate. Note that we show a zoomed in view around phase 0 for visual clarity of each transit event. The full phase-folded flux data sets extend beyond the edges of each panel.}
    \label{fig:newCandLCs1}
\end{figure}

\begin{figure}\ContinuedFloat
    \centering
    \includegraphics[width=\linewidth]{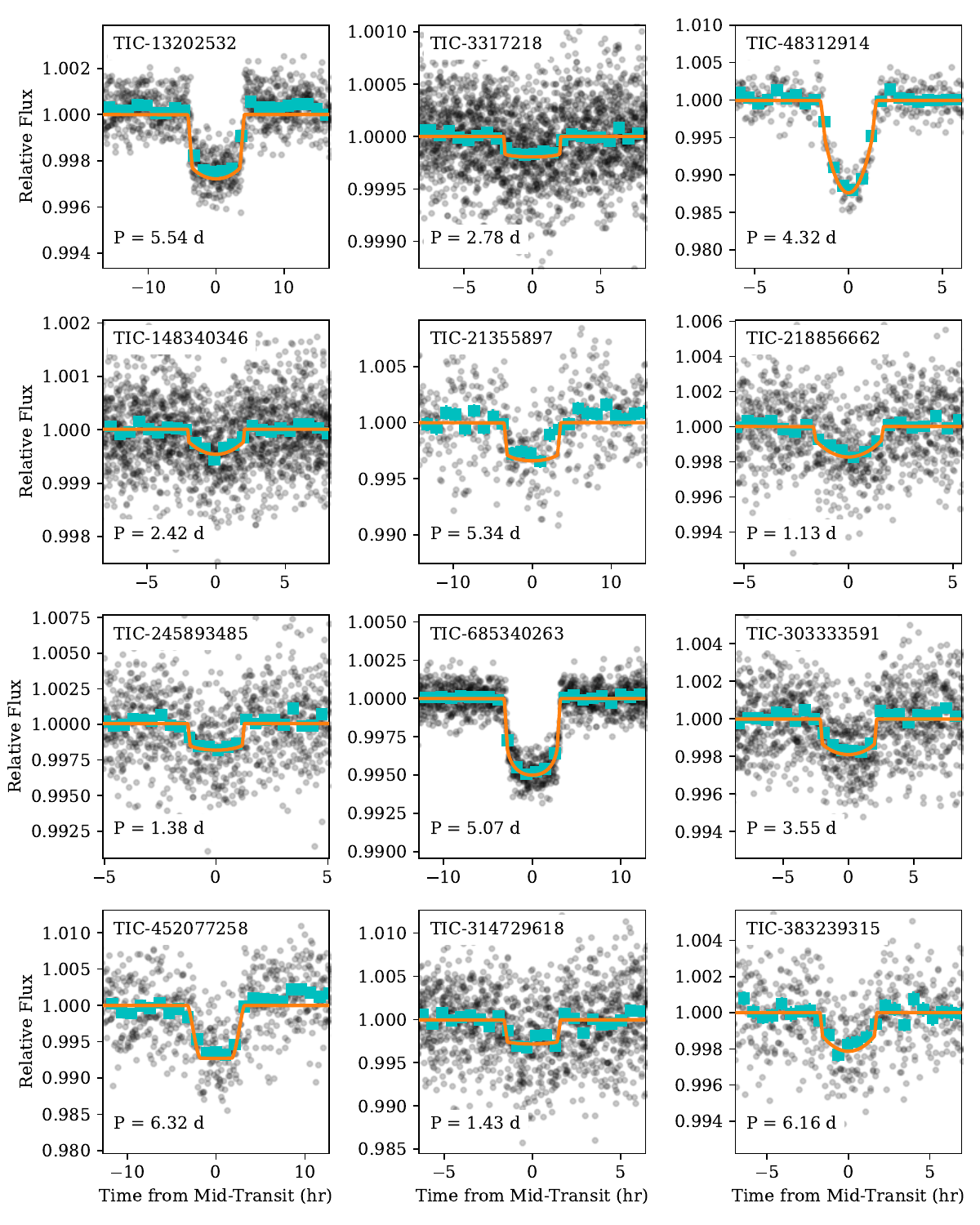}
    \caption{\textbf{Cont.}}
    \label{fig:newCandLCs2}
\end{figure}

\begin{figure}\ContinuedFloat
    \centering
    \includegraphics[width=\linewidth]{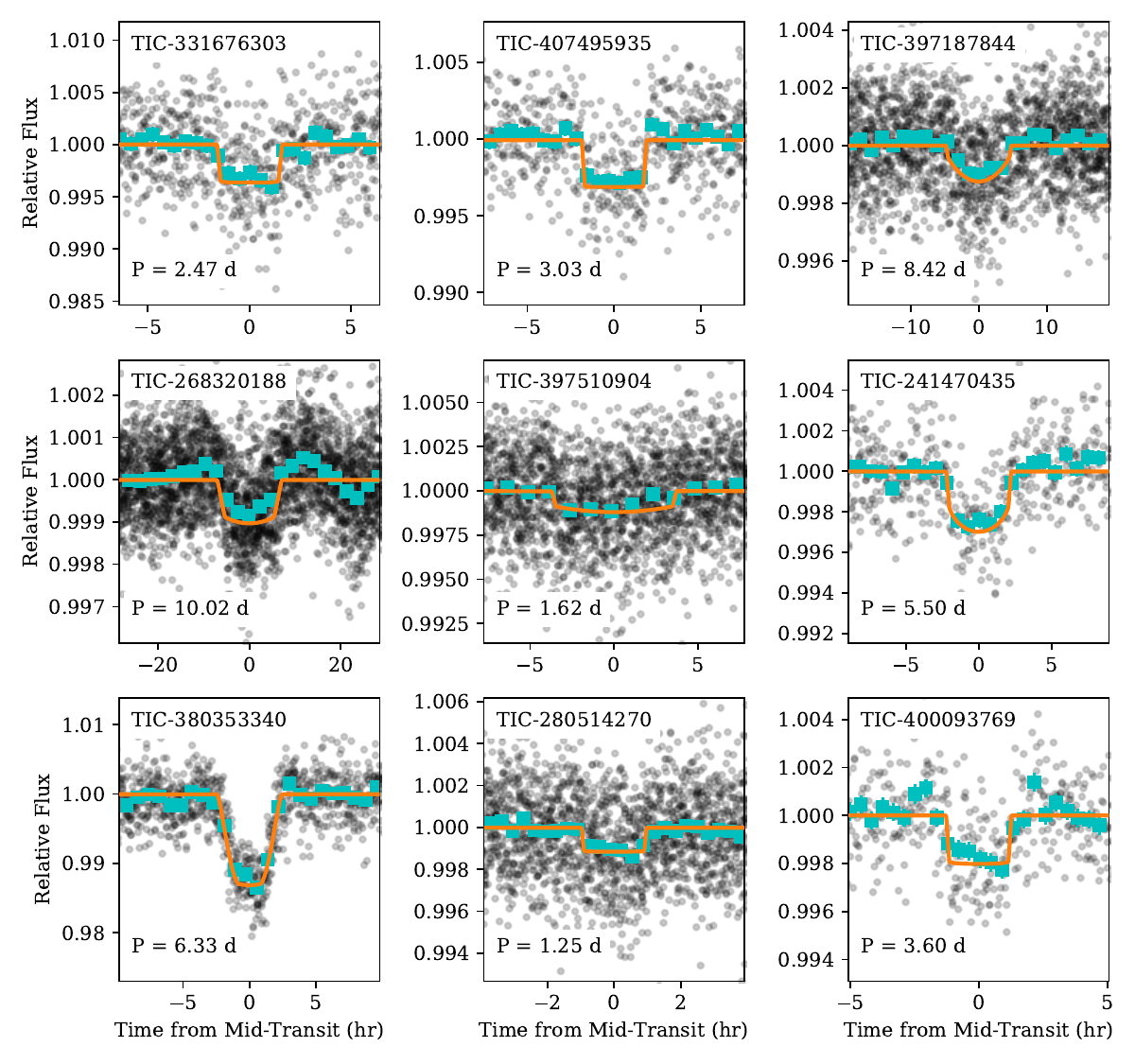}
    \caption{\textbf{Cont.}}
    \label{fig:newCandLCs3}
\end{figure}

\twocolumn
%%%%%%%%%%%%%%%%%%%%%%%%%%%%%%%%%%%%%%%%%%%%%%%%%%

% Don't change these lines
\bsp	% typesetting comment
\label{lastpage}
\end{document}

%% file: commands.tex
\newcommand{\mpl}{\mbox{$M_{\rm P}$}}
\newcommand{\rpl}{\mbox{$R_{\rm P}$}}

\newcommand{\mstar}{\mbox{$M_{\ast}$}}
\newcommand{\rstar}{\mbox{$R_{\ast}$}}

\newcommand{\rjup}{\mbox{$R_{\rm J}$}}

\newcommand{\rearth}{\mbox{R$_{\oplus}$}}
\newcommand{\msun}{\mbox{M$_{\odot}$}}
\newcommand{\rsun}{\mbox{R$_{\odot}$}}

\newcommand{\teff}{$T_{\rm eff}$}
\newcommand{\feh}{\mbox{[Fe/H]}}

\newcommand{\tc}{$T_{\rm C}$}
\newcommand{\rprs}{\mbox{$R_{\rm P}/R_{\ast}$}}
\newcommand{\ars}{\mbox{$a / R_{\ast}$}}

\newcommand{\focc}{\mbox{$f_\text{occ}$}}

%% file: properties.tex
\newcommand{\Ntotal}{456,941} %N stars in initial input catalogue
\newcommand{\NBLShits}{17,096} % N BLS hits (SDE>=7.3; SNR>=8.5; q<=0.25)
\newcommand{\NBLShitsThreeTrans}{14,860} % Include Ncount >= 3.

\newcommand{\NRplReject}{5,339}
\newcommand{\RplRejectPerCent}{35.93}
\newcommand{\NQReject}{2,980}
\newcommand{\QRejectPerCent}{20.05}
\newcommand{\NSecEclipseReject}{5,835}
\newcommand{\SecEclipseRejectPerCent}{39.27}
\newcommand{\NOddEvenReject}{1,982}
\newcommand{\OddEvenRejectPerCent}{13.34}
\newcommand{\NHarmonicReject}{1,036}
\newcommand{\HarmonicRejectPerCent}{6.97}
\newcommand{\NGLSReject}{3,175}
\newcommand{\GLSRejectPerCent}{21.37}
\newcommand{\NExcessRMSReject}{2,816}
\newcommand{\ExcessRMSRejectPerCent}{18.95}
\newcommand{\NPtPSymmReject}{812}
\newcommand{\PtPSymmRejectPerCent}{5.46}
\newcommand{\NPtPInOutReject}{513}
\newcommand{\PtPInOutRejectPerCent}{3.45}

\newcommand{\NLnLRatioReject}{488}
\newcommand{\LnLRatioRejectPerCent}{3.28}
\newcommand{\NRhoTransRejectLow}{2493}
\newcommand{\RhoTransRejectLowPerCent}{16.78}
\newcommand{\NRhoTransRejectUpp}{184}
\newcommand{\RhoTransRejectUppPerCent}{1.24}
\newcommand{\NRhoTransRejectTot}{2677}
\newcommand{\RhoTransRejectTotPerCent}{18.01}

\newcommand{\NTotalVetPass}{3,550}
\newcommand{\NTotalVetFPs}{11,310}
\newcommand{\TotalVetFPsPerCent}{76.11}

\newcommand{\InjRecRplPerCent}{0.03} 
\newcommand{\InjRecQPerCent}{0.34} 
\newcommand{\InjRecSecEclipsePerCent}{0.39} 
\newcommand{\InjRecOddEvenPerCent}{0.34} 
\newcommand{\InjRecHarmonicPerCent}{0.02} 
\newcommand{\InjRecGLSPerCent}{0.15} 
\newcommand{\InjRecExcessRMSPerCent}{0.06} 
\newcommand{\InjRecPtPSymmPerCent}{0.01} 
\newcommand{\InjRecPtPInOutPerCent}{0.003} 
 
\newcommand{\InjRecLnLRatioPerCent}{0.01}
\newcommand{\InjRecRhoTransRejectPerCent}{0.13}

\newcommand{\InjRecTotalPerCent}{2.38}

\newcommand{\NTransitFitPass}{1,364}
\newcommand{\NTransitFitReject}{2,186}

%% file: fitting_priors.tex
\begin{tabular}{cc}
\toprule
\textbf{Symbol} & \textbf{Prior} \\
\toprule
    \hline
    \tc & $\mathcal{U}\left( T_\text{C; BLS} - 0.05 \times P_\text{BLS}, T_\text{C; BLS} + 0.05 \times P_\text{BLS} \right)$ \\
    $P$ & $\mathcal{U}\left( 0.95 \times P_\text{BLS}, 1.05 \times P_\text{BLS} \right)$ \\
    \rprs & $\mathcal{U}\left( 0, 1 \right)$ \\
    \ars & $\mathcal{U}\left( 1.1, \infty \right)$ \\
    $i$ & $\mathcal{U}\left( 0, 90 \right)$ \\
    $q_\text{1}$, $q_\text{2}$ & $\mathcal{U}\left( 0, 1 \right)$ \\
\toprule
\end{tabular}

%% file: planet_param_table_new2.tex
14230289 & $11.61$ & $1.93\pm0.10$ & $6276\pm129$ & SG & $2.41097^{+0.00044}_{-0.00045}$ & $2449.5805^{+0.0026}_{-0.0026}$ & $0.073^{+0.1674}_{-0.0087}$ & $15.3^{+35.2}_{-1.8}$ & $9.1^{+3.3}_{-3.9}$ & $85.5^{+3.4}_{-8.5}$ & $0.72^{+0.41}_{-0.47}$ \\ 
21355897 & $10.28$ & $2.74\pm0.14$ & $6262\pm134$ & SG & $5.3424^{+0.0107}_{-0.006}$ & $2363.483^{+0.017}_{-0.021}$ & $0.0551^{+0.0095}_{-0.0042}$ & $16.5^{+2.8}_{-1.3}$ & $4.82^{+0.88}_{-1.93}$ & $84.7^{+3.8}_{-12.1}$ & $0.46^{+0.4}_{-0.32}$ \\ 
48312914 & $11.95$ & $1.79\pm0.10$ & $6220\pm126$ & SG & $4.31651^{+0.00047}_{-0.00047}$ & $2338.3026^{+0.0016}_{-0.0015}$ & $0.0945^{+0.0109}_{-0.0054}$ & $18.5^{+2.1}_{-1.1}$ & $10.88^{+0.92}_{-1.72}$ & $87.9^{+1.5}_{-2.0}$ & $0.4^{+0.26}_{-0.27}$ \\ 
64329653 & $11.39$ & $1.73\pm0.09$ & $5802\pm125$ & SG & $4.3976^{+0.003}_{-0.0029}$ & $2231.2925^{+0.0058}_{-0.0071}$ & $0.071^{+0.0076}_{-0.0052}$ & $13.36^{+1.42}_{-0.97}$ & $11.6^{+2.2}_{-3.6}$ & $87.3^{+1.9}_{-3.3}$ & $0.55^{+0.28}_{-0.36}$ \\ 
202059363 & $11.28$ & $2.07\pm0.15$ & $6232\pm96$ & SG & $4.24748^{+0.00062}_{-0.00062}$ & $2202.7227^{+0.0017}_{-0.0017}$ & $0.0774^{+0.0018}_{-0.0019}$ & $17.51^{+0.41}_{-0.43}$ & $7.41^{+0.35}_{-0.75}$ & $87.7^{+1.6}_{-2.2}$ & $0.3^{+0.22}_{-0.2}$ \\ 
241470435 & $11.66$ & $2.28\pm0.13$ & $5827\pm118$ & SG & $5.4997^{+0.0028}_{-0.0028}$ & $2334.6001^{+0.0074}_{-0.0074}$ & $0.0513^{+0.0104}_{-0.0041}$ & $12.8^{+2.6}_{-1.0}$ & $7.8^{+1.7}_{-3.3}$ & $85.7^{+3.1}_{-7.3}$ & $0.58^{+0.32}_{-0.39}$ \\ 
241630802 & $11.84$ & $1.63\pm0.09$ & $5937\pm128$ & SG & $8.1552^{+0.003}_{-0.0031}$ & $2122.7835^{+0.0041}_{-0.0043}$ & $0.0581^{+0.0034}_{-0.003}$ & $10.31^{+0.61}_{-0.53}$ & $19.6^{+2.0}_{-4.2}$ & $88.8^{+0.84}_{-1.41}$ & $0.42^{+0.29}_{-0.28}$ \\ 
255661000 & $11.65$ & $2.59\pm0.18$ & $5811\pm129$ & SG & $7.4225^{+0.004}_{-0.0047}$ & $2504.024^{+0.017}_{-0.017}$ & $0.0388^{+0.0046}_{-0.003}$ & $10.97^{+1.3}_{-0.84}$ & $6.4^{+1.4}_{-2.5}$ & $85.0^{+3.6}_{-7.9}$ & $0.57^{+0.31}_{-0.38}$ \\ 
280514270 & $11.36$ & $2.34\pm0.12$ & $6045\pm136$ & SG & $1.24525^{+0.0003}_{-0.0004}$ & $2309.4702^{+0.0097}_{-0.0067}$ & $0.0338^{+0.0845}_{-0.004}$ & $8.6^{+21.6}_{-1.0}$ & $3.1^{+1.2}_{-1.6}$ & $78.1^{+9.0}_{-35.9}$ & $0.67^{+0.39}_{-0.45}$ \\ 
314729618 & $11.80$ & $3.05\pm0.15$ & $6244\pm133$ & SG & $1.42825^{+0.00065}_{-0.00063}$ & $2335.7128^{+0.0062}_{-0.0056}$ & $0.0508^{+0.0042}_{-0.0036}$ & $16.9^{+1.4}_{-1.2}$ & $3.12^{+0.42}_{-0.77}$ & $81.7^{+5.9}_{-10.7}$ & $0.46^{+0.31}_{-0.31}$ \\ 
331676303 & $11.88$ & $1.34\pm0.07$ & $5473\pm126$ & SG & $2.4665^{+0.0012}_{-0.0019}$ & $2336.4933^{+0.0117}_{-0.0071}$ & $0.0576^{+0.0082}_{-0.0046}$ & $8.42^{+1.2}_{-0.66}$ & $4.89^{+0.93}_{-1.9}$ & $84.2^{+4.2}_{-10.9}$ & $0.5^{+0.36}_{-0.34}$ \\ 
380353340 & $11.41$ & $1.69\pm0.08$ & $6143\pm133$ & SG & $6.32978^{+0.00063}_{-0.00065}$ & $2312.4576^{+0.0028}_{-0.0028}$ & $0.1203^{+0.005}_{-0.0039}$ & $22.23^{+0.93}_{-0.72}$ & $7.16^{+0.49}_{-0.4}$ & $83.35^{+0.65}_{-0.56}$ & $0.829^{+0.023}_{-0.034}$ \\ 
383239315 & $11.41$ & $2.76\pm0.14$ & $5912\pm130$ & SG & $6.1594^{+0.0015}_{-0.0018}$ & $2308.2795^{+0.0086}_{-0.0078}$ & $0.0444^{+0.0517}_{-0.0049}$ & $13.4^{+15.6}_{-1.5}$ & $9.8^{+3.3}_{-5.4}$ & $86.1^{+2.9}_{-9.7}$ & $0.68^{+0.33}_{-0.46}$ \\ 
397510904 & $11.64$ & $2.61\pm0.14$ & $6151\pm514$ & SG & $1.616151^{+8.2e-05}_{-0.000112}$ & $2391.175^{+0.011}_{-0.015}$ & $0.034^{+0.0483}_{-0.0033}$ & $9.68^{+13.78}_{-0.95}$ & $1.53^{+0.27}_{-0.35}$ & $64.5^{+18.29}_{-38.29}$ & $0.66^{+0.36}_{-0.43}$ \\ 
407495935 & $11.97$ & $2.47\pm0.13$ & $6254\pm188$ & SG & $3.0258^{+0.0012}_{-0.0014}$ & $2723.2015^{+0.005}_{-0.0043}$ & $0.0531^{+0.0028}_{-0.0027}$ & $14.31^{+0.75}_{-0.72}$ & $5.78^{+0.54}_{-1.11}$ & $86.2^{+2.7}_{-4.5}$ & $0.39^{+0.29}_{-0.27}$ \\ 
452077258 & $11.92$ & $2.09\pm0.10$ & $6220\pm135$ & SG & $6.3226^{+0.004}_{-0.0051}$ & $2311.4728^{+0.0082}_{-0.0078}$ & $0.0866^{+0.0333}_{-0.0073}$ & $19.8^{+7.6}_{-1.7}$ & $5.8^{+2.5}_{-2.0}$ & $82.6^{+5.3}_{-7.9}$ & $0.75^{+0.2}_{-0.44}$ \\ 
453142142 & $11.63$ & $2.42\pm0.14$ & $6221\pm120$ & SG & $1.70454^{+0.00064}_{-0.00057}$ & $2203.0845^{+0.005}_{-0.0061}$ & $0.0349^{+0.0034}_{-0.0027}$ & $9.18^{+0.9}_{-0.7}$ & $5.31^{+0.81}_{-1.52}$ & $84.8^{+3.7}_{-7.0}$ & $0.48^{+0.31}_{-0.33}$ \\ 
466314014 & $11.57$ & $2.14\pm0.19$ & $6008\pm133$ & SG & $2.89972^{+0.00036}_{-0.00034}$ & $2797.432^{+0.002}_{-0.0021}$ & $0.077^{+0.0025}_{-0.002}$ & $17.95^{+0.58}_{-0.47}$ & $5.81^{+0.31}_{-0.64}$ & $86.8^{+2.2}_{-2.9}$ & $0.33^{+0.23}_{-0.22}$ \\ 
466890210 & $11.71$ & $2.59\pm0.14$ & $6018\pm146$ & SG & $5.1871^{+0.0023}_{-0.0021}$ & $2799.5792^{+0.0063}_{-0.0069}$ & $0.0577^{+0.0026}_{-0.0023}$ & $16.31^{+0.74}_{-0.65}$ & $5.37^{+0.41}_{-0.94}$ & $85.8^{+2.9}_{-4.5}$ & $0.39^{+0.28}_{-0.27}$ \\ 
685340263 & $10.31$ & $1.80\pm0.08$ & $6083\pm135$ & SG & $5.06607^{+0.00028}_{-0.00027}$ & $2119.724^{+0.0014}_{-0.0014}$ & $0.0655^{+0.0014}_{-0.001}$ & $12.9^{+0.28}_{-0.21}$ & $6.26^{+0.18}_{-0.49}$ & $87.7^{+1.6}_{-2.4}$ & $0.25^{+0.22}_{-0.17}$ \\ 
13202532 & $10.87$ & $3.48$ & $5259\pm76$ & ERG & $5.5403^{+0.0013}_{-0.0013}$ & $2177.2393^{+0.003}_{-0.0033}$ & $0.0506^{+0.0012}_{-0.0012}$ & $19.17^{+0.47}_{-0.47}$ & $4.74^{+0.65}_{-0.66}$ & $83.7^{+4.0}_{-3.3}$ & $0.52^{+0.16}_{-0.3}$ \\ 
86894845 & $11.28$ & $2.64$ & $4973\pm122$ & ERG & $4.24021^{+0.00021}_{-0.00021}$ & $2475.1994^{+0.0018}_{-0.002}$ & $0.0531^{+0.0016}_{-0.0016}$ & $15.26^{+0.47}_{-0.45}$ & $8.51^{+0.63}_{-1.41}$ & $87.4^{+1.8}_{-2.6}$ & $0.38^{+0.26}_{-0.26}$ \\ 
218856662 & $10.95$ & $2.26$ & $5247\pm122$ & ERG & $1.12942^{+0.00082}_{-0.00091}$ & $2362.9883^{+0.0103}_{-0.0099}$ & $0.0439^{+0.2663}_{-0.0092}$ & $10.8^{+65.6}_{-2.3}$ & $1.82^{+0.67}_{-0.48}$ & $59.25^{+24.43}_{-29.41}$ & $0.84^{+0.41}_{-0.57}$ \\ 
251906970 & $10.97$ & $4.62$ & $4558\pm122$ & ERG & $1.1399^{+0.00075}_{-0.00074}$ & $2770.4044^{+0.0086}_{-0.0096}$ & $0.0439^{+0.3353}_{-0.0076}$ & $22.1^{+169.0}_{-3.8}$ & $2.01^{+1.04}_{-0.62}$ & $64.19^{+20.57}_{-34.0}$ & $0.83^{+0.49}_{-0.56}$ \\ 
268320188 & $10.80$ & $3.46$ & $4917\pm122$ & ERG & $10.02487^{+0.00034}_{-0.00031}$ & $2394.6478^{+0.0098}_{-0.0103}$ & $0.0306^{+0.0027}_{-0.0016}$ & $11.56^{+1.03}_{-0.62}$ & $4.79^{+0.94}_{-1.36}$ & $82.9^{+5.0}_{-6.9}$ & $0.59^{+0.24}_{-0.38}$ \\ 
291494871 & $10.72$ & $4.85$ & $4826\pm122$ & ERG & $1.15626^{+0.00076}_{-0.00078}$ & $2201.892^{+0.0095}_{-0.0123}$ & $0.0244^{+0.003}_{-0.002}$ & $12.9^{+1.6}_{-1.1}$ & $1.75^{+0.24}_{-0.34}$ & $74.39^{+11.03}_{-19.1}$ & $0.48^{+0.33}_{-0.33}$ \\ 
303333591 & $11.43$ & $4.44$ & $5032\pm122$ & ERG & $3.55076^{+0.00065}_{-0.00069}$ & $2282.9612^{+0.0044}_{-0.0045}$ & $0.0408^{+0.0033}_{-0.0023}$ & $19.8^{+1.6}_{-1.1}$ & $5.67^{+0.69}_{-1.48}$ & $85.5^{+3.3}_{-5.9}$ & $0.45^{+0.31}_{-0.31}$ \\ 
397187844 & $11.52$ & $3.26$ & $4968\pm122$ & ERG & $8.4179^{+0.0028}_{-0.0031}$ & $2285.649^{+0.018}_{-0.02}$ & $0.0383^{+0.377}_{-0.0084}$ & $13.6^{+134.1}_{-3.0}$ & $3.9^{+3.0}_{-1.6}$ & $77.15^{+10.59}_{-24.28}$ & $0.86^{+0.51}_{-0.59}$ 

%% file: planet_param_table_TOI2.tex
13737885 & 3382 & $11.55$ & $2.16\pm0.11$ & $6329\pm119$ & SG & $4.4574^{+0.0013}_{-0.0013}$ & $2230.0811^{+0.0025}_{-0.0024}$ & $0.0637^{+0.0015}_{-0.0017}$ & $15.0^{+0.35}_{-0.4}$ & $5.72^{+0.61}_{-0.88}$ & $85.3^{+3.2}_{-3.3}$ & $0.47^{+0.2}_{-0.3}$ \\ 
14156936 & 5340 & $11.71$ & $1.69\pm0.08$ & $6287\pm107$ & SG & $4.93921^{+0.00017}_{-0.00016}$ & $2450.7911^{+0.0015}_{-0.0015}$ & $0.0747^{+0.0016}_{-0.0014}$ & $13.78^{+0.3}_{-0.26}$ & $8.01^{+0.28}_{-0.71}$ & $88.1^{+1.3}_{-2.0}$ & $0.26^{+0.23}_{-0.18}$ \\ 
21132157 & 5365 & $11.04$ & $2.76\pm0.17$ & $6080\pm137$ & SG & $4.00774^{+0.00063}_{-0.00065}$ & $2449.2678^{+0.005}_{-0.0049}$ & $0.0328^{+0.0022}_{-0.0015}$ & $9.86^{+0.67}_{-0.44}$ & $4.33^{+0.53}_{-1.04}$ & $83.6^{+4.5}_{-6.9}$ & $0.49^{+0.28}_{-0.32}$ \\ 
28961316 & 3682 & $11.32$ & $1.91\pm0.10$ & $5501\pm125$ & SG & $3.346539^{+9.6e-05}_{-9.5e-05}$ & $2448.3827^{+0.0012}_{-0.0013}$ & $0.0675^{+0.0025}_{-0.0016}$ & $14.07^{+0.53}_{-0.33}$ & $6.27^{+0.32}_{-0.7}$ & $87.0^{+2.1}_{-2.7}$ & $0.33^{+0.22}_{-0.23}$ \\ 
32443914 & 6777 & $11.97$ & $1.90\pm0.11$ & $6258\pm152$ & SG & $3.36116^{+0.00058}_{-0.00058}$ & $2037.6528^{+0.002}_{-0.0021}$ & $0.0833^{+0.0026}_{-0.0026}$ & $17.28^{+0.54}_{-0.53}$ & $5.47^{+0.44}_{-0.83}$ & $85.7^{+3.0}_{-3.8}$ & $0.41^{+0.24}_{-0.28}$ \\ 
33227166 & 4999 & $10.82$ & $1.99\pm0.10$ & $5698\pm128$ & SG & $5.5832^{+0.0051}_{-0.0073}$ & $2366.0549^{+0.0146}_{-0.0093}$ & $0.0483^{+0.0039}_{-0.0033}$ & $10.49^{+0.85}_{-0.72}$ & $7.2^{+1.0}_{-1.9}$ & $86.3^{+2.6}_{-4.6}$ & $0.46^{+0.31}_{-0.31}$ \\ 
58372253 & 5328 & $11.23$ & $2.54\pm0.12$ & $6120\pm101$ & SG & $5.48889^{+0.00038}_{-0.0004}$ & $2475.4876^{+0.002}_{-0.0019}$ & $0.06732^{+0.00096}_{-0.00103}$ & $18.62^{+0.26}_{-0.28}$ & $6.08^{+0.23}_{-0.47}$ & $87.3^{+1.8}_{-2.2}$ & $0.28^{+0.19}_{-0.19}$ \\ 
81077799 & 3492 & $8.45$ & $2.59\pm0.11$ & $6332\pm134$ & SG & $9.22261^{+0.00035}_{-0.00036}$ & $2286.85431^{+0.00099}_{-0.00098}$ & $0.05056^{+0.00129}_{-0.00078}$ & $14.3^{+0.36}_{-0.22}$ & $14.05^{+0.53}_{-1.44}$ & $88.84^{+0.82}_{-1.2}$ & $0.29^{+0.24}_{-0.2}$ \\ 
85266608 & 5385 & $10.86$ & $2.08\pm0.10$ & $6238\pm124$ & SG & $5.1676^{+0.0012}_{-0.0011}$ & $2613.4016^{+0.0033}_{-0.0033}$ & $0.0653^{+0.0037}_{-0.0032}$ & $14.8^{+0.84}_{-0.73}$ & $5.84^{+0.96}_{-0.98}$ & $84.3^{+3.6}_{-3.2}$ & $0.58^{+0.17}_{-0.33}$ \\ 
86898676 & 5432 & $11.44$ & $1.48\pm0.08$ & $5612\pm124$ & SG & $2.952497^{+8.8e-05}_{-8.8e-05}$ & $2475.1701^{+0.0013}_{-0.0013}$ & $0.0794^{+0.0025}_{-0.0021}$ & $12.83^{+0.41}_{-0.34}$ & $7.41^{+0.42}_{-0.93}$ & $87.4^{+1.8}_{-2.5}$ & $0.34^{+0.24}_{-0.23}$ \\ 
95191643 & 6417 & $11.70$ & $1.75\pm0.09$ & $6089\pm122$ & SG & $3.34565^{+0.00041}_{-0.0004}$ & $2230.9457^{+0.0014}_{-0.0014}$ & $0.0774^{+0.0013}_{-0.0014}$ & $14.74^{+0.25}_{-0.28}$ & $6.8^{+0.27}_{-0.64}$ & $87.6^{+1.7}_{-2.3}$ & $0.29^{+0.22}_{-0.19}$ \\ 
97669493 & 5474 & $11.69$ & $2.45\pm0.14$ & $5874\pm168$ & SG & $8.4336^{+0.0013}_{-0.0013}$ & $2479.1328^{+0.0052}_{-0.0058}$ & $0.0475^{+0.003}_{-0.0021}$ & $12.72^{+0.79}_{-0.57}$ & $6.71^{+0.79}_{-1.34}$ & $86.0^{+2.8}_{-3.6}$ & $0.47^{+0.24}_{-0.31}$ \\ 
97714451 & 4734 & $11.90$ & $1.76\pm0.11$ & $5854\pm149$ & SG & $6.235664^{+6.4e-05}_{-6.6e-05}$ & $2205.1508^{+0.0029}_{-0.0027}$ & $0.055^{+0.0013}_{-0.0013}$ & $10.55^{+0.24}_{-0.25}$ & $8.22^{+0.42}_{-1.02}$ & $87.8^{+1.5}_{-2.3}$ & $0.32^{+0.25}_{-0.22}$ \\ 
131348451 & 3414 & $11.97$ & $1.81\pm0.09$ & $6259\pm139$ & SG & $9.7358^{+0.0025}_{-0.0023}$ & $2233.5667^{+0.0054}_{-0.0054}$ & $0.0548^{+0.0037}_{-0.0028}$ & $10.79^{+0.73}_{-0.54}$ & $10.5^{+1.5}_{-2.7}$ & $87.2^{+2.0}_{-3.0}$ & $0.51^{+0.27}_{-0.34}$ \\ 
141363913 & 601 & $9.92$ & $3.10\pm0.26$ & $6221\pm133$ & SG & $3.4788^{+0.0014}_{-0.0014}$ & $2258.6743^{+0.005}_{-0.005}$ & $0.0354^{+0.0036}_{-0.0026}$ & $11.96^{+1.22}_{-0.87}$ & $10.3^{+1.8}_{-3.3}$ & $87.2^{+2.0}_{-4.0}$ & $0.51^{+0.32}_{-0.34}$ \\ 
143257768 & 5679 & $11.20$ & $1.61\pm0.09$ & $6152\pm131$ & SG & $1.6622^{+0.00016}_{-0.00016}$ & $2037.108^{+0.0012}_{-0.0012}$ & $0.1053^{+0.0027}_{-0.0023}$ & $18.5^{+0.48}_{-0.41}$ & $3.77^{+0.16}_{-0.29}$ & $85.6^{+3.0}_{-3.6}$ & $0.29^{+0.19}_{-0.2}$ \\ 
193754373 & 4487 & $11.49$ & $1.69\pm0.08$ & $6006\pm156$ & SG & $3.954057^{+1.8e-05}_{-1.8e-05}$ & $2422.6466^{+0.0012}_{-0.0012}$ & $0.081^{+0.0026}_{-0.0021}$ & $14.9^{+0.49}_{-0.4}$ & $7.3^{+0.46}_{-0.85}$ & $87.2^{+1.9}_{-2.2}$ & $0.36^{+0.21}_{-0.24}$ \\ 
202941251 & 5891 & $11.39$ & $2.40\pm0.12$ & $6283\pm153$ & SG & $4.178479^{+2.5e-05}_{-2.5e-05}$ & $2422.4978^{+0.0019}_{-0.0018}$ & $0.0703^{+0.0027}_{-0.0037}$ & $18.41^{+0.71}_{-0.97}$ & $5.68^{+1.18}_{-0.75}$ & $83.5^{+3.8}_{-2.3}$ & $0.64^{+0.12}_{-0.32}$ \\ 
215018906 & 5217 & $11.00$ & $2.45\pm0.20$ & $6029\pm204$ & SG & $4.371281^{+4.9e-05}_{-4.7e-05}$ & $2393.0205^{+0.002}_{-0.0019}$ & $0.0606^{+0.0029}_{-0.0022}$ & $16.2^{+0.76}_{-0.6}$ & $6.91^{+0.84}_{-1.62}$ & $86.0^{+2.8}_{-4.2}$ & $0.48^{+0.28}_{-0.33}$ \\ 
221785300 & 3489 & $11.72$ & $1.23\pm0.07$ & $5266\pm128$ & SG & $3.7834^{+0.00038}_{-0.00038}$ & $2363.2037^{+0.0014}_{-0.0014}$ & $0.1011^{+0.0024}_{-0.0023}$ & $13.57^{+0.32}_{-0.31}$ & $8.7^{+0.27}_{-0.53}$ & $88.5^{+1.0}_{-1.5}$ & $0.22^{+0.2}_{-0.16}$ \\ 
232982938 & 1828 & $11.13$ & $1.56\pm0.08$ & $5809\pm127$ & SG & $9.093948^{+4.2e-05}_{-4.2e-05}$ & $2390.9859^{+0.0013}_{-0.0013}$ & $0.05151^{+0.0009}_{-0.00081}$ & $8.78^{+0.15}_{-0.14}$ & $14.79^{+0.5}_{-1.23}$ & $88.98^{+0.7}_{-0.98}$ & $0.26^{+0.21}_{-0.18}$ \\ 
238197638 & 903 & $11.79$ & $2.21\pm0.10$ & $6147\pm111$ & SG & $7.276601^{+5.8e-05}_{-5.7e-05}$ & $2039.8851^{+0.0014}_{-0.0014}$ & $0.0568^{+0.0021}_{-0.0022}$ & $13.69^{+0.51}_{-0.54}$ & $9.7^{+1.9}_{-1.6}$ & $86.4^{+2.4}_{-1.7}$ & $0.6^{+0.15}_{-0.36}$ \\ 
238920875 & 2329 & $11.79$ & $1.33\pm0.08$ & $5218\pm133$ & SG & $6.534667^{+4.2e-05}_{-4.3e-05}$ & $2064.4841^{+0.0011}_{-0.0011}$ & $0.0783^{+0.0014}_{-0.0021}$ & $11.33^{+0.2}_{-0.3}$ & $12.6^{+2.0}_{-1.5}$ & $87.32^{+1.52}_{-0.94}$ & $0.59^{+0.11}_{-0.29}$ \\ 
258510872 & 1885 & $11.97$ & $1.76\pm0.08$ & $6232\pm123$ & SG & $6.544057^{+6e-05}_{-5.8e-05}$ & $2612.6614^{+0.001}_{-0.0011}$ & $0.0742^{+0.0012}_{-0.0013}$ & $14.23^{+0.24}_{-0.25}$ & $13.31^{+0.42}_{-1.01}$ & $88.94^{+0.74}_{-1.07}$ & $0.25^{+0.21}_{-0.17}$ \\ 
282221346 & 5042 & $11.04$ & $2.32\pm0.11$ & $5784\pm129$ & SG & $5.63204^{+0.0006}_{-0.00057}$ & $2339.3784^{+0.0031}_{-0.0032}$ & $0.0458^{+0.0019}_{-0.0014}$ & $11.59^{+0.48}_{-0.36}$ & $6.16^{+0.5}_{-1.03}$ & $86.2^{+2.6}_{-3.6}$ & $0.41^{+0.25}_{-0.27}$ \\ 
291517604 & 5148 & $10.01$ & $1.91\pm0.10$ & $5447\pm127$ & SG & $5.0642^{+0.00041}_{-0.00042}$ & $2362.8529^{+0.0012}_{-0.0011}$ & $0.0666^{+0.0015}_{-0.001}$ & $13.86^{+0.3}_{-0.21}$ & $6.45^{+0.25}_{-0.57}$ & $87.4^{+1.8}_{-2.3}$ & $0.29^{+0.21}_{-0.2}$ \\ 
294768072 & 5062 & $11.81$ & $2.08\pm0.11$ & $5693\pm115$ & SG & $6.2292^{+0.0053}_{-0.0059}$ & $2367.693^{+0.011}_{-0.014}$ & $0.0429^{+0.0078}_{-0.0037}$ & $9.7^{+1.76}_{-0.84}$ & $6.8^{+1.5}_{-2.7}$ & $85.5^{+3.3}_{-7.7}$ & $0.55^{+0.33}_{-0.37}$ \\ 
306248357 & 6538 & $11.93$ & $1.83\pm0.09$ & $6219\pm130$ & SG & $5.7218^{+0.0047}_{-0.0041}$ & $2310.9598^{+0.0071}_{-0.0061}$ & $0.0524^{+0.0037}_{-0.0031}$ & $10.44^{+0.74}_{-0.62}$ & $7.31^{+0.81}_{-1.56}$ & $86.8^{+2.3}_{-3.8}$ & $0.42^{+0.29}_{-0.29}$ \\ 
306357884 & 3322 & $11.56$ & $2.53\pm0.21$ & $6307\pm130$ & SG & $1.05711^{+0.00016}_{-0.00018}$ & $2362.5405^{+0.002}_{-0.0021}$ & $0.0438^{+0.0057}_{-0.0034}$ & $12.12^{+1.58}_{-0.93}$ & $5.4^{+1.1}_{-1.9}$ & $84.3^{+4.1}_{-8.3}$ & $0.55^{+0.31}_{-0.37}$ \\ 
315096343 & 2978 & $11.67$ & $1.88\pm0.10$ & $5734\pm126$ & SG & $4.3497^{+0.0022}_{-0.002}$ & $2258.1449^{+0.0058}_{-0.0065}$ & $0.0476^{+0.0047}_{-0.0032}$ & $9.75^{+0.96}_{-0.65}$ & $6.8^{+1.1}_{-2.0}$ & $85.8^{+3.0}_{-5.2}$ & $0.51^{+0.29}_{-0.34}$ \\ 
334305570 & 777 & $9.37$ & $1.60\pm0.08$ & $6187\pm137$ & SG & $8.30158^{+0.00086}_{-0.00089}$ & $2314.5557^{+0.0012}_{-0.0012}$ & $0.05003^{+0.00072}_{-0.00073}$ & $8.75^{+0.13}_{-0.13}$ & $13.43^{+0.4}_{-1.07}$ & $88.94^{+0.73}_{-1.06}$ & $0.25^{+0.21}_{-0.17}$ \\ 
346667887 & 5181 & $11.72$ & $1.66\pm0.09$ & $5943\pm137$ & SG & $3.892222^{+2.3e-05}_{-2.4e-05}$ & $2392.347^{+0.0017}_{-0.0016}$ & $0.0739^{+0.0048}_{-0.0043}$ & $13.39^{+0.87}_{-0.79}$ & $9.6^{+1.8}_{-1.8}$ & $86.5^{+2.4}_{-2.2}$ & $0.59^{+0.18}_{-0.36}$ \\ 
368713985 & 6562 & $11.76$ & $2.20\pm0.12$ & $6301\pm133$ & SG & $4.09423^{+0.00068}_{-0.00069}$ & $2338.8733^{+0.0021}_{-0.0021}$ & $0.0747^{+0.0019}_{-0.0018}$ & $17.9^{+0.46}_{-0.43}$ & $6.08^{+0.26}_{-0.62}$ & $87.3^{+1.9}_{-2.7}$ & $0.28^{+0.23}_{-0.2}$ \\ 
369474541 & 6843 & $11.19$ & $1.73\pm0.09$ & $6281\pm126$ & SG & $5.49696^{+0.00034}_{-0.00034}$ & $2451.4593^{+0.002}_{-0.002}$ & $0.0791^{+0.0022}_{-0.0018}$ & $14.9^{+0.42}_{-0.34}$ & $8.48^{+0.5}_{-0.93}$ & $87.6^{+1.6}_{-1.8}$ & $0.35^{+0.2}_{-0.23}$ \\ 
371864043 & 3041 & $11.26$ & $1.84\pm0.09$ & $6268\pm121$ & SG & $2.960003^{+5.3e-05}_{-5.3e-05}$ & $2285.13638^{+0.00076}_{-0.00076}$ & $0.0829^{+0.0016}_{-0.002}$ & $16.6^{+0.31}_{-0.4}$ & $5.95^{+0.59}_{-0.5}$ & $85.2^{+2.6}_{-1.7}$ & $0.5^{+0.12}_{-0.25}$ \\ 
379824738 & 5810 & $10.29$ & $1.87\pm0.09$ & $6142\pm132$ & SG & $5.68536^{+0.00042}_{-0.00042}$ & $2797.7146^{+0.0012}_{-0.0012}$ & $0.07319^{+0.0009}_{-0.00088}$ & $14.9^{+0.18}_{-0.18}$ & $7.44^{+0.15}_{-0.32}$ & $88.57^{+0.98}_{-1.39}$ & $0.19^{+0.16}_{-0.13}$ \\ 
397763985 & 5765 & $11.54$ & $1.57\pm0.08$ & $5798\pm214$ & SG & $3.8782^{+0.001}_{-0.001}$ & $2722.4911^{+0.0031}_{-0.003}$ & $0.0696^{+0.0062}_{-0.0049}$ & $11.92^{+1.07}_{-0.83}$ & $6.4^{+1.3}_{-1.5}$ & $84.6^{+3.6}_{-3.9}$ & $0.6^{+0.2}_{-0.37}$ \\ 
404467699 & 857 & $10.27$ & $1.63\pm0.08$ & $6143\pm131$ & SG & $3.90849^{+0.00023}_{-0.00022}$ & $2118.2899^{+0.00068}_{-0.00067}$ & $0.0872^{+0.0018}_{-0.0028}$ & $15.52^{+0.32}_{-0.5}$ & $7.12^{+0.6}_{-0.4}$ & $84.45^{+1.1}_{-0.72}$ & $0.689^{+0.046}_{-0.09}$ \\ 
409594381 & 4214 & $11.01$ & $1.59\pm0.08$ & $5806\pm174$ & SG & $3.49122^{+0.00016}_{-0.00016}$ & $2203.1546^{+0.0011}_{-0.0011}$ & $0.0662^{+0.002}_{-0.0017}$ & $11.45^{+0.35}_{-0.29}$ & $8.47^{+0.46}_{-1.16}$ & $87.7^{+1.6}_{-2.4}$ & $0.34^{+0.26}_{-0.23}$ \\ 
420268415 & 4436 & $11.89$ & $2.00\pm0.09$ & $6250\pm136$ & SG & $1.7590562^{+2.5e-06}_{-2.5e-06}$ & $2392.93207^{+0.00038}_{-0.00038}$ & $0.09803^{+0.0009}_{-0.0008}$ & $21.38^{+0.2}_{-0.17}$ & $4.137^{+0.086}_{-0.15}$ & $86.8^{+2.1}_{-2.0}$ & $0.23^{+0.13}_{-0.15}$ \\ 
436102447 & 2359 & $11.03$ & $2.70\pm0.14$ & $6130\pm120$ & SG & $5.18706^{+0.00094}_{-0.00093}$ & $2176.3256^{+0.0025}_{-0.0025}$ & $0.0639^{+0.0017}_{-0.0015}$ & $18.8^{+0.49}_{-0.45}$ & $5.32^{+0.36}_{-0.67}$ & $85.8^{+2.9}_{-3.3}$ & $0.39^{+0.22}_{-0.26}$ \\ 
436637677 & 4705 & $11.46$ & $1.77\pm0.09$ & $6266\pm148$ & SG & $11.91384^{+0.00014}_{-0.00013}$ & $2185.5327^{+0.0031}_{-0.0031}$ & $0.0714^{+0.002}_{-0.0019}$ & $13.79^{+0.39}_{-0.37}$ & $16.62^{+0.87}_{-2.21}$ & $88.89^{+0.77}_{-1.21}$ & $0.32^{+0.26}_{-0.22}$ \\ 
440872576 & 3160 & $11.93$ & $1.62\pm0.12$ & $6200\pm125$ & SG & $3.97095^{+0.0004}_{-0.00043}$ & $2335.9688^{+0.0015}_{-0.0015}$ & $0.0874^{+0.0053}_{-0.0035}$ & $15.44^{+0.94}_{-0.61}$ & $10.5^{+1.2}_{-2.2}$ & $87.4^{+1.8}_{-2.5}$ & $0.47^{+0.26}_{-0.32}$ \\ 
448435566 & 4951 & $11.84$ & $1.92\pm0.09$ & $5823\pm125$ & SG & $4.1027^{+0.0013}_{-0.0013}$ & $2337.3539^{+0.0049}_{-0.0043}$ & $0.0599^{+0.0031}_{-0.0028}$ & $12.55^{+0.65}_{-0.59}$ & $5.51^{+0.88}_{-1.33}$ & $84.4^{+4.1}_{-5.2}$ & $0.54^{+0.25}_{-0.37}$ \\ 
446370039 & 3499 & $11.52$ & $3.12\pm0.18$ & $5695\pm132$ & ERG & $4.59876^{+0.0006}_{-0.00059}$ & $2284.9679^{+0.0038}_{-0.0039}$ & $0.0559^{+0.0014}_{-0.0015}$ & $19.04^{+0.48}_{-0.5}$ & $4.02^{+0.44}_{-0.61}$ & $83.1^{+4.7}_{-4.7}$ & $0.49^{+0.2}_{-0.31}$

%% file: planet_param_table_KP2.tex
17746821 & HAT-P-50 b & $11.32$ & $1.73\pm0.11$ & $6322\pm143$ & SG & $3.121989^{+1.1e-05}_{-1e-05}$ & $2229.96168^{+0.00088}_{-0.0009}$ & $0.0788^{+0.0023}_{-0.0033}$ & $14.86^{+0.44}_{-0.63}$ & $5.76^{+0.82}_{-0.53}$ & $84.0^{+2.8}_{-1.7}$ & $0.606^{+0.097}_{-0.231}$ & \citet{hartman2015} \\ 
20096620 & HAT-P-13 b & $9.96$ & $1.83\pm0.09$ & $5720\pm119$ & SG & $2.916382^{+8.3e-05}_{-8.3e-05}$ & $2582.97935^{+0.0004}_{-0.00039}$ & $0.0859^{+0.0015}_{-0.0018}$ & $17.11^{+0.3}_{-0.37}$ & $5.6^{+0.33}_{-0.23}$ & $82.66^{+0.94}_{-0.68}$ & $0.716^{+0.034}_{-0.055}$ & \citet{bakos2009} \\ 
44792534 & TOI-954 b & $9.78$ & $1.93\pm0.10$ & $5756\pm127$ & SG & $3.68425^{+0.00033}_{-0.00033}$ & $2145.2174^{+0.0012}_{-0.0012}$ & $0.04616^{+0.00109}_{-0.00084}$ & $9.73^{+0.23}_{-0.18}$ & $5.91^{+0.23}_{-0.56}$ & $87.2^{+1.9}_{-2.6}$ & $0.29^{+0.22}_{-0.2}$ & \citet{sha2021} \\ 
51234631 & XO-4 b & $10.17$ & $1.62\pm0.09$ & $6304\pm140$ & SG & $4.12527^{+0.00014}_{-0.00014}$ & $2584.51491^{+0.00041}_{-0.00041}$ & $0.08787^{+0.00068}_{-0.00061}$ & $15.55^{+0.12}_{-0.11}$ & $7.7^{+0.13}_{-0.27}$ & $88.57^{+0.98}_{-1.09}$ & $0.19^{+0.13}_{-0.13}$ & \citet{mccullough2008} \\ 
68007716 & TOI-2587 A b & $10.97$ & $1.72\pm0.10$ & $5776\pm137$ & SG & $5.4574^{+0.0012}_{-0.0012}$ & $2233.8368^{+0.0022}_{-0.0021}$ & $0.0615^{+0.0018}_{-0.0017}$ & $11.52^{+0.34}_{-0.32}$ & $10.53^{+0.49}_{-1.15}$ & $88.4^{+1.1}_{-1.7}$ & $0.3^{+0.24}_{-0.2}$ & \citet{yee2023} \\ 
69356857 & TOI-5232 b & $11.78$ & $1.74\pm0.08$ & $6212\pm128$ & SG & $4.096625^{+1.9e-05}_{-1.9e-05}$ & $2390.9111^{+0.0014}_{-0.0014}$ & $0.0709^{+0.0017}_{-0.0015}$ & $13.46^{+0.32}_{-0.28}$ & $6.67^{+0.28}_{-0.6}$ & $87.5^{+1.7}_{-2.3}$ & $0.3^{+0.21}_{-0.2}$ & \citet{schulte2024} \\ 
70524163 & TOI-2421 b & $10.69$ & $1.79\pm0.10$ & $5618\pm132$ & SG & $4.34781^{+0.00058}_{-0.00058}$ & $2117.8801^{+0.0019}_{-0.0018}$ & $0.0556^{+0.003}_{-0.002}$ & $10.86^{+0.58}_{-0.39}$ & $6.18^{+0.8}_{-1.09}$ & $85.3^{+3.2}_{-3.4}$ & $0.51^{+0.21}_{-0.32}$ & \citet{yee2022} \\ 
77156657 & WASP-159 b & $11.88$ & $2.04\pm0.09$ & $5977\pm130$ & SG & $3.84053^{+0.00019}_{-0.00019}$ & $2146.6463^{+0.0015}_{-0.0015}$ & $0.0736^{+0.0019}_{-0.0017}$ & $16.36^{+0.41}_{-0.38}$ & $4.93^{+0.4}_{-0.51}$ & $85.1^{+3.3}_{-2.8}$ & $0.42^{+0.17}_{-0.27}$ & \citet{hellier2019} \\ 
88992642 & TOI-2145 b & $8.56$ & $2.78\pm0.13$ & $6202\pm130$ & SG & $10.261132^{+3.3e-05}_{-3.6e-05}$ & $2392.9417^{+0.001}_{-0.00098}$ & $0.04367^{+0.0007}_{-0.00089}$ & $13.25^{+0.21}_{-0.27}$ & $8.75^{+1.08}_{-0.69}$ & $85.99^{+1.4}_{-0.89}$ & $0.612^{+0.076}_{-0.163}$ & \citet{rodriguez2023} \\ 
92352620 & WASP-94 A b & $9.64$ & $1.71\pm0.07$ & $6073\pm106$ & SG & $3.95011^{+0.00012}_{-0.00012}$ & $2039.33562^{+0.00036}_{-0.00036}$ & $0.10633^{+0.00073}_{-0.00072}$ & $19.84^{+0.14}_{-0.13}$ & $7.349^{+0.067}_{-0.123}$ & $89.11^{+0.62}_{-0.91}$ & $0.115^{+0.113}_{-0.079}$ & \citet{neveuvanmalle2014} \\ 
97735908 & KELT-2 A b & $8.22$ & $1.88\pm0.09$ & $6327\pm132$ & SG & $4.113764^{+2e-05}_{-1.9e-05}$ & $2475.42662^{+0.0002}_{-0.0002}$ & $0.06933^{+0.00037}_{-0.00026}$ & $14.195^{+0.077}_{-0.053}$ & $6.607^{+0.083}_{-0.162}$ & $88.48^{+1.0}_{-1.02}$ & $0.17^{+0.11}_{-0.11}$ & \citet{beatty2012} \\ 
99869022 & TOI-1199 b & $10.42$ & $1.54\pm0.08$ & $5618\pm122$ & SG & $3.671434^{+1.7e-05}_{-1.7e-05}$ & $2420.53868^{+0.00064}_{-0.00062}$ & $0.0656^{+0.0018}_{-0.0019}$ & $10.98^{+0.3}_{-0.31}$ & $7.74^{+1.07}_{-0.6}$ & $83.86^{+1.23}_{-0.79}$ & $0.827^{+0.033}_{-0.074}$ & \citet{serranobell2024} \\ 
100566492 & WASP-88 b & $11.12$ & $1.99\pm0.09$ & $6282\pm106$ & SG & $4.953^{+0.0009}_{-0.00092}$ & $2040.9075^{+0.0015}_{-0.0015}$ & $0.0863^{+0.0012}_{-0.0012}$ & $18.74^{+0.26}_{-0.26}$ & $6.74^{+0.15}_{-0.33}$ & $88.4^{+1.1}_{-1.7}$ & $0.19^{+0.18}_{-0.13}$ & \citet{delrez2014} \\ 
122612091 & WASP-72 b & $10.46$ & $2.66\pm0.13$ & $5773\pm111$ & SG & $2.216706^{+4.9e-05}_{-4.9e-05}$ & $2117.14139^{+0.00067}_{-0.00067}$ & $0.0665^{+0.0012}_{-0.0016}$ & $19.3^{+0.36}_{-0.48}$ & $3.59^{+0.31}_{-0.2}$ & $78.9^{+2.2}_{-1.4}$ & $0.694^{+0.045}_{-0.086}$ & \citet{gillon2013} \\ 
138294130 & HAT-P-4 b & $10.65$ & $1.60\pm0.07$ & $5981\pm117$ & SG & $3.056495^{+0.0001}_{-0.0001}$ & $2671.14452^{+0.00061}_{-0.00061}$ & $0.0849^{+0.0019}_{-0.0014}$ & $14.8^{+0.33}_{-0.25}$ & $5.68^{+0.21}_{-0.4}$ & $87.0^{+2.0}_{-2.2}$ & $0.29^{+0.18}_{-0.19}$ & \citet{kovacs2007} \\ 
139528693 & WASP-78 b & $11.83$ & $2.06\pm0.10$ & $6291\pm130$ & SG & $2.175226^{+5.1e-05}_{-5.1e-05}$ & $2145.13632^{+0.00074}_{-0.00071}$ & $0.0829^{+0.00082}_{-0.0008}$ & $18.65^{+0.19}_{-0.18}$ & $3.819^{+0.058}_{-0.13}$ & $87.4^{+1.8}_{-2.5}$ & $0.17^{+0.15}_{-0.12}$ & \citet{smalley2012}\\ 
157230659 & NGTS-12 b & $11.66$ & $1.53\pm0.07$ & $5869\pm124$ & SG & $7.53206^{+0.00056}_{-0.00057}$ & $2288.4659^{+0.0018}_{-0.0017}$ & $0.0682^{+0.0011}_{-0.0011}$ & $11.37^{+0.18}_{-0.19}$ & $10.59^{+0.26}_{-0.61}$ & $88.83^{+0.8}_{-1.1}$ & $0.22^{+0.18}_{-0.15}$ & \citet{bryant2020} \\ 
159742538 & TrES-4 b & $11.11$ & $2.01\pm0.10$ & $6184\pm138$ & SG & $3.5539341^{+7.3e-06}_{-7.2e-06}$ & $2391.21043^{+0.00051}_{-0.00053}$ & $0.0966^{+0.0015}_{-0.0022}$ & $21.16^{+0.33}_{-0.48}$ & $5.97^{+0.43}_{-0.27}$ & $82.63^{+1.03}_{-0.67}$ & $0.765^{+0.031}_{-0.061}$ & \citet{mandushev2007} \\ 
166836920 & WASP-99 b & $8.93$ & $1.71\pm0.08$ & $6079\pm132$ & SG & $5.752698^{+7.7e-05}_{-7.9e-05}$ & $2089.77512^{+0.0004}_{-0.0004}$ & $0.06914^{+0.00034}_{-0.00032}$ & $12.891^{+0.063}_{-0.059}$ & $8.785^{+0.062}_{-0.151}$ & $89.26^{+0.51}_{-0.74}$ & $0.113^{+0.109}_{-0.078}$ & \citet{hellier2014} \\ 
176220787 & WASP-136 b & $9.48$ & $2.39\pm0.12$ & $6067\pm120$ & SG & $5.215357^{+1.1e-05}_{-1.2e-05}$ & $2092.52423^{+0.00046}_{-0.00047}$ & $0.06915^{+0.00054}_{-0.00048}$ & $18.03^{+0.14}_{-0.13}$ & $7.6^{+0.13}_{-0.33}$ & $88.5^{+1.0}_{-1.3}$ & $0.19^{+0.16}_{-0.13}$ & \citet{lam2017}\\ 
211438925 & WASP-20 b & $10.25$ & $1.50$ & $5776$ & SG & $4.89965^{+0.00024}_{-0.00024}$ & $2091.51188^{+0.00063}_{-0.00062}$ & $0.0983^{+0.0025}_{-0.0035}$ & $16.13^{+0.41}_{-0.57}$ & $10.49^{+1.09}_{-0.72}$ & $86.82^{+1.14}_{-0.71}$ & $0.581^{+0.081}_{-0.17}$ & \citet{anderson2015}\\ 
219015370 & TOI-1294 b & $10.95$ & $1.58\pm0.08$ & $5714\pm126$ & SG & $3.91527^{+2e-05}_{-2e-05}$ & $2393.00789^{+0.0009}_{-0.00091}$ & $0.0606^{+0.0018}_{-0.0016}$ & $10.45^{+0.31}_{-0.27}$ & $6.45^{+0.68}_{-0.45}$ & $82.48^{+1.12}_{-0.82}$ & $0.843^{+0.027}_{-0.049}$ & \citet{polanski2024toi320} \\ 
219854185 & TOI-1296 b & $10.80$ & $1.75\pm0.09$ & $5494\pm126$ & SG & $3.9443915^{+6.6e-06}_{-6.6e-06}$ & $2392.24607^{+0.00048}_{-0.00048}$ & $0.07677^{+0.00093}_{-0.00063}$ & $14.69^{+0.18}_{-0.12}$ & $6.53^{+0.12}_{-0.25}$ & $88.1^{+1.3}_{-1.4}$ & $0.21^{+0.14}_{-0.14}$ & \citet{montou2021}\\
229510866 & TOI-1181 b & $10.08$ & $1.93\pm0.09$ & $6122\pm125$ & SG & $2.1031922^{+1.2e-06}_{-1.2e-06}$ & $2391.07951^{+0.00016}_{-0.00016}$ & $0.07785^{+0.00053}_{-0.00048}$ & $16.36^{+0.11}_{-0.10}$ & $4.14^{+0.10}_{-0.10}$ & $86.3^{+1.9}_{-1.3}$ & $0.27^{+0.09}_{-0.14}$ & \citet{kabath2022toi1181} \\
231670397 & WASP-73 b & $9.86$ & $2.22\pm0.10$ & $6124\pm134$ & SG & $4.08758^{+0.00032}_{-0.00033}$ & $2038.86393^{+0.00095}_{-0.00093}$ & $0.05755^{+0.00101}_{-0.0009}$ & $13.93^{+0.24}_{-0.22}$ & $5.56^{+0.37}_{-0.5}$ & $86.0^{+2.6}_{-2.3}$ & $0.39^{+0.16}_{-0.24}$ & \citet{delrez2014} \\ 
237104103 & TOI-1298 b & $10.96$ & $1.49\pm0.07$ & $5731\pm125$ & SG & $4.537122^{+9.3e-06}_{-9.3e-06}$ & $2392.37499^{+0.00057}_{-0.00058}$ & $0.06271^{+0.00079}_{-0.00063}$ & $10.17^{+0.13}_{-0.1}$ & $8.89^{+0.27}_{-0.55}$ & $88.3^{+1.1}_{-1.3}$ & $0.27^{+0.17}_{-0.17}$ & \citet{montou2021}\\ 
257060897 & TIC 257060897 b & $11.26$ & $1.89\pm0.09$ & $6019\pm123$ & SG & $3.6600368^{+8.1e-06}_{-8e-06}$ & $2393.42489^{+0.00048}_{-0.00047}$ & $0.0852^{+0.0024}_{-0.0027}$ & $17.54^{+0.49}_{-0.55}$ & $5.66^{+0.51}_{-0.42}$ & $84.9^{+2.4}_{-1.7}$ & $0.5^{+0.12}_{-0.22}$ & \citet{montalto2022} \\ 
258920431 & TOI-2567 b & $11.75$ & $1.72\pm0.09$ & $5609\pm130$ & SG & $5.983912^{+2.5e-05}_{-2.5e-05}$ & $2396.7393^{+0.0012}_{-0.0011}$ & $0.05926^{+0.00081}_{-0.00076}$ & $11.1^{+0.15}_{-0.14}$ & $8.78^{+0.18}_{-0.44}$ & $88.75^{+0.87}_{-1.28}$ & $0.19^{+0.18}_{-0.13}$ & \citet{yee2022} \\ 
271893367 & TOI-150 b & $10.90$ & $1.68\pm0.07$ & $5983\pm109$ & SG & $5.857429^{+1.3e-05}_{-1.3e-05}$ & $2040.88514^{+0.00042}_{-0.00042}$ & $0.07889^{+0.00068}_{-0.00075}$ & $14.46^{+0.12}_{-0.14}$ & $7.63^{+0.28}_{-0.23}$ & $86.86^{+0.78}_{-0.57}$ & $0.419^{+0.061}_{-0.092}$ & \citet{kossakowski2019} \\ 
281731203 & K2-261 b & $9.96$ & $1.73\pm0.09$ & $5466\pm126$ & SG & $11.633722^{+7.1e-05}_{-7.3e-05}$ & $2267.9521^{+0.0016}_{-0.0015}$ & $0.05283^{+0.00119}_{-0.00095}$ & $10.0^{+0.23}_{-0.18}$ & $17.34^{+0.61}_{-1.52}$ & $89.13^{+0.61}_{-0.85}$ & $0.26^{+0.21}_{-0.18}$ & \citet{johnson2018}\\ 
284475976 & WASP-48 b & $11.15$ & $1.77\pm0.09$ & $6148\pm141$ & SG & $2.1436361^{+2.5e-06}_{-2.4e-06}$ & $2392.44558^{+0.00033}_{-0.00035}$ & $0.0944^{+0.0015}_{-0.0023}$ & $18.24^{+0.29}_{-0.45}$ & $4.84^{+0.32}_{-0.22}$ & $82.9^{+1.6}_{-1.0}$ & $0.601^{+0.054}_{-0.103}$ & \citet{enoch2011} \\ 
286923464 & HD 118203 b & $7.46$ & $2.05\pm0.11$ & $5837\pm140$ & SG & $6.13515^{+0.00025}_{-0.00025}$ & $2645.17912^{+0.00046}_{-0.00046}$ & $0.05667^{+0.00041}_{-0.00036}$ & $12.679^{+0.093}_{-0.082}$ & $8.65^{+0.11}_{-0.3}$ & $88.91^{+0.76}_{-1.04}$ & $0.16^{+0.15}_{-0.11}$ & \citet{dasilva2006}\\ 
322307342 & HATS-68 b & $11.78$ & $2.00\pm0.11$ & $5705\pm125$ & SG & $3.58615^{+0.00015}_{-0.00015}$ & $2063.657^{+0.0012}_{-0.0012}$ & $0.0757^{+0.0015}_{-0.0027}$ & $16.54^{+0.32}_{-0.59}$ & $5.94^{+1.17}_{-0.42}$ & $82.39^{+2.5}_{-0.95}$ & $0.786^{+0.036}_{-0.154}$ & \citet{hartman2019} \\ 
339672028 & TOI-481 b & $9.39$ & $1.73\pm0.09$ & $5803\pm134$ & SG & $10.33114^{+2.6e-05}_{-2.6e-05}$ & $2038.53073^{+0.00052}_{-0.0005}$ & $0.06309^{+0.00048}_{-0.00037}$ & $11.893^{+0.091}_{-0.07}$ & $14.61^{+0.2}_{-0.52}$ & $89.31^{+0.48}_{-0.61}$ & $0.18^{+0.14}_{-0.12}$ & \citet{brahm2020} \\ 
342642208 & TOI-2158 b & $10.17$ & $1.50\pm0.08$ & $5389\pm123$ & SG & $8.600729^{+5.8e-05}_{-6.3e-05}$ & $2397.3563^{+0.0018}_{-0.0015}$ & $0.0654^{+0.003}_{-0.0028}$ & $10.67^{+0.49}_{-0.46}$ & $17.6^{+2.4}_{-3.3}$ & $88.3^{+1.1}_{-1.2}$ & $0.52^{+0.21}_{-0.32}$ & \citet{knudstrup2022}\\ 
375942197 & WASP-176 b & $11.54$ & $1.86\pm0.10$ & $6152\pm134$ & SG & $3.89946^{+0.00046}_{-0.00046}$ & $2797.7001^{+0.0015}_{-0.0015}$ & $0.0786^{+0.0014}_{-0.0015}$ & $15.97^{+0.29}_{-0.31}$ & $6.0^{+0.23}_{-0.51}$ & $87.3^{+1.8}_{-2.4}$ & $0.28^{+0.21}_{-0.19}$ & \citet{cooke2020}\\ 
386259537 & WASP-169 b & $11.33$ & $2.21\pm0.11$ & $6158\pm130$ & SG & $5.6102^{+0.0013}_{-0.0013}$ & $2234.5457^{+0.0024}_{-0.0024}$ & $0.0745^{+0.002}_{-0.0018}$ & $17.98^{+0.49}_{-0.44}$ & $6.92^{+0.36}_{-0.74}$ & $87.3^{+1.8}_{-2.4}$ & $0.32^{+0.22}_{-0.22}$ & \citet{nielsen2019} \\ 
393414358 & WASP-63 b & $10.44$ & $1.79\pm0.10$ & $5715\pm129$ & SG & $4.378028^{+7.6e-05}_{-7.6e-05}$ & $2174.56842^{+0.00057}_{-0.00056}$ & $0.0798^{+0.00095}_{-0.00072}$ & $15.6^{+0.19}_{-0.14}$ & $6.56^{+0.14}_{-0.29}$ & $88.0^{+1.3}_{-1.5}$ & $0.22^{+0.15}_{-0.15}$ & \citet{hellier2012}\\ 
394722182 & TOI-2236 b & $10.87$ & $1.62\pm0.07$ & $6209\pm127$ & SG & $3.5315806^{+7.7e-06}_{-7.6e-06}$ & $2039.71045^{+0.00053}_{-0.00054}$ & $0.08614^{+0.00114}_{-0.00098}$ & $15.26^{+0.2}_{-0.17}$ & $6.84^{+0.34}_{-0.29}$ & $83.5^{+0.56}_{-0.5}$ & $0.774^{+0.025}_{-0.032}$ & \citet{yee2022} \\ 
422655579 & WASP-71 b & $10.16$ & $2.14\pm0.10$ & $6180\pm113$ & SG & $2.90354^{+9.3e-05}_{-9.4e-05}$ & $2449.75839^{+0.0008}_{-0.0008}$ & $0.06602^{+0.00103}_{-0.00086}$ & $15.43^{+0.24}_{-0.2}$ & $4.49^{+0.19}_{-0.33}$ & $86.0^{+2.7}_{-2.7}$ & $0.32^{+0.17}_{-0.2}$ & \citet{smith2013} \\ 
454248975 & TOI-3023 b & $11.48$ & $1.69\pm0.09$ & $5512\pm127$ & SG & $3.90158^{+0.00016}_{-0.00016}$ & $2313.0843^{+0.0011}_{-0.0011}$ & $0.09^{+0.0017}_{-0.0013}$ & $16.56^{+0.3}_{-0.24}$ & $6.36^{+0.15}_{-0.32}$ & $88.1^{+1.3}_{-1.8}$ & $0.21^{+0.18}_{-0.14}$ & \citet{yee2023} \\ 
459730973 & WASP-90 b & $11.07$ & $1.84\pm0.11$ & $6228\pm129$ & SG & $3.91669^{+0.0006}_{-0.00058}$ & $2799.3644^{+0.0019}_{-0.0018}$ & $0.0843^{+0.0043}_{-0.0056}$ & $16.91^{+0.87}_{-1.13}$ & $6.3^{+1.65}_{-0.79}$ & $82.8^{+2.7}_{-1.7}$ & $0.791^{+0.067}_{-0.169}$ & \citet{west2016} \\ 
176956893 & TOI-2184 b & $11.41$ & $3.15\pm0.17$ & $5721\pm123$ & ERG & $6.906686^{+8.6e-05}_{-8.2e-05}$ & $2036.6195^{+0.0023}_{-0.0024}$ & $0.04194^{+0.00124}_{-0.00092}$ & $14.41^{+0.43}_{-0.32}$ & $7.26^{+0.74}_{-1.18}$ & $86.4^{+2.4}_{-2.7}$ & $0.45^{+0.21}_{-0.29}$ & \citet{saunders2022} \\ 
230001847 & TOI-2337 b & $11.23$ & $3.23$ & $4784\pm122$ & ERG & $2.994327^{+3.7e-05}_{-3.8e-05}$ & $2392.6794^{+0.0033}_{-0.0034}$ & $0.0382^{+0.0018}_{-0.0022}$ & $13.44^{+0.65}_{-0.77}$ & $2.15^{+0.47}_{-0.25}$ & $67.7^{+7.6}_{-4.9}$ & $0.813^{+0.056}_{-0.152}$ & \citet{grunblatt2022tgt2} \\ 
394918211 & TOI-4377 b & $10.79$ & $3.41$ & $4913\pm122$ & ERG & $4.37702^{+0.00056}_{-0.00053}$ & $2337.6436^{+0.0036}_{-0.0038}$ & $0.03894^{+0.0012}_{-0.00099}$ & $14.5^{+0.45}_{-0.37}$ & $4.0^{+0.23}_{-0.49}$ & $85.0^{+3.5}_{-4.6}$ & $0.35^{+0.23}_{-0.24}$ & \citet{pereira2024}

%% file: planet_param_table_small_V3.tex
3317218 &  &  & $9.18$ & $1.85\pm0.09$ & $5646\pm204$ & SG & $2.77493^{+0.00075}_{-0.00066}$ & $2475.2606^{+0.0086}_{-0.0109}$ & $0.01298^{+0.00134}_{-0.00097}$ & $2.61^{+0.27}_{-0.2}$ & $4.08^{+0.68}_{-1.15}$ & $83.2^{+4.8}_{-9.0}$ & $0.49^{+0.31}_{-0.33}$ \\ 
169765334 &  & TOI-329 b & $10.69$ & $1.61\pm0.08$ & $5560\pm123$ & SG & $5.70446^{+0.00021}_{-0.0002}$ & $2090.7948^{+0.0045}_{-0.0053}$ & $0.0324^{+0.0027}_{-0.0022}$ & $5.69^{+0.47}_{-0.38}$ & $9.4^{+1.6}_{-3.0}$ & $86.9^{+2.2}_{-4.3}$ & $0.51^{+0.31}_{-0.34}$ \\ 
198186769 & 1291 &  & $10.87$ & $1.81\pm0.08$ & $6229\pm124$ & SG & $7.161999^{+0.000111}_{-9.3e-05}$ & $2397.8154^{+0.0033}_{-0.0039}$ & $0.03048^{+0.00117}_{-0.00077}$ & $6.02^{+0.23}_{-0.15}$ & $6.37^{+0.48}_{-1.17}$ & $86.5^{+2.5}_{-3.9}$ & $0.39^{+0.28}_{-0.27}$ \\ 
245893485 &  &  & $11.26$ & $1.70\pm0.09$ & $5559\pm131$ & SG & $1.38348^{+0.00083}_{-0.00083}$ & $2363.2784^{+0.0067}_{-0.0082}$ & $0.0408^{+0.0066}_{-0.0036}$ & $7.58^{+1.23}_{-0.67}$ & $3.5^{+0.64}_{-1.24}$ & $81.6^{+5.9}_{-13.9}$ & $0.52^{+0.34}_{-0.35}$ \\ 
298189384 &  &  & $11.45$ & $1.54\pm0.08$ & $5876\pm138$ & SG & $9.06049^{+0.00059}_{-0.00052}$ & $2425.744^{+0.015}_{-0.015}$ & $0.0388^{+0.0049}_{-0.0032}$ & $6.5^{+0.83}_{-0.54}$ & $8.3^{+1.8}_{-3.0}$ & $86.4^{+2.6}_{-5.7}$ & $0.53^{+0.32}_{-0.36}$ \\ 
350020859 &  & HD 89345 b & $8.75$ & $1.79\pm0.09$ & $5609\pm129$ & SG & $11.81458^{+0.00081}_{-0.00081}$ & $2532.3756^{+0.0016}_{-0.0016}$ & $0.0368^{+0.00115}_{-0.00079}$ & $7.17^{+0.22}_{-0.15}$ & $15.6^{+1.2}_{-2.5}$ & $88.57^{+0.99}_{-1.34}$ & $0.39^{+0.24}_{-0.26}$ \\ 
350033870 & 5177 &  & $8.85$ & $1.68\pm0.08$ & $6183\pm128$ & SG & $2.20703^{+0.00086}_{-0.0008}$ & $2526.907^{+0.011}_{-0.012}$ & $0.01375^{+0.00205}_{-0.0009}$ & $2.52^{+0.38}_{-0.17}$ & $3.01^{+0.5}_{-1.08}$ & $79.6^{+7.4}_{-16.8}$ & $0.55^{+0.33}_{-0.37}$ \\ 
366804698 & 5646 &  & $10.88$ & $1.93\pm0.11$ & $5999\pm134$ & SG & $2.4278^{+0.00013}_{-0.00011}$ & $2553.855^{+0.0041}_{-0.005}$ & $0.0334^{+0.0028}_{-0.0019}$ & $7.02^{+0.58}_{-0.39}$ & $4.74^{+0.62}_{-1.31}$ & $84.3^{+4.0}_{-7.5}$ & $0.47^{+0.31}_{-0.32}$ \\ 
381360757 & 5069 &  & $9.64$ & $1.87\pm0.09$ & $6332\pm126$ & SG & $8.8608^{+0.0027}_{-0.0029}$ & $2448.236^{+0.018}_{-0.014}$ & $0.0188^{+0.0025}_{-0.0015}$ & $3.84^{+0.51}_{-0.31}$ & $7.8^{+1.5}_{-2.9}$ & $86.2^{+2.8}_{-6.0}$ & $0.53^{+0.32}_{-0.36}$ \\ 
400093769 &  &  & $11.17$ & $1.69$ & $5843$ & SG & $3.6023^{+0.0017}_{-0.0016}$ & $2745.8379^{+0.0066}_{-0.0063}$ & $0.0425^{+0.0043}_{-0.0031}$ & $7.84^{+0.79}_{-0.57}$ & $8.3^{+1.6}_{-2.9}$ & $86.4^{+2.6}_{-5.3}$ & $0.53^{+0.31}_{-0.35}$ \\ 
408618999 &  & TOI-1736 b & $8.33$ & $1.41\pm0.07$ & $5656\pm123$ & SG & $7.0731^{+0.003}_{-0.0028}$ & $2726.4431^{+0.0037}_{-0.0036}$ & $0.0222^{+0.0018}_{-0.0012}$ & $3.42^{+0.28}_{-0.18}$ & $11.8^{+1.9}_{-4.4}$ & $87.6^{+1.8}_{-4.1}$ & $0.51^{+0.34}_{-0.35}$ \\ 
148340346 &  &  & $10.53$ & $2.73$ & $5441\pm122$ & ERG & $2.42332^{+0.00075}_{-0.00061}$ & $2204.2^{+0.0067}_{-0.0101}$ & $0.0195^{+0.0025}_{-0.0015}$ & $5.82^{+0.75}_{-0.45}$ & $4.03^{+0.57}_{-1.1}$ & $83.3^{+4.8}_{-8.9}$ & $0.48^{+0.31}_{-0.33}$